\documentclass[12pt]{amsart}
\usepackage{amssymb}
\usepackage{amsmath}
\usepackage{amsfonts}
\usepackage{mathrsfs}
\usepackage{graphicx}
\usepackage{float}
\usepackage{color}
\usepackage[onehalfspacing]{setspace}
\usepackage{fix-cm}
\usepackage{epigraph}
\usepackage{natbib}
\usepackage{placeins}
\usepackage{afterpage}
\usepackage[utf8]{inputenc}
\usepackage{charter}
\usepackage{epigraph}
\usepackage[colorlinks=true,citecolor=blue,urlcolor=blue,pdfpagemode=UseNone,pdfstartview=FitH]{hyperref}
\usepackage{apptools}
\usepackage{chngcntr}
\AtAppendix{\counterwithin{lemma}{section}}

\makeatletter
\def\section{\@startsection{section}{1}
	\z@{1.0\linespacing\@plus\linespacing}{.8\linespacing}{\Large}}

\def\subsection{\@startsection{subsection}{2}
	\z@{.8\linespacing\@plus.7\linespacing}{.7\linespacing}{\large}}

\def\subsubsection{\@startsection{subsubsection}{3}
	\z@{.5\linespacing\@plus.7\linespacing}{-.5em}{\normalfont\bfseries}}
\makeatother

\numberwithin{equation}{section}

\newtheorem{theorem}{Theorem}[section]
\newtheorem{lemma}{Lemma}[section]
\newtheorem{corollary}{Corollary}[section]

\theoremstyle{definition}
\newtheorem{definition}{Definition}[section]

\theoremstyle{definition}
\newtheorem{assumption}{Assumption}[section]

\theoremstyle{definition}

\title{}
\begin{document}
	\vspace*{5ex minus 1ex}
	\begin{center}
		\LARGE \textsc{Ordering-Free Inference from Locally Dependent Data}
		\bigskip
	\end{center}
	
	\date{%
		\today%
	}

	\vspace*{3ex minus 1ex}
	\begin{center}
		\medskip
		
		Kyungchul Song\\
		\textit{Vancouver School of Economics, University of British Columbia}\\
		\medskip
	\end{center}
	
	\fontsize{13}{14} \selectfont
	
	\begin{abstract}
		{\footnotesize
			This paper focuses on a data-rich environment where the data set has a very large cross-sectional dimension, is likely to exhibit local dependence, and yet is hard to determine the dependence ordering. Such a situation arises, for example, when the data set is collected from the Internet, through a method of web crawling. This paper proposes an approach of randomized subsampling inference, where one constructs a test statistic by aggregating many randomized test statistics using random draws of subsamples, and uses for inference the conditional distribution of the test statistic given data. This paper explores two approaches of such inference: one based on an M-type statistic constructed from randomized mean statistics and the other based on a U-type statistic constructed from randomized U-statistics. This paper provides conditions for local dependence, the number of the random draws, and the subsample size, under which randomized subsampling inference is asymptotically valid. From the Monte Carlo simulation studies, this paper finds that the randomized subsampling inference based on the U-type statistics performs better than that based on the M-type statistics.}\bigskip\
		
		{\footnotesize \noindent \textsc{Key words.} Randomized Subsampling Inference; Local Dependence; Cross-Sectional Dependence; Ordering-Free Inference; Randomized Tests \bigskip\ }
		
		{\footnotesize \noindent \textsc{JEL Classification: C01, C12, C13}}
	\end{abstract}
	
	\thanks{I thank Peter Phillips for encouragements in this line of research and for valuable comments and advice. I also thank the Co-Editor and the referees for their detailed criticisms and suggestions. I am grateful to Xiaohong Chen, Christian Gourieroux, Michael Jansson, and Jim Powell for their valuable comments. I thank seminar participants at the University of British Columbia (Department of Statistics), University of California at Berkeley, the University of Toronto, and Yale University for questions and comments. I thank Denis Kojevnikov for excellent research assistance. All errors are mine. I acknowledge financial support from Social Sciences and Humanities Research Council of Canada. Corresponding Address: Kyungchul Song, Vancouver School of Economics, University of British Columbia, 6000 Iona Drive, Vancouver, V6T 1L4, Canada.}
	\maketitle
	
	\normalsize
	
	\epigraph{\footnotesize \textit{At Google, for example, I have found that random samples on the order of 0.1 percent work fine for analysis of business data.}}{\footnotesize \textit{\cite{Varian:14:JEP}}}
	
	\section{Introduction}
	
	Collecting data from the Internet has recently been among the common practices in empirical research. Typically, such data sets have a huge cross-sectional dimension, and are collected using a sampling scheme that is far from random sampling. The sampling often relies on various methods of web crawling where a computer code automatically searches through numerous websites following a set of protocols. In this set-up, if one simply assumes that the observations are i.i.d., the inference may give nearly zero standard errors, making it irrelevant to perform statistical inference. (See \cite{Granger:88:SN}.) On the other hand, incorporating a dependence structure in statistical inference is far from trivial due to the use of a non-random sampling scheme often consisting of complex, and ad hoc protocols.
	
	The issue with unknown dependence ordering is not confined to such online data. For example, if the cross-sectional units are people in a social network, their observed actions are likely to be correlated, but it is hard for the researcher to determine the very network which shapes the cross-sectional dependence ordering of the observations. Another example is a situation where the cross-sectional units are differentiated products and observations are their sales in the market. Due to substitutability among the products, the sales should be correlated, and yet, the precise substitution pattern of the products is notoriously hard to determine from data.
	
	To the best of the author's knowledge, there does not exist a method of formal statistical inference that is generally available in such a situation. The existing methods of asymptotic inference based on normal approximation do not apply, because it is not possible to make the test statistic asymptotically pivotal without knowing the pattern of local dependence ordering. Inference based on various existing bootstrap methods such as nonparametric bootstrap, residual-based bootstrap or multiplier bootstrap does not apply here.\footnote{Nonparametric bootstrap and multiplier bootstrap do not apply due to the local dependence (and potentially distributional heterogeneity) of the cross-sectional observations. Residual bootstrap does not either unless one has a way to recover dependence ordering from data.} The method of subsampling does not apply either, because without properly incorporating the dependence pattern in the subsamples, there is no way the distribution of subsamples can imitate the distribution of the whole sample.
	
	In such a situation, this paper proposes using what this paper calls \textit{randomized subsampling inference}. The main idea is to construct a randomized test statistic by aggregating random subsamples. Randomized subsampling inference is a general idea which is easy to use. It is \textit{ordering-free inference} in the sense that it enables one to perform statistical inference without committing oneself to a particular model of the way cross-sectional dependence ordering arises among the sample units. For example, suppose that the cross-sectional observations have the same mean and we are interested in testing whether the mean is zero or not. For each randomly drawn subsample, we construct a two-sided t-test statistic. To enhance the power of the test, we build a randomized test statistic by aggregating such t-test statistics. Finally, one constructs many such randomized test statistics and uses their empirical distribution to construct critical values.
	
	The intuition behind this approach is straightforward. As long as the cross-sectional dimension is high and the dependence is local (in the sense that each cross-sectional unit has only a small set of other units that it is strongly correlated with), the observations in each subsample will be likely to be nearly independent, and under proper conditions, this independence will be maintained even after a certain level of aggregation of those subsamples, regardless of what dependence ordering the local dependence of the original observations follows. Hence the distribution of the randomized test statistic may be approximated by a distribution in a way that does not invoke knowledge of the local dependence ordering.
	
	The main goal of this paper is to formally study the method of randomized subsampling inference, focusing on the simple task of testing for the population mean, though the results can be directly extended to more general set-ups. As a first step, this paper formalizes the notion of \textit{local dependence} in a way that is proper to this type of study. This paper introduces a local dependence measure that is \textit{ordering-free} in the sense that the set of observations has the same local dependence measure as any permutation of the observations. Roughly speaking, the local dependence measure is based on the likelihood that any randomly chosen small subset of observations turn out to be strongly dependent with one another. When the observations are dependent only locally, this likelihood should be small, yielding a low measure of local dependence. As shown in this paper, a bound for the ordering-free measure of local dependence can be derived from various notions of temporal and spatial weak dependence such as strong mixing time series or random fields, $m$-dependent series, observations with a dependency graph, and observations with a group-dependence structure such that within-group dependence is permitted but between-group dependence is not.
	
	This paper studies two approaches of constructing a test statistic. The first approach uses an M-type statistic which is based on the sample mean of randomized subsamples. The second approach uses a U-type statistic which is based on the U-statistic of randomized subsamples. The main results of this paper are two-fold. First, we establish conditions for local dependence, the number of the randomly drawn subsamples, and the subsample sizes such that the test statistics constructed without knowledge of local dependence ordering are asymptotically pivotal. Second, we introduce the notion of \textit{rate-dominance with size control} to compare the two approaches. The comparison is not trivial because their size properties are potentially different. This rate-dominance result compares the two approaches in terms of power after the size is controlled up to a higher order term. This paper shows that the U-type statistic approach rate-dominates the M-type statistic approach when permutation-based critical values are used.
	
	This paper studies the performance of the two types of the test statistics through Monte Carlo simulations. The Monte Carlo study considers two types of local dependence; dependency graph and approximate dependency graphs. The study finds the superior performance of randomized subsampling inference based on the U-type statistic approach. In particular, as compared to the normal approximation, it performs more stably across a wide range of local dependence configurations. When the finite sample size is controlled, using the U-type statistic shows higher power than using the M-type statistic.
	
	There are two potential drawbacks of the randomized subsampling inference approach. First, the randomness of the test statistic in the use of inference is primarily from the randomness of the randomized subsamples, which means that given the same sample, the method can yield two different results though with a small probability. To overcome this issue, one may perform randomized inference many times and report the distribution of such results. More specifically, this paper follows \cite{Geyer/Meeden:05:SS} and proposes what it calls a \textit{randomized confidence function} which traces out the average non-rejection probabilities of many randomized tests across different parameter values. Another potential limitation comes from the fact that the convergence rate of the randomized subsampling inference is slower than the convergence rate of the estimation (e.g. the convergence rate of the sample mean). This paper regards the slower rate as a price to pay for performing inference that is robust to any form of local dependence ordering.	When the cross-sectional dimension is high, some loss of power in statistical inference to gain more robustness may not be so detrimental.
	\medskip
	
    \noindent \textbf{Literature Review: } The early literature of robust inference under weak dependence and heteroskedasticity has focused on consistent estimation of asymptotic covariance matrix (\citet*{White:80:Eca}, \citet*{Newey/West:87:Eca} and \citet*{Andrews:91:Eca}). A more recent strand of literature uses inconsistent scale normalizer that uses a fixed smoothing parameter (\citet*{Kiefer/Vogelsang/Bunzel:2000:Eca} and \citet*{Kiefer/Vogelsang:2002:Eca}).  For example, \citet*{Phillips/Sun/Jin:07:JSPI} and \citet*{Sun/Phillips/Jin:11:ET} considered power kernels. \citet*{Jansson:72:Eca} and \citet*{Sun/Phillips/Jin:08:Eca} explored higher order accuracy of fixed smoothing asymptotics over increasing smoothing asymptotics. \citet*{Sun:14:Eca} established related results in a more general context of two-step GMM estimation. In the meanwhile, \citet*{Muller:07:JOE} showed non-robustness of HAC estimators to local perturbations of the DGP and proposed a class of quadratic long-run variance estimators. \citet*{Sun/Kim:15:ReStat} proposed asymptotic F tests based on fixed smoothing asymptotics in the GMM framework with weakly dependent random fields. Also apart from this econometrics literature, it is worth noting that \citet*{Shao/Politis:13:JRSS} applied the fixed smoothing approach to subsampling inference that uses a calibration method.
    
    Robust estimation of asymptotic covariance matrix has received a great deal of attention in the literature of linear panel models and spatial models as well. \citet*{Arellano:87:OBES} proposed HAC estimation in linear panel models with fixed effects. \citet*{Driscoll/Kraay:98:ReStat} suggested a simple approach of HAC estimation in linear panel models based on cross-sectional averages of moment functions. \citet*{Vogelsang:12:JOE} compares the two approaches using fixed smoothing asymptotics. See also \citet*{Kelejian/Prucha:2007:JOE} and \citet*{Kim/Sun:2011:JOE} for HAC estimation in spatial models.
    
    A closely related approach in dealing with cross-sectional dependence is linear or nonlinear modeling of spatial autoregressive models and the approach of clustered errors. (See \cite{Lee:04:Eca}, and \cite{Lee/Liu/Lin:10:EJ} for contributions to the literature of spatial autoregressive models, among many others, and references therein.) As for models with clustered errors, \citet*{Cameron/Gelbach/Miller:08:ReStat} proposed and studied bootstrap approaches to deal with clustered errors. This literature typically assumes many independent clusters in a linear set-up. \citet*{Ibragimov/Muller:10:JBES} proposed a novel approach based on t-statistics where the observations are divided into multiple clusters that are (approximately) independent from each other and the inference is based on within-cluster estimators that are asymptotically normal. \citet*{Bester/Conley/Hansen:11:JOE} elaborated this multiple-cluster approach in linear panel models and provided conditions that are more primitive than those of \citet*{Ibragimov/Muller:10:JBES}. The t-statistic approach of \citet*{Ibragimov/Muller:10:JBES} allows for the clusters to be few and to be heterogeneous on various dimensions such as size or within-cluster dependence strength. However, the approach requires knowledge of this group structure of (at least approximately) independent clusters.
    
    A strand of literature has focused on developing tests for cross-sectional dependence using mainly cross-sectional variations. \citet*{Pesaran:04:WP} developed a general test for cross-sectional dependence in linear panel models with a short time series dimension. See also \citet*{Hsiao/Pesaran/Pick:12:OBES} for an extension to limited dependent models. \citet*{Robinson:08:JOE} proposed a correlation test that can be applied for testing cross-sectional dependence in a spatial model. \citet*{Kuersteiner/Prucha:2013:JOE} considered a linear panel model with a large cross-sectional dimension which does not require a long time series. Adopting a sequential exogeneity condition, and assuming conditional moment type restrictions, they obtained a limit theory for GMM estimators that accommodate unknown common shocks and various latent cross-sectional dependence structures. (See also \citet*{Kuersteiner/Prucha:2015:WP} for a more general framework which includes linear quadratic moment restrictions in dynamic panel models which accommodate social interactions and networks models.)
    
    The use of permutations and subsamples in this paper is different from permutation tests and subsampling-based inference in the literature. Most importantly, the main use of permutations or subsamples in this paper is for constructing test statistics rather than for finding critical values. It is more like a Monte Carlo test than a standard permutation test. Also, randomized subsampling inference is fundamentally different from the subsampling inference of \cite{Politis/Romano:94:AS}. Here, subsamples are used primarily to construct a test statistic rather than critical values.
    
    Using randomly drawn subsamples for data analysis is a common practice in empirical research when the data set size is huge. (See \cite{Varian:14:JEP}.) A closely related approach is bootstrap aggregating (or bagging) proposed by \cite{Breiman:96:ML} which is used to obtain stable predictions using many bootstrap samples. Recently \cite{Kleiner/Talwalker/Sarkar/Jordan:2014:JRSS} proposed using random subsamples to measure the quality of the estimates adn showed the measure is consistent as the subsample size increases to infinity. The main motivation for their proposal is to reduce the computational costs when the sample size is large. Unlike their approach, the focus here is on inference with observations that are locally dependent but the dependence ordering is not known to the researcher. It does not seem to have received attention in the literature that using randomly drawn subsamples one may obtain inference that is robust to a wide range of configurations of local dependence ordering. 
    \medskip
    
    \noindent \textbf{Organization of the Paper:} Section 2 introduces the main idea of randomized subsampling inference, ordering-free local dependence measure, and illustrates its meaning through examples. The section also establishes conditions for asymptotic validity, and provide results that compare the M-type statistic and the U-type statistic approaches through the notion of rate-dominance with size control. In Section 3, this paper presents and discusses results from a Monte Carlo simulation study. Technical proofs of part of the main results are found in the appendix. Supplemental Note to this paper contains extension to inference from moment-based restrictions and proofs of the other results in the paper.
    
	\section{Randomized Subsampling Inference on the Mean}
	\subsection{The Basic Set-Up}
	Consider the simple set-up of estimating the population mean from locally dependent data. Suppose that we are given observed random vectors $X_{1,n},...,X_{n,n} \in \mathbf{R}^d$ and a (potentially latent) common shock $\mathcal{C}_n$ such that
	\begin{eqnarray*}
		\mathbf{E} [X_{i,n}] = \mathbf{E} [X_{i,n}|\mathcal{C}_n],
	\end{eqnarray*}
	for all $i \in \{1,...,n\}$. Thus we assume that $X_{i,n}$ is conditionally mean independent of the common shock $\mathcal{C}_n$. Let us assume that $X_{i,n}$'s have the same mean:
	\begin{eqnarray*}
	\mu_0 \equiv \mathbf{E} X_{i,n}.	
	\end{eqnarray*}
	The main goal here is to develop a procedure to yield an asymptotically valid confidence set for $\mu_0$ as the sample size $n$ goes to infinity. For this, consider the following testing problem of null and alternative hypotheses.
	\begin{eqnarray*}
		&& H_0: \mu_0 = \bar \mu, \textnormal{ against }\\
		&& H_1: \mu_0 \neq \bar \mu,
	\end{eqnarray*} 
	for a given vector $\bar \mu$.
	
	Without knowing the local dependence ordering, it is not possible to use the usual t-statistic. The t-statistic (in the case of $d=1$) takes the following form:
	\begin{eqnarray*}
		\left|\frac{\sqrt{n} \overline X_n}{\hat \sigma}\right|,
	\end{eqnarray*}
where $\overline X_n = \frac{1}{n}\sum_{i=1}^n X_{i,n}$ and $\hat \sigma^2$ is a consistent estimator of $\sigma^2$ such that
\begin{eqnarray*}
	\sqrt{n} (\overline X_n - \mu_0) \rightarrow_d N(0,\sigma^2).
\end{eqnarray*} 
The difficulty here is that without knowing which observations are strongly correlated with $X_{i,n}$ for each $i \in \{1,...,n\}$, it is hard to find a consistent estimator of $\sigma^2$. In fact, we can write
\begin{eqnarray*}
	\sigma^2 = \lim_{n \rightarrow \infty} \sigma_n^2,
\end{eqnarray*}
where
\begin{eqnarray*}
	\sigma_n^2 = \frac{1}{n}\sum_{i=1}^n \text{Var}(X_{i,n})
	+ \frac{1}{n}\sum_{i=1}^n \sum_{j=1:j \ne i}^n \text{Cov}(X_{i,n},X_{j,n}).
\end{eqnarray*}
Under local dependence, one can consistently estimate the leading term in the definition of $\sigma_n^2$ without knowledge of dependence ordering. The problem is the second term for which, to the best of the author's knowledge, there is no existing method to consistently estimate it without knowing which set of $X_{j,n}$'s are nearly uncorrelated with $X_{i,n}$ for each $i$. Thus performing statistical inference without knowing local dependence ordering is fundamentally a nontrivial task, even for this simple testing problem for a population mean.
	
	\subsection{Two Approaches of Randomized Subsampling Inference}
	
	\subsubsection{The M-Type Statistic Approach}
	Let $\Pi$ be the set of permutations on $\{1,...,n\}$ and $\pi_1,...,\pi_R$ be i.i.d. draws from the uniform distribution on $\Pi$. Given each $\pi_r \in \Pi$, $r=1,...,R_n$, and $\bar \mu \in \mathbf{R}^d$, we define
	\begin{eqnarray*}
		M_n(\bar \mu;\pi_r) = \frac{1}{\sqrt{d b_n}} \sum_{i = 1}^{b_n} {\hat \Sigma}^{-1/2} \left(X_{\pi_r(i),n} - \bar \mu \right),
	\end{eqnarray*}
	where
	\begin{eqnarray}
	    \label{sample cov}
		\hat \Sigma = \frac{1}{n} \sum_{i=1}^n (X_{i,n} - \overline{X}_n)(X_{i,n} - \overline{X}_n)',
	\end{eqnarray}
	and $b_n \le n$. We call $\{X_{\pi_r(i),n}\}_{i =1 }^{b_n}$ \textit{a randomized subsample} and the number $b_n$ \textit{the subsample size}. A randomized subsample is obtained by first permuting the sample $\{X_{i,n}\}_{i=1}^n$ using a permutation randomly drawn from $\Pi$ and by taking the first $b_n$ observations from the permuted sample. The normalization by the covariance matrix $\hat \Sigma$ controls only for the ``short-run" covariance among the entries of the random vector $X_{i,n}$, not the cross-sectional dependence. Then we define
	\begin{eqnarray*}
		S_{M,n}(\bar \mu;\boldsymbol{\pi}) =  \left\|\frac{1}{\sqrt{R_n}} \sum_{r = 1}^{R_n} M_n(\bar \mu;\pi_r) \right\|^2,
	\end{eqnarray*}
	where $\boldsymbol{\pi} = (\pi_1,...,\pi_{R_n})$ and $||a||^2 = a'a$ for a vector $a$. We introduce the test statistic as follows:	
	\begin{eqnarray*}
		T_{M,n}(\bar \mu;\boldsymbol{\pi}) = S_{M,n}(\bar \mu;\boldsymbol{\pi}) - \frac{R_n b_n}{n}.		
	\end{eqnarray*}
	The additive term $R_n b_n /n$ is a bias adjustment term which we will explain later.
	
	Let us consider the following method of constructing a critical value. 
	For given $L$, we draw $ \boldsymbol{\pi}_l$, $l=1,...,L$, i.i.d., where $\boldsymbol{\pi}_l = (\pi_{1,l},...,\pi_{R_n,l})$ and $\pi_{r,l}$'s are i.i.d. draws from the uniform distribution on $\Pi$. Define for $\alpha \in (0,1)$,
	\begin{eqnarray*}
		c_{M,\alpha}= \inf \left\{c>0: \frac{1}{L} \sum_{l=1}^{L} 1\left\{S_{M,n}(\bar \mu;\boldsymbol{\pi}_l) > c\right\} \le \alpha \right\}.
	\end{eqnarray*}
	In other words, the critical values are read from the $1-\alpha$ percentile of the empirical distribution of $\{S_{M,n}(\bar \mu;\boldsymbol{\pi}_l)\}_{l=1}^L$. Note that we use the sample mean $\overline{X}_n$ in place of $\bar \mu$ to ensure that the test may have power when $\mu_0 \ne \bar \mu$. 
	
	\subsubsection{The U-Type Statistic Approach}
	As before, let $\Pi$ be the space of permutations on $\{1,...,n\}$ and $\pi_1,...,\pi_R$ be i.i.d. draws from the uniform distribution on $\Pi$. Given $\pi_r \in \Pi$ and $\bar \mu \in \mathbf{R}^d$, we define
	\begin{eqnarray*}
		U_n(\bar \mu;\pi_r) = \frac{1}{d b_n} \sum_{i,j =1: i \ne j}^{b_n} \left(X_{\pi_r(i),n} - \bar \mu \right)'{\hat \Sigma}^{-1} \left(X_{\pi_r(j),n} - \bar \mu \right),
	\end{eqnarray*}
	where $\hat \Sigma$ is as defined in (\ref{sample cov}), and let
	\begin{eqnarray*}
		S_{U,n}(\bar \mu;\boldsymbol{\pi}) = \frac{1}{\sqrt{R_n}}\sum_{r=1}^{R_n} U_n(\bar \mu;\pi_r).
	\end{eqnarray*}
	We introduce the test statistic as follows:	
	\begin{eqnarray*}
		T_{U,n}(\bar \mu;\boldsymbol{\pi}) = S_{U,n}(\bar \mu;\boldsymbol{\pi}) - \frac{\sqrt{R_n} (b_n-1)}{n}.		
	\end{eqnarray*}
	Again, the term $\sqrt{R_n}(b_n-1)/n$ is a bias adjustment term.
	
	For given $L$, we draw $\boldsymbol{\pi}_l$, $l=1,...,L$, i.i.d., similarly as before, where $\boldsymbol{\pi}_l = (\pi_{1,l},...,\pi_{R_n,l})$ and $\pi_{r,l}$'s are i.i.d. draws from the uniform distribution on $\Pi$. Define for $\alpha \in (0,1)$,
	\begin{eqnarray*}
		c_{U,\alpha} = \inf \left\{c>0: \frac{1}{L} \sum_{l=1}^{L} 1\{S_{U,n}(\overline{X}_n;\boldsymbol{\pi}_l)>c\} \le \alpha \right\}.
	\end{eqnarray*}
	
	\subsection{Ordering-Free Local Dependence Measure}
	We first formulate the notion of ``locality" in local dependence without invoking dependence ordering. The main idea in this paper is to measure the locality by quantifying the likelihood that a random selection of a small subset of $\{X_{i,n}\}_{i=1}^n$ gives a set that is partitioned into two nearly independent subsets. If the likelihood is large, the underlying dependence is deemed local. This notion of locality does not invoke any underlying dependence ordering.
	
	To formalize this intuition, suppose that for any subset $A \subset \{1,2,...,n\}$, $c_n(A)$ measures the strength of the joint conditional dependence of $X_{A,n} = (X_{i,n})_{i \in A}$ given a common shock $\mathcal{C}_n$. We will introduce one definition of $c_n(\cdot)$ later. Then, the ordering-free local dependence measure of the triangular array $\{X_{i,n}\}_{i=1}^n$ is defined as follows: for each $k \in \{1,...,n\}$,
	\begin{eqnarray}
	\label{lambda coeff}
	\lambda_n(k) = \frac{1}{|\Pi|} \sum_{\pi \in \Pi} c_n(\{\pi(1),...,\pi(k)\}).
	\end{eqnarray}
    The local dependence of a large set of  observations is measured by the convergence rate of $\lambda_n(k)$ (with each fixed $k$) to zero as the sample size $n$ goes to infinity. We call $\lambda$ the $\lambda$-\textit{coefficient} of $\{X_{i,n}\}_{i=1}^n$ (with respect to a given common shock $\mathcal{C}_n$). The $\lambda$-coefficient is \textit{ordering-free}. Indeed, two triangular arrays $\{X_{i,n}\}_{i=1}^n$ and $\{Y_{i,n}\}_{i=1}^n$ have the same $\lambda$-coefficient if $X_{i,n} = Y_{\pi(i),n}$ for all $i \in \{1,...,n\}$, for some $\pi \in \Pi$.
	
	Let us introduce $c_n(\cdot)$. First, let for each $A \subset \{1,...,n\}$
	\begin{eqnarray*}
		\mathcal{P}(A) \equiv \left\{\{A_1,A_2\}: A_1 \cup A_2 = A, A_1 \cap A_2 = \varnothing, \textnormal{ and } A_1,A_2 \ne \varnothing \right\},
	\end{eqnarray*}
	i.e., the collection of the sets $\{A_1,A_2\}$ of two nonempty subsets $A_1,A_2$ of $\{1,...,n\}$ which constitute a partition of $A$. Let $\mathcal{F}$ be the union of $\mathcal{F}_l$'s over $l \in \{1,...,n\}$, where $\mathcal{F}_l$ is a given class of real measurable functions on $\mathbf{R}^{ld}$. (For this paper's proposal, it suffices to take the class of functions $\mathcal{F}_l$ as a finite set of eighth-order polynomials. See a remark prior to Assumption \ref{assump: local dep M} below.) For each $A \subset \{1,2,...,n\}$, we take
	\begin{eqnarray*}
		c_n(A) = \underset{\{A_1,A_2\} \in \mathcal{P}(A)}{\min} c_n(A_1,A_2),
	\end{eqnarray*}
	where
	\begin{eqnarray*}
		c_n(A_1,A_2) = \sup_{(f_1,f_2)\in \mathcal{F}_{|A_1|} \times \mathcal{F}_{|A_2|}} \left| Corr \left(f_1(X_{A_1,n}),f_2(X_{A_2,n})|\mathcal{C}_n\right) \right|,
	\end{eqnarray*}
    and $Corr(X,Y|\mathcal{C}_n)$ represents the conditional (Pearson) correlation coefficient between $X$ and $Y$ given $\mathcal{C}_n$ if $Var(X|\mathcal{C}_n)Var(Y|\mathcal{C}_n)>0$ and zero otherwise. (The minimum over an empty set in the above expression is taken to be one.) For example, when there exists $i \in A$ such that $X_{i,n}$ is conditionally independent of $\{X_{j,n}: j \in A \backslash \{i\} \}$ given $\mathcal{C}_n$, we have $c_n(A) = 0$. Hence when $X_{i,n}$'s are conditionally independent, $\lambda_n(k)=0$, for any $k\ge 1$.
	
	The notion of $\lambda$-coefficient here is inspired by $\psi$-weak dependence notion proposed by \citet*{Doukhan/Louhichi:99:SPA} for time series. The major distinction is that the dependence measure is made invariant to the permutations of the observations, making unnecessary any reference to the underlying dependence ordering.
	\medskip
	
	\noindent \textbf{Example 1: Triangular Arrays with a Dependency Graph} Let a graph $G_n = (\{1,...,n\},E_n)$ over $\{1,...,n\}$ be given, where $E_n$ denotes the collection of pairs $ij$, with $i,j \in \{1,...,n\},$ representing an edge (or a link) between vertices (or nodes) $i$ and $j$. We exclude loops, i.e., for all $ij \in E_n$, $i \ne j$. Let us assume that the graph is undirected so that whenever $ij \in E_n$, $ji \in E_n$. Let $(X_{i,n})_{i =1}^n$ be a given triangular array of random variables. Let us say that $G_n$ is a \textit{conditional dependency graph} for $(X_{i,n})_{i =1}^n$ given a $\sigma$-field $\mathcal{C}_n$, if for any two subsets $A_1$ and $A_2$ of $\{1,...,n\}$ such that $\{ij \in E_n: i \in A_1,j \in A_2\} = \varnothing$, $(X_{i,n})_{i \in A_1}$ and $(X_{i,n})_{i \in A_2}$ are conditionally independent given $\mathcal{C}_n$. (See e.g. \cite{Penrose:03:RGP}, p.22.) 
	
	The case of a triangular array with a dependency graph having a bounded maximum degree includes $m$-dependent time series as a special case.\footnote{A \textit{degree} of a node $i$, denoted by $d_n(i)$, is the size of its neighborhood, i.e., $d_n(i)=|\{j\in \{1,...,n\}: ij \in E_n\}|$. The \textit{maximum degree} is $\max_{1 \le i \le n}d_n(i)$, i.e., the maximum of $d_n(i)$ over $i \in \{1,...,n\}$.} Also dependence with clusters, where there is a partition of the observations into clusters and dependence is restricted to within-cluster observations, not between clusters, is a special case of local dependence with a dependency graph.
	
	A simple combinatoric argument gives a bound for the $\lambda$-coefficient of the triangular array $(X_{i,n})_{i =1}^n$. (The proof is found in Supplemental Note.)
	
	\begin{lemma}
		\label{lemma: dep graph}
		Suppose that $\{X_{i,n}\}_{i=1}^n$ is a triangular array of random variables having $G_n = (\{1,...,n\},E_n)$ as a conditional dependency graph given $\mathcal{C}_n$.
		
		Then for each integer $k \ge 2$, there exist $C_k>0$ and $n_k\ge 1$ depending only on $k$ such that for all $n \ge n_k$,
		\begin{eqnarray*}
			\lambda_n(k) \le C_k n^{-k+1}d_{n}^{k-1},
		\end{eqnarray*}
		where $d_{n}$ denotes the maximum degree of $G_n$.
	\end{lemma}
	
	Suppose that $d_n$ is bounded. Then at any fixed $k \ge 2$, the $\lambda$-coefficient converges to zero as $n \rightarrow \infty$ at the rate of $n^{-k+1}$. The bound in Lemma \ref{lemma: dep graph} is conveniently simple, as it depends on the graph through only its maximum degree. It is possible to obtain a more sophicated bound that involves other features of the network. 
	\medskip
	
	\noindent \textbf{Example 2: Weakly Dependent Random Fields}
	Suppose that a random field $(Y_{j,n})_{j \in \mathbb{Z}'_d}$ is given, where $\mathbb{Z}'_d$ is a subset of $\mathbb{Z}_d$ such that $n = |\mathbb{Z}_d'|$ and $\mathbb{Z}_d$ is a lattice as a subset of $\mathbf{R}^d$. For each pair $(j_1,j_2) \in \mathbb{Z}_d \times \mathbb{Z}_d$, we define distance $d(j_1,j_2) = \max_{1\le k \le d} |j_{1,k} - j_{2,k}|$, where $j_{1,k}$ and $j_{2,k}$ are the $k$-th entry of $j_1$ and $j_2$ respectively. We assume that the lattice $\mathbb{Z}_d$ is infinite countable having $d_0>0$ such that for all $j_1,j_2 \in \mathbb{Z}_d$, we have $d(j_1,j_2) \ge d_0$. Hence as in \citet*{Conley:99:JOE} and \citet*{Jenish/Prucha:09:JOE}, we exclude the infill asymptotics where the sampling points become dense in a given domain as the sample size increases.
	
	To map the random field to a triangular array, let $\mu:\{1,...,n\} \rightarrow \mathbb{Z}'_d$ be a one-to-one map, so that we let $X_{i,n} = Y_{\mu(i),n}$, $i=1,...,n$, and define $d_\mu(i,j) = d(\mu(i),\mu(j))$ for  simplicity. Also, for given subsets $A,A'$ of $\{1,...,n\}$, let
	\begin{eqnarray*}
		d_\mu(A,A') = \underset{i \in A,j\in A'}{\min} d(\mu(i),\mu(j)).
	\end{eqnarray*}
	
	For simplicity, let us assume that there is no common shock. Define for $m \ge 0$,
	\begin{eqnarray*}
		\bar{c}_{m,n}(A) = \min_{\{A_1,A_2\} \in \mathcal{P}(A):m \le d_\mu(A_1,A_2)<m+1} c_n(A_1,A_2).
	\end{eqnarray*}
	Hence $\bar{c}_{m,n}(A)$ measures stochastic dependence among $X_{i,n}$'s with $i \in A$, when there exists a partition $(A_1,A_2)$ of $A$ such that $m \le d_\mu(A_1,A_2) \le m+1$. The following lemma characterizes a bound for the $\lambda$-coefficient of $\{X_{i,n}\}_{i=1}^n$ in terms of $\bar{c}_{m,n}(A)$. (The proof is found in Supplemental Note.)
	
	\begin{lemma}
		\label{lemma: random fields}
		Suppose that $(Y_{j,n})_{j \in \mathbb{Z}'_d}$ is a random field and let $X_{i,n} = Y_{\mu(i),n}$, $i=1,...,n.$ Suppose further that for some $q \ge 2$, and for each integer $2 \le k \le q$, there exists $C_{k,d}>0$ such that
		\begin{eqnarray}
		\label{cond}
		\sum_{m=1}^{\infty} m^{(k-1)(d-1)} \max_{\pi \in \Pi} \bar c_{m,n}(\{\pi(1),...,\pi(k)\}) \le C_{k,d}, \textit{ for all } n \ge 1.	
		\end{eqnarray}
		
		Then, for each integer $2 \le k \le q$, there exist $C_{k,d}'>0$ and $n_{k,d} \ge 1$ such that
		\begin{eqnarray*}
			\lambda_n(k) \le C_{k,d}' n^{-k+1}, \textit{ for all } n \ge n_{k,d},
		\end{eqnarray*}
	where $C_{k,d}'$ and $n_{k,d}$ are constants depending only on $k,d$ and $C_{k,d}$.
	\end{lemma}
	
	When the random field is a strong-mixing random field, the condition (\ref{cond}) can be verified in terms of the strong-mixing coefficient. To see this, define
	\begin{eqnarray}
	\label{strong mixing}
	\alpha(A_1,A_2) = \sup_{i \in A_1}\sup_{j \in A_2} \alpha(\sigma(X_{i,n}),\sigma(X_{j,n})),
	\end{eqnarray}
	where $\alpha(\sigma(X_{i,n}),\sigma(X_{j,n}))$ denotes the strong mixing coefficient between the two $\sigma$-fields $\sigma(X_{i,n}),\sigma(X_{j,n})$ generated by $X_{i,n}$ and $X_{j,n}$, i.e.,
	\begin{eqnarray*}
		\alpha(\sigma(X_{i,n}),\sigma(X_{j,n})) = \sup_{A,B} \left|P\{X_{i,n} \in A,X_{j,n} \in B\} - P\{X_{i,n} \in A\}P\{X_{j,n} \in B\}\right|, 
	\end{eqnarray*}
	with the supremum being that over all the Borel sets $A$ and $B$. Let for $p \ge 1$
	\begin{eqnarray}
	\label{H}
	H_{p,\mathcal{F}}(k) = \max_{\pi \in \Pi} \max_{1 \le a \le k} \sup_{f \in \mathcal{F}_a} \left(\mathbf{E}[|f(X_{\pi(1)},...,X_{\pi(a),n})|^p]\right)^{1/p}.
	\end{eqnarray}
	Suppose that for some $p>2$ and $q \ge 2$, $H_{p,\mathcal{F}}(k)<\infty$ for all $k \in \{1,...,q\}$. Then by covariance inequality (e.g. Corollary A.2 of \cite{Hall/Heyde:80:MartingaleLimitTheory}, p.278), for any disjoint subsets $A_1$ and $A_2$ of $\{1,...,n\}$ such that $k = |A_1|+|A_2|$, we have for some constant $C>0$,
	\begin{eqnarray*}
		c_n(A_1,A_2) \le C H_{p,\mathcal{F}}^2(k) \alpha(A_1,A_2)^{1-(2/p)},
	\end{eqnarray*} 
 The condition (\ref{cond}) is reduced to the requirement of the existence of a constant $C>0$ such that for all $n \ge 1$,
	\begin{eqnarray*}
		\sum_{m=1}^\infty m^{(k-1)(d-1)} \max_{\pi \in \Pi}\bar{\alpha}_{m,\pi}(k)^{1-(2/p)}<C,
	\end{eqnarray*}
	where
	\begin{eqnarray*}
		\bar{\alpha}_{m,\pi}(k)= \textnormal{  } \min_{\{A_1,A_2\} \in \mathcal{P}(\{\pi(1),...,\pi(k)\}):m \le d_\mu(A_1,A_2)<m+1} \alpha(A_1,A_2).
	\end{eqnarray*}
	
	\subsection{Asymptotic Theory}
	
	\subsubsection{Asymptotic Validity}
	
	We take the class of functions $\mathcal{F}$ to be the union of $\mathcal{F}_l$ over $l \in \{1,...,n\}$, where $\mathcal{F}_l$ is the collection of real maps $\phi$ on $\mathbf{R}^{ld}$ of the form: $\phi(x_1,...,x_l) =x_1^{k_1} x_2^{k_2}\cdot \cdot \cdot x_l^{k_l}$, where $|k_1|+\cdot \cdot \cdot+|k_l| \le 8$ with $k_v = (k_{v,1},...,k_{v,d})$, $x_v^{k_v} = x_{v,1}^{k_{v,1}}\cdot \cdot \cdot x_{v,d}^{k_{v,d}}$, $|k_v| = \sum_{k=1}^d k_{v,k}$, and $k_{v,1},...,k_{v,d} \in \{0,1,...,8\}$.  Define
	\begin{eqnarray*}
		\bar \lambda_n(k) = \mathbf{E}[\lambda_n(k)].
	\end{eqnarray*}
	We introduce two pairs of conditions in the assumption below, one for the M-type statistic approach and the other for the U-type statistic approach.
	\begin{assumption}
		\label{assump: local dep M} For a sequence $\varepsilon_n \rightarrow 0$ and a constant $C>0$ which do not depend on the joint distribution of $(X_{i,n})_{i =1}^n$, it is satisfied that for all $n \ge 1$,
		\medskip
		
		\noindent M-(i) $R_n b_n\{\bar \lambda_n(2)+n^{-1}\} \le \varepsilon_n $, and
		
		\noindent M-(ii) $b_n^k \lambda_n(k) \le C$, for each $2 \le k \le 4$.
	\end{assumption}
	
	\begin{assumption}
		\label{assump: local dep U} For a sequence $\varepsilon_n \rightarrow 0$ and a constant $C>0$ which do not depend on the joint distribution of $(X_{i,n})_{i =1}^n$, it is satisfied that for all $n \ge 1$,
		\medskip
		
		\noindent U-(i) $R_n b_n^2\{\bar \lambda_n(4) + n^{-1}(\bar \lambda_n(3)+\bar \lambda_n(2))+n^{-2}\} \le \varepsilon_n$, and
		
		\noindent U-(ii) $b_n^k \lambda_n(k) \le C$, for each $2 \le k \le 8$.
	\end{assumption}
	
	Assumptions \ref{assump: local dep M} and \ref{assump: local dep U} specify the requirement on the local dependence of the triangular array $(X_{i,n})_{i =1}^n$. For example in the case of a dependency graph with a bounded maximum degree, Assumption \ref{assump: local dep M} requires that
	\begin{eqnarray*}
		R_n b_n/n \rightarrow 0, \textnormal{ and } b_n^2/n = O(1).
	\end{eqnarray*}
	For Assumption \ref{assump: local dep M}, it suffices to take $R_n=\sqrt{n}$ and $b_n$ such that $b_n^2/n \rightarrow 0$ as $n \rightarrow \infty$. On the other hand, Assumption \ref{assump: local dep U} requires that
	\begin{eqnarray*}
		R_n b_n^2 /n^2 \rightarrow 0, \textnormal{ and } b_n^2/n = O(1).
	\end{eqnarray*}
	It suffices to take $R_n=n$ and $b_n$ such that $b_n^2/n \rightarrow 0$ as $n \rightarrow \infty$. Assumptions \ref{assump: local dep M} and \ref{assump: local dep U} do not require that $b_n \rightarrow \infty$. In fact $b_n$ can be chosen to be a fixed constant as $n \rightarrow \infty$.
	
	\begin{assumption}
		\label{assump: moment cond} There exist constants $C,c>0$ which do not depend on the joint distribution of $(X_{i,n})_{i =1}^n$ and satisfy the following conditions for all $n \ge 1$.
		\medskip
		
		\noindent (i) $\max_{1\le i \le n} \mathbf{E}[||X_{i,n}||^8|\mathcal{C}_n] < C$.
		
		\noindent (ii) The minimum eigenvalue of $\Sigma_n$ is greater than $c$, where
		\begin{eqnarray*}
			\Sigma_n \equiv \frac{1}{n} \sum_{i=1}^n \mathbf{E}[(X_{i,n} - \mathbf{E}[X_{i,n}|\mathcal{C}_n])(X_{i,n} - \mathbf{E}[X_{i,n}|\mathcal{C}_n])'|\mathcal{C}_n].
		\end{eqnarray*}
	\end{assumption}
	
	In Assumption \ref{assump: moment cond}, we require bounded moments and nondegenerate variances. The following theorem is the main result of this paper. Let
	\begin{eqnarray*}
		\mathscr{Z}_n = \sigma(\{X_{i,n}\}_{i=1}^n) \vee \mathcal{C}_n,
	\end{eqnarray*}
	i.e., the $\sigma$-field generated by $\{X_{i,n}\}_{i=1}^n$ and $\mathcal{C}_n$.
	
	\begin{theorem}
		\label{thm: Type U}
		\noindent (i) Suppose that Assumptions \ref{assump: local dep M} and \ref{assump: moment cond} hold. Then, as $n \rightarrow \infty$,
		\begin{eqnarray*}
		    \sup_{t \in \mathbf{R}}\left| P\left\{T_{M,n}(\mu_0;\boldsymbol{\pi}) \le t |\mathscr{Z}_n\right\} - G_0(t) \right| \rightarrow_P 0,
		\end{eqnarray*}
		where $G_0$ is the CDF of $\mathbb{Z}'\mathbb{Z}/d$, and $\mathbb{Z} \sim_d N(0,I_d)$.
		
		Furthermore, for each $\alpha \in (0,1)$, we have as $n,L \rightarrow \infty$,
		\begin{eqnarray*}
			P\{T_{M,n}(\mu_0;\boldsymbol{\pi})> c_{M,\alpha}\} \rightarrow \alpha.
		\end{eqnarray*}
		
		\noindent (ii) Suppose that Assumptions \ref{assump: local dep U} and \ref{assump: moment cond} hold. Then, as $n \rightarrow \infty$,
		\begin{eqnarray*}
			\sup_{t \in \mathbf{R}}\left| P\left\{T_{U,n}(\mu_0;\boldsymbol{\pi}) \le t |\mathscr{Z}_n\right\} - \Phi(t) \right| \rightarrow_P 0,
		\end{eqnarray*}
where $\Phi$ is the CDF of $N(0,1)$.
	
	Furthermore, for each $\alpha \in (0,1)$, we have as $n,L \rightarrow \infty$,
	\begin{eqnarray*}
		P\{T_{U,n}(\mu_0;\boldsymbol{\pi})> c_{U,\alpha}\} \rightarrow \alpha.
	\end{eqnarray*}
	\end{theorem}
	Both tests are asymptotically pivotal under the stated conditions, despite that they were constructed without knowledge about the underlying dependence ordering.
	
	\subsubsection{Heuristics and Discussions}
	\label{sec: heuristics}
	Let us give heuristics on the asymptotic validity of the randomized subsampling approach. We show how the statistics $S_{M,n}(\overline X_n;\boldsymbol{\pi})$ and $S_{U,n}(\overline X_n;\boldsymbol{\pi})$ which are used to generate critical values is linked to the test statistics $T_{M,n}(\mu_0;\boldsymbol{\pi})$ and $T_{U,n}(\mu_0;\boldsymbol{\pi})$. In doing so, we will see how the bias adjustment term arises. For simplicity, we assume that there is no common shock $\mathcal{C}_n$ and we know $\Sigma_n$. 
		
	    First, let us consider the M-type statistic approach.  Let $Z_{i,n} = \Sigma_n^{-1/2}(X_{i,n} - \mu_0)/\sqrt{d}$ and write 
	    \begin{eqnarray}
	    \label{S_M}
	    	S_{M,n}(\overline{X}_n;\boldsymbol{\pi}) &=& S_{M,n}(\mu_0;\boldsymbol{\pi}) + B_{M,n} \\ \notag
	    	&=& \frac{1}{R_n b_n} \sum_{r_1=1}^{R_n} \sum_{r_2=1}^{R_n} \sum_{i_1=1}^{b_n} \sum_{i_2=1}^{b_n} Z_{\pi_{r_1}(i_1),n}' Z_{\pi_{r_2}(i_2),n} + B_{M,n},
	    \end{eqnarray}
	    where $\overline{Z}_n = n^{-1} \sum_{i=1}^n Z_{i,n}$ and
	    \begin{eqnarray}
	    \label{B_Mn}
	    B_{M,n} = R_n b_n \overline{Z}_n'\overline{Z}_n - \frac{\overline{Z}_n'}{R_n b_n} \sum_{r_1=1}^{R_n} \sum_{r_2=1}^{R_n} \sum_{i_1=1}^{b_n} \sum_{i_2=1}^{b_n} \left(Z_{\pi_{r_1}(j),n} + Z_{\pi_{r_2}(i),n} \right).
	    \end{eqnarray}
	    As for the last term in the definition of $B_{M,n}$, note that
	    \begin{eqnarray*}
	    	\frac{\overline{Z}_n'}{R_n b_n} \sum_{r_1=1}^{R_n} \sum_{r_2=1}^{R_n} \sum_{i_1=1}^{b_n} \sum_{i_2=1}^{b_n} Z_{\pi_{r_1}(i_1)} &=& \overline{Z}_n' \sum_{r_1=1}^{R_n} \sum_{i_1=1}^{b_n} Z_{\pi_{r_1}(i_1)}\\
	    	&=& R_n b_n \overline{Z}_n'\overline{Z}_n + A_{M,n},
	    \end{eqnarray*}
	    where $A_{M,n} = \overline{Z}_n' \sum_{r_1=1}^{R_n} \sum_{i_1=1}^{b_n} (Z_{\pi_{r_1}(i_1)} - \overline{Z}_n).$ One can show that the conditional expectation of $A_{M,n}^2$ given $(Z_{i,n})_{i=1}^n$ is $O_P(R_n b_n/n)$. Thus,
	    \begin{eqnarray*}
	    	B_{M,n} = - R_n b_n \overline{Z}_n'\overline{Z}_n + O_P(\sqrt{R_n b_n/n}).
	    \end{eqnarray*}
	    We conclude that
	    \begin{eqnarray*}
	    	S_{M,n}(\overline{X}_n;\boldsymbol{\pi}) &=& S_{M,n}(\mu_0;\boldsymbol{\pi}) - R_n b_n\overline{Z}_n'\overline{Z}_n + O_P(\sqrt{R_n b_n/n}) \\
	    	&=& T_{M,n}^*(\mu_0;\boldsymbol{\pi}) - (R_n b_n /n) (n\overline{Z}_n'\overline{Z}_n -1 - n\lambda_n(2)) + O_P(\sqrt{R_n b_n/n}),
	    \end{eqnarray*}
	    where
	    \begin{eqnarray}
	    \label{T_M^*}
	    T_{M,n}^*(\mu_0;\boldsymbol{\pi}) = S_{M,n}(\mu_0;\boldsymbol{\pi}) - \frac{R_n b_n(1 + n\lambda_n(2))}{n}. 
	    \end{eqnarray}
	    Since $n\overline{Z}_n'\overline{Z}_n -1 = O_P(1)$, by Assumption \ref{assump: local dep M} M-(i), it follows that
	    \begin{eqnarray}
	    \label{approxM}
	    S_{M,n}(\overline{X}_n;\boldsymbol{\pi}) = T_{M,n}^*(\mu_0;\boldsymbol{\pi}) +o_P(1).
	    \end{eqnarray}
	    Since $S_{M,n}(\overline{X}_n;\boldsymbol{\pi})$ is a sum of i.i.d. random variables conditional on $(Z_{i,n})_{i=1}^n$ (due to the i.i.d. property of random permutations), we can apply the central limit theorem to establish the asymptotic distribution of $T_{M,n}^*(\mu_0;\boldsymbol{\pi})$.
	    
	    Using $T_{M,n}^*(\mu_0;\boldsymbol{\pi})$ as a test statistic is not feasible, because the last term $R_n b_n (1 + n\lambda_n(2))/n$ in (\ref{T_M^*}) is not consistently estimable without knowledge of the dependence ordering. Thus we take the case of $X_{i,n}$'s being independent as a benchmark in which case $\lambda_n(2)=0$, and replace $\lambda_n(2)$ by zero. We obtain the following form of a test statistic:
	    \begin{eqnarray}
	    \label{TM}
	    T_{M,n}(\mu_0;\boldsymbol{\pi}) = S_{M,n}(\mu_0;\boldsymbol{\pi}) - \frac{R_n b_n}{n}. 
	    \end{eqnarray}
	    Under Assumption \ref{assump: local dep M}, the bias adjustment term $R_n b_n/n$ is asymptotically negligible, as $n \rightarrow \infty$. However, including it stablizes the finite sample size properties of the test. The relation (\ref{approxM}) shows why using the conditional distribution of $S_{M,n}(\overline{X}_n;\boldsymbol{\pi})$ given data is an alternative way of obtaining the critical values, i.e., permutation-based ones $c_{M,\alpha}$.
	    
	    Now let us turn to the U-type statistic approach. Similarly as before, we write
		\begin{eqnarray*}
			S_{U,n}(\overline{X}_n;\boldsymbol{\pi}) = S_{U,n}(\mu_0;\boldsymbol{\pi}) + B_{U,n} = \frac{1}{\sqrt{R_n}} \sum_{r=1}^{R_n} \frac{1}{b_n} \sum_{i=1}^{b_n} \sum_{j=1,j \ne i}^{b_n} Z_{\pi_r(i),n}' Z_{\pi_r(j),n} + B_{U,n},
		\end{eqnarray*}
		where
		\begin{eqnarray}
		\label{B_Un}
		B_{U,n} = \sqrt{R_n} (b_n-1) \overline{Z}_n'\overline{Z}_n - \frac{\overline{Z}_n'}{\sqrt{R_n}} \sum_{r=1}^{R_n} \frac{1}{b_n} \sum_{i=1}^{b_n} \sum_{j=1,j \ne i}^{b_n} \left(Z_{\pi_r(j),n} + Z_{\pi_r(i),n} \right).
		\end{eqnarray}
		As for the last term in the definition of $B_{U,n}$, note that
		\begin{eqnarray*}
			\frac{\overline{Z}_n'}{\sqrt{R_n}} \sum_{r=1}^{R_n} \frac{1}{b_n} \sum_{i=1}^{b_n} \sum_{j=1,j \ne i}^{b_n} Z_{\pi_r(i)} = \overline{Z}_n'\overline{Z}_n\sqrt{R_n}(b_n-1) + A_{U,n},
		\end{eqnarray*}
		where
		\begin{eqnarray*}
		A_{U,n} = \frac{\overline{Z}_n'\sqrt{b_n}(b_n-1)}{b_n}\frac{1}{\sqrt{R_n b_n}} \sum_{r=1}^{R_n} \sum_{i=1}^{b_n} (Z_{\pi_r(i)} - \overline{Z}_n).	
		\end{eqnarray*}
		One can show that the conditional expectation of $A_{U,n}^2$ given $(Z_{i,n})_{i=1}^n$ is $O_P(b_n/n)$. Thus,
		\begin{eqnarray*}
			B_{U,n} = - \sqrt{R_n} (b_n-1) \overline{Z}_n'\overline{Z}_n + O_P(\sqrt{b_n/n}).
		\end{eqnarray*}
		Therefore, if we define
		\begin{eqnarray}
		\label{T_U^*}
		T_{U,n}^*(\mu_0;\boldsymbol{\pi}) = S_{U,n}(\mu_0;\boldsymbol{\pi}) - \frac{\sqrt{R_n} (b_n-1)(1 + n\lambda_n(2))}{n},
		\end{eqnarray}
		we can write 
		\begin{eqnarray*}
			S_{U,n}(\overline{X}_n;\boldsymbol{\pi}) = T_{U,n}^*(\mu_0;\boldsymbol{\pi}) - (\sqrt{R_n} (b_n-1)/n) (n\overline{Z}_n'\overline{Z}_n -1 - n\lambda_n(2)) + O_P(\sqrt{b_n/n}).
		\end{eqnarray*}
		By Assumptions \ref{assump: local dep U}, it follows that
		\begin{eqnarray}
		   \label{approxU}
			S_{U,n}(\overline{X}_n;\boldsymbol{\pi}) = T_{U,n}^*(\mu_0;\boldsymbol{\pi}) +o_P(1).
		\end{eqnarray}
		Again, since $S_{U,n}(\overline{X}_n;\boldsymbol{\pi})$ is a sum of i.i.d. random variables conditional on $(Z_{i,n})_{i=1}^n$, we can apply the central limit theorem to establish its asymptotic normality of $T_{U,n}^*(\mu_0;\boldsymbol{\pi})$.
		
		Similarly as before, we replace $\lambda_n(2)$ by zero and use the following as our test statistic:
		\begin{eqnarray}
		\label{TU}
		T_{U,n}(\mu_0;\boldsymbol{\pi}) = S_{U,n}(\mu_0;\boldsymbol{\pi}) - \frac{\sqrt{R_n} (b_n-1)}{n}.
		\end{eqnarray}
        Under Assumption \ref{assump: local dep U}, the bias adjustment term $\sqrt{R_n}(b_n-1)/n$ is asymptotically negligible, as $n \rightarrow \infty$.
		
	\subsubsection{Local Power Analysis}
	One might wonder whether the use of randomized subsampling inference leads to a test that achieves the $\sqrt{n}$ convergence rate, i.e., the same convergence rate that can be achieved with knowledge of local dependence ordering. Here we show that the answer is negative. Consider the following Pitman local alternatives:
	\begin{eqnarray*}
		&& H_{\delta,M}: \mu_0 = \mu_{\delta,M}, \textnormal{ and }\\
		&& H_{\delta,U}: \mu_0 = \mu_{\delta,U},
	\end{eqnarray*} 
	where $\delta \in \mathbf{R}^d \backslash \{0\}$ is a constant vector, 
	\begin{eqnarray*}
		\mu_{\delta,M} = \bar \mu + \frac{\delta}{R_n^{1/2}b_n^{1/2}}, \text{ and } 
		\mu_{\delta,U} = \bar \mu + \frac{\delta}{R_n^{1/4}b_n^{1/2}}.
	\end{eqnarray*}
    We maintain the assumption that $\mathbf{E}[X_{i,n}] = \mathbf{E}[X_{i,n}|\mathcal{C}_n]$.
	\begin{theorem}
		\label{thm: local power analysis}
		
		\noindent Suppose that Assumption \ref{assump: moment cond} holds and $\Sigma_n \rightarrow_{a.s.} \Sigma$ for some positive definite matrix $\Sigma$.
		
		\noindent (i) Suppose further that Assumption \ref{assump: local dep M} holds. Then under $H_{\delta,M}$,
		\begin{eqnarray*}
			\sup_{t \in \mathbf{R}}|P\{T_{M,n}(\bar \mu;\boldsymbol \pi) > t|\mathscr{Z}_n\} -(1-G_\delta (t))| \rightarrow_P 0,
		\end{eqnarray*}
		where $G_\delta$ is the conditional CDF of $(\mathbb{Z} + \Sigma^{-1/2}\delta)'(\mathbb{Z} + \Sigma^{-1/2}\delta)/d$, given $\Sigma$, and $\mathbb{Z} \sim_d N(0,I_d)$.
	
		\noindent (ii) Suppose further that Assumption \ref{assump: local dep U} holds. Then under $H_{\delta,U}$,
		\begin{eqnarray*}
			\sup_{t \in \mathbf{R}} |P\{T_{U,n}(\bar \mu;\boldsymbol \pi) > t|\mathscr{Z}_n\} - 1-\Phi (t - \delta'\Sigma^{-1}\delta/d)| \rightarrow_P 0.
		\end{eqnarray*}
	\end{theorem}
	
	The local power function for the U-type statistic depends on the quadratic form of the drift term $\delta$, showing that the test is for two-sided testing. The rate of convergence of the test is $R_n^{-1/4} b_n^{-1/2}$ which is slower than $\sqrt{n}$ by Assumption \ref{assump: local dep U}(i). A similar remark applies to the M-type statistic approach.
	
	\subsubsection{Rate-Dominance with Size Control}
	
	In this section, we compare the M-type test and the U-type test. Since the tests have different finite sample size distortions depending on the choice of $R_n$ and $b_n$, we cannot directly compare them solely based on the rate of convergence of Pitman drifts for which the tests have nontrivial power. For proper comparison, we introduce the notion of \textit{rate-dominance with size control}. 
	
	Let $\mathscr{P}_n$ be the collection of the joint distributions of the random vector $(X_{1,n},...,X_{n,n})$. Let $\mu(P) \in \mathbf{R}^d$ be the parameter of interest, which is a map from $P \in \mathscr{P}_n$ to $\mathbf{R}^d$. We are interested in testing
	\begin{eqnarray*}
		H_0: \mu(P) = \bar \mu, \text{ against } H_1: \mu(P) \ne \bar \mu.
	\end{eqnarray*}
	Corresponding to the hypothesis testing problem, we partition $\mathscr{P}_n = \mathscr{P}_{n,0} \cup \mathscr{P}_{n,1}$, where
	\begin{eqnarray*}
		\mathscr{P}_{n,0} &\equiv& \left\{ P \in \mathscr{P}_n: \mu(P) = \bar \mu \right\}, \text{ and }\\
		\mathscr{P}_{n,1} &\equiv& \left\{ P \in \mathscr{P}_n: \mu(P) \ne \bar \mu \right\}.
	\end{eqnarray*}
	For each sequence $\delta_n \rightarrow 0$, we define
	\begin{eqnarray*}
		\mathscr{P}_{n,1}(\delta_n) \equiv \left\{ P \in \mathscr{P}_{n,1}: \mu(P) = \bar \mu + \delta_n \right\}.
	\end{eqnarray*}
	Thus a sequence of probabilities $P_n \in \mathscr{P}_{n,1}(\delta_n)$ constitutes Pitman local alternatives at the rate $\delta_n$. Let us introduce the following definitions.
	
	\begin{definition}
		A sequence of tests $(T_n,c_n)$ is said to be \textit{asymptotically exact at the rate $\{\omega_n\}$}, if 
		\begin{eqnarray*}
			\sup_{P \in \mathscr{P}_{n,0}} P \{T_n > c_n\} = \alpha + O(\omega_n),
		\end{eqnarray*}
		as $n \rightarrow \infty$, and the sequence $O(\omega_n)$ is not $o(\omega_n)$.
	\end{definition}
	
	\begin{definition}
		The \textit{rate of a sequence of tests $(T_n,c_n)$} is defined to be a nonstochastic sequence $\delta_n^*$ such that for any sequence $\delta_n$ such that $\delta_n/\delta_n^* \rightarrow 0$ as $n \rightarrow \infty$,
		\begin{eqnarray*}
			P_n\{T_n > c_n\} \le \alpha + o(1), \text{ along any sequence } P_n \in \mathscr{P}_{n,1}(\delta_n),
		\end{eqnarray*}
		and for any sequence $\delta_n$ such that $\delta_n/\delta_n^* \rightarrow \infty$ as $n \rightarrow \infty$,
		\begin{eqnarray*}
			P_n\{T_n > c_n\} \rightarrow 1, \text{ along any sequence } P_n \in \mathscr{P}_{n,1}(\delta_n).
		\end{eqnarray*}
	\end{definition}
	\medskip
	
	It is not hard to see that the rate of tests $(T_n,c_n)$ is unique in the equivalence class of sequences where two sequences $x_n$ and $y_n$ are defined to be equivalent if $x_n = O(y_n)$ and $y_n = O(x_n)$. Similarly the rate at which the tests $(T_n,c_n)$ are asymptotically exact is unique in the equivalence class of sequences.
	
	Finally, we introduce the notion of rate-dominance of one sequence of test statistics over another.
	
	\begin{definition}
		Given two sequences of tests $(T_n,c_n)$ and $(\tilde T_n,\tilde c_n)$, we say that $(T_n,c_n)$ \textit{rate-dominates $(\tilde T_n,\tilde c_n)$ with size control at} $\{\omega_n\}$, if $(T_n,c_n)$ and $(\tilde T_n,\tilde c_n)$ are asymptotically exact at the rate $\{\omega_n\}$, and
		\begin{eqnarray*}
			\lim_{n \rightarrow \infty}\frac{\delta_{n}^*(T_n,c_n)}{\delta_{n}^*(\tilde T_n,\tilde c_n)} = 0,
		\end{eqnarray*}
		where $\delta_{n}^*(T_n,c_n)$ and $\delta_{n}^*(\tilde T_n,\tilde c_n)$ are the rates of $(T_n,c_n)$ and $(\tilde T_n,\tilde c_n)$.
	\end{definition}  
	
	When $(T_n,c_n)$ rate-dominates $(\tilde T_n,\tilde c_n)$ with size control at $\{\omega_n\}$, it means that using the test $(T_n,c_n)$, one can detect a deviation from the null hypothesis more sensitively than the test $(\tilde T_n,\tilde c_n)$, given that both tests control the size asymptotically up to $O(\omega_n)$. Since the rate at which the tests $(T_n,c_n)$ are asymptotically exact is unique in the equivalence class of sequences, the rate dominance relation among the sequences of tests is transitive. 
	
	We introduce a theorem which shows U-type statistic-based tests rate-dominates M-type statistic-based tests when permutation-based critical values are used. Suppose that $(X_{i,n})_{i=1}^n$ is a continuous random vector, and let
	\begin{eqnarray}
	\label{sets}
		&& \{M_{j,n}: j= 1,...,p_n\} = \{M_n(\overline X_n; \pi): \pi \in \Pi\},\text{ and }\\
		&& \{U_{j,n}: j= 1,...,p_n\} = \{U_n(\overline X_n; \pi): \pi \in \Pi\},
	\end{eqnarray}
	where $p_n = {n \choose b_n}$. Define 
	\begin{eqnarray*}
		\Omega_{M,n} &=& \frac{1}{|\Pi|}\sum_{\pi \in \Pi} \mathbf{E}[M_n(\overline X_n;\pi)M_n(\overline X_n;\pi)'|\mathcal{C}_n], \text{ and }\\
		\Omega_{U,n} &=& \frac{1}{|\Pi|}\sum_{\pi \in \Pi} \mathbf{E}[U_n(\overline X_n;\pi)^2|\mathcal{C}_n].
	\end{eqnarray*}
    We make the following assumptions.
	\begin{assumption}
		\label{assump: abs cont}
		The joint distributions of $\{M_{j,n}: j= 1,...,p_n\}$ and $\{U_{j,n}: j= 1,...,p_n\}$ are absolutely continuous with respect to Lebesgue measure.
	\end{assumption}

    \begin{assumption}
	\label{assump: moment rate dominance}
	There exist $n' \ge 1$, $q \ge 2$, $s \ge 3$, and $M_1,\varepsilon_1>0$ such that for all $n \ge n'$ and all $P \in \mathscr{P}_n$, the following conditions hold.
	
	(i) The minimum eigenvalues of $\Omega_{M,n}$ and $\Omega_{U,n}$ are larger than $\varepsilon_1$. 
	
	(ii) $\max_{1 \le i \le n} \mathbf{E}\|X_{i,n}\|^{2q(s+1)} \le M_1$. 
	
	(iii) $n \lambda_n(2) + n^{k+1 -q} \lambda_n(k) \le M_1$ for all $2 \le k \le 2q$.
	
	(iv) $R_{M,n}^{-(2s-3)/4} = O(\sqrt{R_{M,n} b_{M,n}/n})$ and $R_{U,n}^{-(2s-3)/4} = O(\sqrt{R_{U,n}} b_{U,n}/n+\sqrt{b_{U,n}/n})$. 
	
    \end{assumption}

	Our rate-dominance result uses a modulus of continuity for the empirical measures of the sets in (\ref{sets}) up to a higher order. One can use an Edgeworth expansion for the empirical measure, but the classical Edgeworth expansion is not applicable here, because the classical Cram\'{e}r condition for the empirical measure does not apply. To overcome this, we use \cite{Angst/Poly:17:EJP} and \cite{Song:18:WP}. Assumptions \ref{assump: abs cont} and \ref{assump: moment rate dominance} are used primarily for this. The following theorem gives a result of rate-dominance with size control between the M-type statistic and the U-type statistic approaches. (The proof is found in Supplemental Note.) 
	
	\begin{theorem}
		\label{thm: comparison}
		Let $(R_{M,n},b_{M,n})$ be $(R_n,b_n)$ in the definition of the test $(T_{M,n},c_{M,\alpha})$, and $(R_{U,n},b_{U,n})$ be $(R_n,b_n)$ in the definition of the test $(T_{U,n},c_{U,\alpha})$. Suppose that Assumption \ref{assump: local dep M} is satisfied with $(R_n,b_n) = (R_{M,n},b_{M,n})$ and Assumption \ref{assump: local dep U} with $(R_n,b_n) = (R_{U,n},b_{U,n})$, and that Assumptions \ref{assump: moment cond} - \ref{assump: moment rate dominance} hold.
		
		Then, for each $\alpha \in (0,1)$, $(T_{U,n},c_{U,\alpha})$ rate-dominates $(T_{M,n},c_{M,\alpha})$ with size control at $\{\omega_n\}$, if we choose $R_{M,n},R_{U,n}$ such that
		\begin{eqnarray*}
			\omega_n = \sqrt{\frac{R_{M,n}b_{M,n}}{n}} =  \frac{\sqrt{R_{U,n}}b_{U,n}}{n}+\sqrt{\frac{b_{U,n}}{n}}.
		\end{eqnarray*}
	\end{theorem}
	\medskip
	
	Let us give a heuristic for Theorem \ref{thm: comparison}. Using the Edgeworth expansion of the conditional distribution of $T_{U,n}$ given $\mathscr{Z}_n$, we find that
	\begin{eqnarray*}
		P\{ T_{U,n} \le c_{U,\alpha} |\mathscr{Z}_n\} = \alpha + O_P\left(\frac{\sqrt{R_{U,n}}b_{U,n}}{n} + \sqrt{\frac{b_{U,n}}{n}}\right).
	\end{eqnarray*} 
	(In fact, the last rate comes from $B_{U,n}$, and is sharp under the regularity conditions, and the error in the Edgeworth expansion is dominated by this rate.) Similarly, we also obtain that
	\begin{eqnarray*}
		P\{ T_{M,n} \le c_{M,\alpha} |\mathscr{Z}_n\} = \alpha + O_P\left(\sqrt{\frac{R_{M,n}b_{M,n}}{n}}\right).
	\end{eqnarray*} 
	We normalize the rate by choosing $(R_{M,n},b_{M,n})$ and $(R_{U,n},b_{U,n})$ such that
	\begin{eqnarray*}
		\omega_n = \frac{\sqrt{R_{U,n}}b_{U,n}}{n} + \sqrt{\frac{b_{U,n}}{n}} = \sqrt{\frac{R_{M,n}b_{M,n}}{n}},
	\end{eqnarray*}
	for some sequence $\omega_n \rightarrow 0$. From the local power result in Theorem \ref{thm: local power analysis} and using Assumption \ref{assump: moment rate dominance}(iv), we find that
	\begin{eqnarray*}
		\frac{\delta_n^*(T_{U,n},c_{U,n})}{\delta_n^*(T_{M,n},c_{M,n})} = \frac{\sqrt{n} \omega_n}{\sqrt{n}\omega_n x_n} \rightarrow 0,
	\end{eqnarray*}
	as $n \rightarrow \infty$, where $x_n \equiv R_{U,n}^{1/4}/(\sqrt{n} \omega_n) \rightarrow \infty$, by Assumption \ref{assump: local dep U}. Thus $(T_{U,n},c_{U,\alpha})$ rate-dominates $(T_{M,n},c_{M,\alpha})$ with size control.
	
\subsection{Non-randomized Inference}
	The tests based on $(T_{M,n},c_{M,\alpha})$ and $(T_{U,n},c_{U,\alpha})$ are randomized tests, where there is randomness of the test statistic apart from that of the samples. Hence different researchers may have different results using the same data and the same model though with a small probability. To address this issue, this paper proposes the following approach of constructing confidence intervals. For a given large positive integer $S$, for each $s=1,2,...,S$, we let $\boldsymbol{\pi}_s = (\pi_{1,s},...,\pi_{R_n,s})$, where $\pi_{r,s}$'s are i.i.d. draws from the uniform distribution on $\Pi$. Then we define for $\tau \in \{M,U\}$
	\begin{eqnarray*}
		q_\tau(\bar \mu;\alpha) = \frac{1}{S} \sum_{s=1}^S 1\{T_{\tau,n}(\bar \mu;\boldsymbol{\pi}_s) \le c_{\tau,\alpha} \}.
	\end{eqnarray*}
	We call $q_\tau(\cdot;\alpha)$ the \textit{randomized confidence function}.
	
	\begin{corollary}
		\label{cor: rand conf interval}
		Suppose that Assumption \ref{assump: moment cond} holds. Then for each $\bar \mu \in \mathbf{R}^d$, the following statements hold as ${n,S \rightarrow \infty}$ jointly.
		
		(i) Under Assumption \ref{assump: local dep M}, $q_M(\bar \mu;\alpha) \rightarrow_p 1-\alpha$ if $\bar \mu = \mu_0$, and $q_M(\bar \mu;\alpha) \rightarrow_p 0$ otherwise.
	    
		(ii) Under Assumption \ref{assump: local dep U}, $q_U(\bar \mu;\alpha) \rightarrow_p 1-\alpha$ if $\bar \mu = \mu_0$, and $q_U(\bar \mu;\alpha) \rightarrow_p 0$ otherwise.
	\end{corollary}
	
	The convergence of $q_M(\bar \mu;\alpha)$ or $q_U(\bar \mu;\alpha)$ to 0 at $\bar \mu \ne \mu_0$ reflects the consistency property of the randomized test. The result of Corollary \ref{cor: rand conf interval} can be shown by slightly modifying the proof of Theorem \ref{thm: Type U}.
	
	\begin{figure}[t]
		\begin{center}
			\includegraphics[scale=0.5]{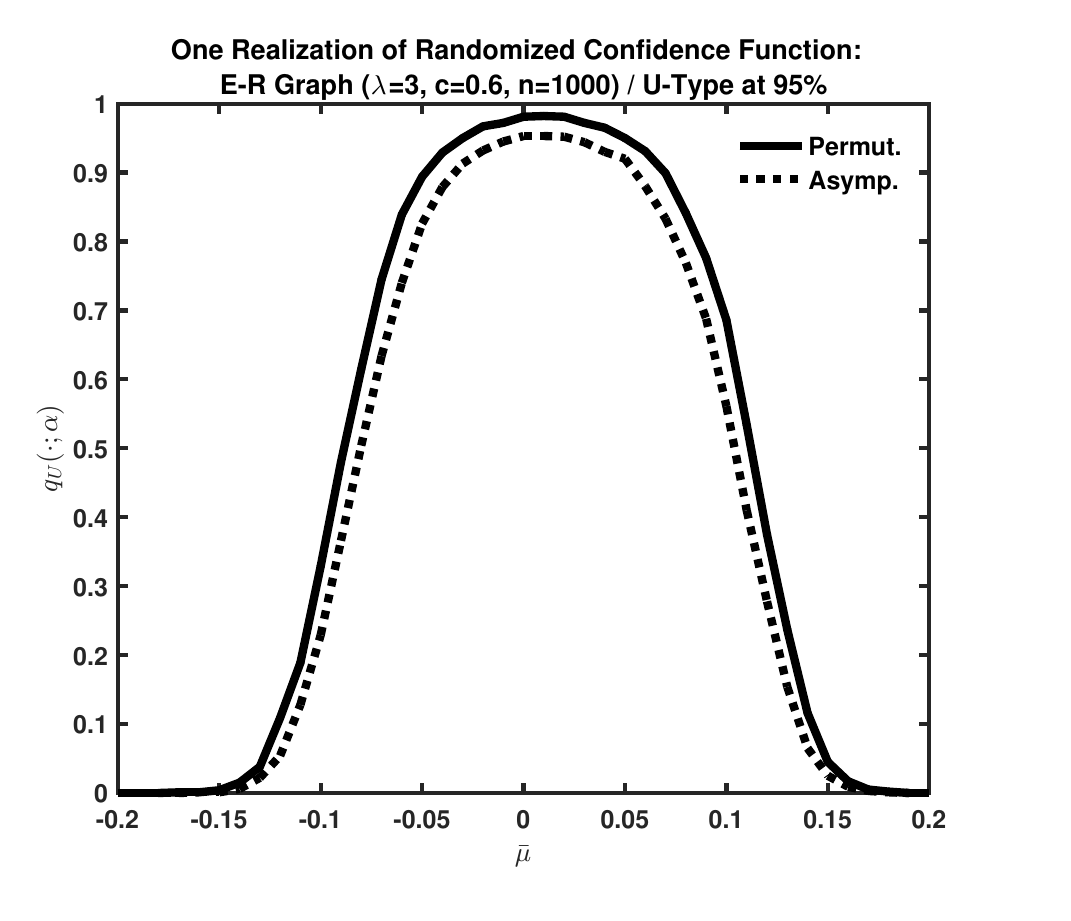}
			 \caption{ \small Illustration of a Randomized Confidence Interval at 95\% When $\mu_0 = 0$ : \footnotesize The result is based on network Dependent observations on one realization of an Erd\"{o}s-R\'{e}ny random graph. See Section 3 for the meaning of the parameter values.}
			\label{fig:figure_illustration}
		\end{center}
	\end{figure}
	\normalsize
	\afterpage{\FloatBarrier}
	
	The randomized confidence function is not a familiar concept in econometrics.\footnote{The notion of the randomized confidence function in this paper coincides with what \citet*{Geyer/Meeden:05:SS} referred to as the membership function of fuzzy confidence intervals. The way randomized tests arise in this paper is different. They do here because the asymptotic pivotalness of the test (thus permitting ordering-free inference) prevents us from drawing an arbitrarily large number of random permutations.} Instead, one may use a nonrandomized confidence set of the following form. Take $\beta \in (0,\alpha)$ and define for $\tau \in \{M,U\}$
	\begin{eqnarray}
	    \label{CS1}
		C_{\tau,\alpha} = \{\mu: q_\tau(\bar \mu;\alpha-\beta) \ge 1 - \alpha\}.
	\end{eqnarray}
	Then it is not hard to see from Corollary \ref{cor: rand conf interval} that
	\begin{eqnarray*}
		\liminf_{n \rightarrow \infty} P\{\mu_0 \in C_{\tau,\alpha}\} \ge 1 - \alpha.
	\end{eqnarray*}
	as $n \rightarrow \infty$. For example, we may take $\beta = 0.005$, which is used in simulation studies in this paper. (In simulation studies unreported in this paper, the choice of $\beta = 0$ was also used, and the results were not very different.) 
			     
\section{Simulation Studies}
\subsection{Data Generating Process}
This section presents and discusses a Monte Carlo simulation study which investigates the finite sample properties of the randomized subsampling approach in various situations with local dependence. As for local dependence, this study considered three kinds of data generating processes: (i)  i.i.d. variables, (ii) variables having a dependency graph, (iii) network dependent variables. Both asymptotic and permutation critical values are considered.

As for the dependency graph case, we use two kinds of graphs. One is based on Erd\"{o}s-R\'{e}nyi graphs, and the other is based on Barab\'{a}si-Albert graphs of preferential attachment. In an Erd\"{o}s-R\'{e}nyi random graph, each pair of the vertices form an edge with equal probability $p=\lambda /(n-1)$. The simulation study here chose $\lambda $ from $\{1,3,5\}$. Thus each vertex from this random graph has degree $\lambda $ on average, and the degree distribution is approximately a Poisson distribution with parameter $\lambda $ when the graph is large. For a Barab\'{a}si-Albert random graph of preferential
attachment, we first began with an Erd\"{o}s-R\'{e}nyi random graph of size $20$ with $\lambda =1$. Then we let the graph grow by adding each vertex sequentially and let the vertex form edges with $m $ other existing vertices. (We chose $m$ from $\{1,2,3\}$ for this study.) The probability of a new vertex forming an edge with an existing vertex is proportional to the number of the neighbors of the existing vertex. We keep adding new vertices until the size of the graph becomes $n$.

As for the generation of the random variables, we follow the design in \citet*{Song:15:WP}. We first generate $\{Y_i^{\ast }\}_{i=1}^{n}$ i.i.d. from $N(0,1)$. Let $E=\{e_{1},...,e_{S}\}$ be the set of edges in the graph with redundant edges removed from $E$ (i.e., remove $ji$ with $j<i$%
) and let $M$ be two-column matrix whose entries are of the form $%
[i_{s},j_{s}]$ for $e_{s}=i_{s}j_{s}$. Let $M$ be sorted on the first column
so that $i_{s}\leq i_{s+1}$.\bigskip

\noindent \textsc{Step 1:}\textbf{\ }For $s=1$, such that $e_{1}=i_{1}j_{1}$%
, we draw $Z_{1}\sim N(0,1)$ and set%
\begin{equation*}
(Y_{i_{1}},Y_{j_{1}})=\sqrt{1-c^{2}}\times (Y_{i_{1}}^{\ast
},Y_{j_{1}}^{\ast })+c\times Z_{1},
\end{equation*}%
where $c$ is a parameter that determines the strength of graph dependence.
We replace $(Y_{i_{1}}^{\ast },Y_{j_{1}}^{\ast })$ by $(Y_{i_{1}},Y_{j_{1}})$%
, and redefine the series $\{Y_i^{\ast }\}_{i=1}^{n}$.\bigskip

\noindent \textsc{Step }$s=2,...,S$\textbf{: }For $s>1$ such that $%
e_{s}=(i_{s},j_{s})$, we draw $Z_{s}\sim N(0,1)$ and set%
\begin{equation*}
(Y_{i_{s}},Y_{j_{s}})=\sqrt{1-c^{2}}\times (Y_{i_{s}}^{\ast
},Y_{j_{s}}^{\ast })+c\times Z_{s}.
\end{equation*}%
We replace $(Y_{i_{s}}^{\ast },Y_{j_{s}}^{\ast })$ by $(Y_{i_{s}},Y_{j_{s}})$%
, and redefine the series $\{Y_i^{\ast }\}_{i=1}^{n}$.\bigskip

Turning now to the third data generation design involving network dependent observations, they are drawn from a jointly normal random vector with mean zero and a covariance matrix such that the correlation between two random variables at the distance $D$ is set to be $\exp(-\rho D)$, where $\rho$ is a parameter that determines the strength of the correlation. The parameter $\rho$ was taken from $\{1.0,1.5,2.0\}$. (Note that $\exp(-1)$ is roughly 0.3674 and $\exp(-2)$ is roughly 0.1353.) For the graph underlying the network-dependent observations, the simulation study considered a realization of a Erd\"{o}s-Re\'{e}nyi random graph with $\lambda \in \{1,2\}$.

The size $n$ of the networks was taken from $\{500,1000,2000\}$. As for $R$ and $b_n$, we chose $R_n=\sqrt{n}$ and $b_n = n^{1/3}$ for the M-type statistic and $R_n=n$ and $b_n = n^{1/3}$ for the U-type statistic. The Monte Carlo simulation number and the permutation number (i.e., $L$) were set to be 1000, and the tuning parameter $\beta$ 0.005.

\begin{table}[t]
	\small
	\caption{\small The Empirical Coverage Probability of Confidence Interval: Independent Observations}
	\begin{center}
		U-Type Statistics
		
		\begin{tabular}{cccc|ccc}
			\hline\hline
			&  & Asymp. &  &  & Permut. &  \\ \cline{2-7}
			\multicolumn{1}{l}{} & $99\%$ & $95\%$ & $90\%$ & $99\%$ & $95\%$ & $90\%$ \\ 
			\hline
			\multicolumn{1}{l|}{$n = 500$} & \multicolumn{1}{|l}{0.968} & \multicolumn{1}{c}{0.919} & \multicolumn{1}{l|}{0.875} & \multicolumn{1}{l}{0.990} & \multicolumn{1}{c}{0.946} & \multicolumn{1}{l}{0.895} \\ 
			\multicolumn{1}{l|}{$n = 1000$} & \multicolumn{1}{|l}{0.967} & \multicolumn{1}{c}{0.914} & \multicolumn{1}{l|}{0.868} & \multicolumn{1}{l}{0.991} & \multicolumn{1}{c}{0.947} & \multicolumn{1}{l}{0.896} \\ 
			\multicolumn{1}{l|}{$n = 2000$} & \multicolumn{1}{|l}{0.963} & \multicolumn{1}{c}{0.909} & \multicolumn{1}{l|}{0.861} & \multicolumn{1}{l}{0.991} & \multicolumn{1}{c}{0.945} & \multicolumn{1}{l}{0.894} \\ 
			\hline
		\end{tabular}
		\bigskip
		
		M-Type Statistics
		
		\begin{tabular}{cccc|ccc}
			\hline\hline
			&  & Asymp. &  &  & Permut. &  \\ \cline{2-7}
			\multicolumn{1}{l}{} & $99\%$ & $95\%$ & $90\%$ & $99\%$ & $95\%$ & $90\%$ \\ 
			\hline
			\multicolumn{1}{l|}{$n = 500$} & \multicolumn{1}{|l}{0.980} & \multicolumn{1}{c}{0.928} & \multicolumn{1}{l|}{0.874} & \multicolumn{1}{l}{0.986} & \multicolumn{1}{c}{0.931} & \multicolumn{1}{l}{0.876} \\ 
			\multicolumn{1}{l|}{$n = 1000$} & \multicolumn{1}{|l}{0.981} & \multicolumn{1}{c}{0.929} & \multicolumn{1}{l|}{0.876} & \multicolumn{1}{l}{0.987} & \multicolumn{1}{c}{0.933} & \multicolumn{1}{l}{0.878} \\ 
			\multicolumn{1}{l|}{$n = 2000$} & \multicolumn{1}{|l}{0.980} & \multicolumn{1}{c}{0.928} & \multicolumn{1}{l|}{0.875} & \multicolumn{1}{l}{0.987} & \multicolumn{1}{c}{0.933} & \multicolumn{1}{l}{0.878} \\ 
			\hline
		\end{tabular}
	\end{center}
	\bigskip \bigskip
\end{table}

\begin{table}[t]
	\small
	\begin{center}
		\small
		\caption{\small Empirical Coverage Probability: 95\% for Dependency Graphs}
		
		U-Type Statistics
		
		\begin{tabular}{cccccc|ccc}
			\hline\hline
			&  &  &  & E-R &  &  & B-A &  \\ \cline{4-9}
			\multicolumn{1}{l}{}&  & \multicolumn{1}{l}{} & \multicolumn{1}{l}{$\lambda=1$} & \multicolumn{1}{l}{$\lambda =3$} & \multicolumn{1}{l|}{$\lambda =5$} & \multicolumn{1}{|l}{$m=1$} & \multicolumn{1}{l}{$m=2$} & \multicolumn{1}{l}{$m=3$}\\
			
			\hline
			\multicolumn{1}{l|}{}&  & \multicolumn{1}{l|}{$n = 500$} & \multicolumn{1}{|l}{0.919} & \multicolumn{1}{l}{0.916} & \multicolumn{1}{l|}{0.915} & \multicolumn{1}{|l}{0.914} & \multicolumn{1}{l}{0.913} & \multicolumn{1}{l}{0.914} \\ 
			\multicolumn{1}{l|}{}&  {$c=0.3$} & \multicolumn{1}{l|}{$n =1000$} & \multicolumn{1}{|l}{0.911} & \multicolumn{1}{l}{0.916} & \multicolumn{1}{l|}{0.914} & \multicolumn{1}{|l}{0.912} & \multicolumn{1}{l}{0.917} & \multicolumn{1}{l}{0.914} \\ 
			\multicolumn{1}{l|}{}&    & \multicolumn{1}{l|}{$n =2000$} & \multicolumn{1}{|l}{0.908} & \multicolumn{1}{l}{0.911} & \multicolumn{1}{l|}{0.909} & \multicolumn{1}{|l}{0.907} & \multicolumn{1}{l}{0.909} & \multicolumn{1}{l}{0.912} \\ 
			\multicolumn{1}{l|}{Asymp.}&   & \multicolumn{1}{l|}{} & & & \multicolumn{1}{l|}{} \\
			\multicolumn{1}{l|}{}&  & \multicolumn{1}{l|}{$n = 500$} & \multicolumn{1}{|l}{0.917} & \multicolumn{1}{l}{0.919} & \multicolumn{1}{l|}{0.919} & \multicolumn{1}{|l}{0.916} & \multicolumn{1}{l}{0.917} & \multicolumn{1}{l}{0.916} \\ 
			\multicolumn{1}{l|}{}&  {$c=0.6$} & \multicolumn{1}{l|}{$n =1000$} & \multicolumn{1}{|l}{0.916} & \multicolumn{1}{l}{0.912} & \multicolumn{1}{l|}{0.910} & \multicolumn{1}{|l}{0.917} & \multicolumn{1}{l}{0.915} & \multicolumn{1}{l}{0.914} \\ 
			\multicolumn{1}{l|}{}&    & \multicolumn{1}{l|}{$n =2000$} & \multicolumn{1}{|l}{0.910} & \multicolumn{1}{l}{0.913} & \multicolumn{1}{l|}{0.912} & \multicolumn{1}{|l}{0.908} & \multicolumn{1}{l}{0.913} & \multicolumn{1}{l}{0.911} \\ 
			\hline
			\multicolumn{1}{l|}{}&  & \multicolumn{1}{l|}{$n = 500$} & \multicolumn{1}{|l}{0.946} & \multicolumn{1}{l}{0.944} & \multicolumn{1}{l|}{0.942} & \multicolumn{1}{|l}{0.942} & \multicolumn{1}{l}{0.941} & \multicolumn{1}{l}{0.942} \\ 
			\multicolumn{1}{l|}{}&  {$c=0.3$} & \multicolumn{1}{l|}{$n =1000$} & \multicolumn{1}{|l}{0.945} & \multicolumn{1}{l}{0.948} & \multicolumn{1}{l|}{0.947} & \multicolumn{1}{|l}{0.945} & \multicolumn{1}{l}{0.949} & \multicolumn{1}{l}{0.947} \\ 
			\multicolumn{1}{l|}{}&    & \multicolumn{1}{l|}{$n =2000$} & \multicolumn{1}{|l}{0.945} & \multicolumn{1}{l}{0.947} & \multicolumn{1}{l|}{0.946} & \multicolumn{1}{|l}{0.945} & \multicolumn{1}{l}{0.945} & \multicolumn{1}{l}{0.948} \\ 
			\multicolumn{1}{l|}{Permut.}&   & \multicolumn{1}{l|}{} & & & \multicolumn{1}{l|}{} \\
			\multicolumn{1}{l|}{}&  & \multicolumn{1}{l|}{$n = 500$} & \multicolumn{1}{|l}{0.944} & \multicolumn{1}{l}{0.946} & \multicolumn{1}{l|}{0.946} & \multicolumn{1}{|l}{0.944} & \multicolumn{1}{l}{0.944} & \multicolumn{1}{l}{0.943} \\ 
			\multicolumn{1}{l|}{}&  {$c=0.6$} & \multicolumn{1}{l|}{$n =1000$} & \multicolumn{1}{|l}{0.948} & \multicolumn{1}{l}{0.945} & \multicolumn{1}{l|}{0.943} & \multicolumn{1}{|l}{0.949} & \multicolumn{1}{l}{0.947} & \multicolumn{1}{l}{0.946} \\ 
			\multicolumn{1}{l|}{}&    & \multicolumn{1}{l|}{$n =2000$} & \multicolumn{1}{|l}{0.946} & \multicolumn{1}{l}{0.949} & \multicolumn{1}{l|}{0.949} & \multicolumn{1}{|l}{0.945} & \multicolumn{1}{l}{0.949} & \multicolumn{1}{l}{0.947} \\ 
			\hline
		\end{tabular}
		\bigskip
		
		M-Type Statistics
		
		\begin{tabular}{cccccc|ccc}
			\hline\hline
			&  &  &  & E-R &  &  & B-A &  \\ \cline{4-9}
			\multicolumn{1}{l}{}&  & \multicolumn{1}{l}{} & \multicolumn{1}{l}{$\lambda=1$} & \multicolumn{1}{l}{$\lambda =3$} & \multicolumn{1}{l|}{$\lambda =5$} & \multicolumn{1}{|l}{$m=1$} & \multicolumn{1}{l}{$m=2$} & \multicolumn{1}{l}{$m=3$}\\		
			\hline
			\multicolumn{1}{l|}{}&  & \multicolumn{1}{l|}{$n = 500$} & \multicolumn{1}{|l}{0.928} & \multicolumn{1}{l}{0.926} & \multicolumn{1}{l|}{0.925} & \multicolumn{1}{|l}{0.923} & \multicolumn{1}{l}{0.923} & \multicolumn{1}{l}{0.924} \\ 
			\multicolumn{1}{l|}{}&  {$c=0.3$} & \multicolumn{1}{l|}{$n =1000$} & \multicolumn{1}{|l}{0.926} & \multicolumn{1}{l}{0.930} & \multicolumn{1}{l|}{0.929} & \multicolumn{1}{|l}{0.926} & \multicolumn{1}{l}{0.930} & \multicolumn{1}{l}{0.929} \\ 
			\multicolumn{1}{l|}{}&    & \multicolumn{1}{l|}{$n =2000$} & \multicolumn{1}{|l}{0.927} & \multicolumn{1}{l}{0.929} & \multicolumn{1}{l|}{0.927} & \multicolumn{1}{|l}{0.926} & \multicolumn{1}{l}{0.928} & \multicolumn{1}{l}{0.930} \\ 
			\multicolumn{1}{l|}{Asymp.}&   & \multicolumn{1}{l|}{} & & & \multicolumn{1}{l|}{} \\
			\multicolumn{1}{l|}{}&  & \multicolumn{1}{l|}{$n = 500$} & \multicolumn{1}{|l}{0.927} & \multicolumn{1}{l}{0.928} & \multicolumn{1}{l|}{0.927} & \multicolumn{1}{|l}{0.925} & \multicolumn{1}{l}{0.926} & \multicolumn{1}{l}{0.925} \\ 
			\multicolumn{1}{l|}{}&  {$c=0.6$} & \multicolumn{1}{l|}{$n =1000$} & \multicolumn{1}{|l}{0.929} & \multicolumn{1}{l}{0.927} & \multicolumn{1}{l|}{0.926} & \multicolumn{1}{|l}{0.931} & \multicolumn{1}{l}{0.930} & \multicolumn{1}{l}{0.928} \\ 
			\multicolumn{1}{l|}{}&    & \multicolumn{1}{l|}{$n =2000$} & \multicolumn{1}{|l}{0.928} & \multicolumn{1}{l}{0.931} & \multicolumn{1}{l|}{0.931} & \multicolumn{1}{|l}{0.927} & \multicolumn{1}{l}{0.931} & \multicolumn{1}{l}{0.929} \\ 
			\hline
			\multicolumn{1}{l|}{}&  & \multicolumn{1}{l|}{$n = 500$} & \multicolumn{1}{|l}{0.932} & \multicolumn{1}{l}{0.929} & \multicolumn{1}{l|}{0.929} & \multicolumn{1}{|l}{0.927} & \multicolumn{1}{l}{0.927} & \multicolumn{1}{l}{0.927} \\ 
			\multicolumn{1}{l|}{}&  {$c=0.3$} & \multicolumn{1}{l|}{$n =1000$} & \multicolumn{1}{|l}{0.930} & \multicolumn{1}{l}{0.934} & \multicolumn{1}{l|}{0.932} & \multicolumn{1}{|l}{0.930} & \multicolumn{1}{l}{0.934} & \multicolumn{1}{l}{0.932} \\ 
			\multicolumn{1}{l|}{}&    & \multicolumn{1}{l|}{$n =2000$} & \multicolumn{1}{|l}{0.931} & \multicolumn{1}{l}{0.933} & \multicolumn{1}{l|}{0.931} & \multicolumn{1}{|l}{0.931} & \multicolumn{1}{l}{0.932} & \multicolumn{1}{l}{0.935} \\ 
			\multicolumn{1}{l|}{Permut.}&   & \multicolumn{1}{l|}{} & & & \multicolumn{1}{l|}{} \\
			\multicolumn{1}{l|}{}&  & \multicolumn{1}{l|}{$n = 500$} & \multicolumn{1}{|l}{0.930} & \multicolumn{1}{l}{0.931} & \multicolumn{1}{l|}{0.930} & \multicolumn{1}{|l}{0.928} & \multicolumn{1}{l}{0.929} & \multicolumn{1}{l}{0.928} \\ 
			\multicolumn{1}{l|}{}&  {$c=0.6$} & \multicolumn{1}{l|}{$n =1000$} & \multicolumn{1}{|l}{0.933} & \multicolumn{1}{l}{0.931} & \multicolumn{1}{l|}{0.930} & \multicolumn{1}{|l}{0.934} & \multicolumn{1}{l}{0.933} & \multicolumn{1}{l}{0.932} \\ 
			\multicolumn{1}{l|}{}&    & \multicolumn{1}{l|}{$n =2000$} & \multicolumn{1}{|l}{0.932} & \multicolumn{1}{l}{0.935} & \multicolumn{1}{l|}{0.935} & \multicolumn{1}{|l}{0.932} & \multicolumn{1}{l}{0.935} & \multicolumn{1}{l}{0.933} \\ 
			\hline
		\end{tabular}
		
	\end{center}
	\par
	\medskip 
	\parbox{6.2in}{\footnotesize
		
		Notes: The E-R represents Erd\"{o}s-R\'{e}nyi Random Graph with probability equal to $p=\lambda /(n-1)$, and $\lambda $ chosen from ${1,3,5}$, and the B-A represents Barab\'{a}si-Albert random graph of preferential attachment, with $m$ refering to the number of links each new node forms with other existing nodes. A larger parameter $c$ represents a stronger correlation between two linked observations.
		
		\bigskip \bigskip}
\end{table}

\begin{table}[t]
	\caption{\small Empirical Coverage Probability at 95\% for Network Dependent Observations}
	
	\begin{center}
		\small
		
		U-Type Statistics
		
		\begin{tabular}{cccccccc}
			\hline\hline
			\multicolumn{1}{l}{} & & \multicolumn{1}{l}{} & \multicolumn{1}{l}{Asymp.} & \multicolumn{1}{l}{} & \multicolumn{1}{l}{} & \multicolumn{1}{l}{Permut.} & \multicolumn{1}{l}{} \\
			\multicolumn{1}{l}{} & & \multicolumn{1}{l}{$\rho =2.0$} & \multicolumn{1}{l}{$\rho =1.5$} & \multicolumn{1}{l}{$\rho =1.0$} & \multicolumn{1}{|l}{$\rho=2.0$} & \multicolumn{1}{l}{$\rho =1.5$} & \multicolumn{1}{l}{$\rho =1.0$}\\
			\hline 
			\multicolumn{1}{l|}{}  & \multicolumn{1}{l|}{$n = 500$} & \multicolumn{1}{|l}{0.912} & \multicolumn{1}{l}{0.904} & \multicolumn{1}{l|}{0.889} & \multicolumn{1}{|l}{0.940} & \multicolumn{1}{l}{0.934} & \multicolumn{1}{l}{0.921} \\ 
			\multicolumn{1}{l|}{$\lambda=1$} & {$n =1000$} & \multicolumn{1}{|l}{0.908} & \multicolumn{1}{l}{0.901} & \multicolumn{1}{l|}{0.883} & \multicolumn{1}{|l}{0.942} & \multicolumn{1}{l}{0.935} & \multicolumn{1}{l}{0.921} \\ 
			\multicolumn{1}{l|}{}  & {$n =2000$} & \multicolumn{1}{|l}{0.902} & \multicolumn{1}{l}{0.897} & \multicolumn{1}{l|}{0.884} & \multicolumn{1}{|l}{0.941} & \multicolumn{1}{l}{0.936} & \multicolumn{1}{l}{0.926} \\
			\hline
			\multicolumn{1}{l|}{}  & \multicolumn{1}{l|}{$n = 500$} & \multicolumn{1}{|l}{0.900} & \multicolumn{1}{l}{0.881} & \multicolumn{1}{l|}{0.817} & \multicolumn{1}{|l}{0.930} & \multicolumn{1}{l}{0.914} & \multicolumn{1}{l}{0.854} \\ 
			\multicolumn{1}{l|}{$\lambda=2$} & {$n =1000$} & \multicolumn{1}{|l}{0.904} & \multicolumn{1}{l}{0.877} & \multicolumn{1}{l|}{0.811} & \multicolumn{1}{|l}{0.939} & \multicolumn{1}{l}{0.914} & \multicolumn{1}{l}{0.857} \\ 
			\multicolumn{1}{l|}{}  & {$n =2000$} & \multicolumn{1}{|l}{0.893} & \multicolumn{1}{l}{0.873} & \multicolumn{1}{l|}{0.805} & \multicolumn{1}{|l}{0.934} & \multicolumn{1}{l}{0.918} & \multicolumn{1}{l}{0.857} \\
			\hline
		\end{tabular}
		\bigskip
		
		M-Type Statistics
		
		\begin{tabular}{cccccccc}
			\hline\hline
			\multicolumn{1}{l}{} & & \multicolumn{1}{l}{} & \multicolumn{1}{l}{Asymp.} & \multicolumn{1}{l}{} & \multicolumn{1}{l}{} & \multicolumn{1}{l}{Permut.} & \multicolumn{1}{l}{} \\
			\multicolumn{1}{l}{} & & \multicolumn{1}{l}{$\rho =2.0$} & \multicolumn{1}{l}{$\rho =1.5$} & \multicolumn{1}{l}{$\rho =1.0$} & \multicolumn{1}{|l}{$\rho=2.0$} & \multicolumn{1}{l}{$\rho =1.5$} & \multicolumn{1}{l}{$\rho =1.0$}\\
			\hline 
			\multicolumn{1}{l|}{}  & \multicolumn{1}{l|}{$n = 500$} & \multicolumn{1}{|l}{0.922} & \multicolumn{1}{l}{0.917} & \multicolumn{1}{l|}{0.907} & \multicolumn{1}{|l}{0.926} & \multicolumn{1}{l}{0.920} & \multicolumn{1}{l}{0.911} \\ 
			\multicolumn{1}{l|}{$\lambda=1$} & {$n =1000$} & \multicolumn{1}{|l}{0.924} & \multicolumn{1}{l}{0.919} & \multicolumn{1}{l|}{0.908} & \multicolumn{1}{|l}{0.929} & \multicolumn{1}{l}{0.923} & \multicolumn{1}{l}{0.913} \\ 
			\multicolumn{1}{l|}{}  & {$n =2000$} & \multicolumn{1}{|l}{0.923} & \multicolumn{1}{l}{0.920} & \multicolumn{1}{l|}{0.911} & \multicolumn{1}{|l}{0.928} & \multicolumn{1}{l}{0.925} & \multicolumn{1}{l}{0.915} \\
			\hline
			\multicolumn{1}{l|}{}  & \multicolumn{1}{l|}{$n = 500$} & \multicolumn{1}{|l}{0.915} & \multicolumn{1}{l}{0.902} & \multicolumn{1}{l|}{0.860} & \multicolumn{1}{|l}{0.918} & \multicolumn{1}{l}{0.906} & \multicolumn{1}{l}{0.865} \\ 
			\multicolumn{1}{l|}{$\lambda=2$} & {$n =1000$} & \multicolumn{1}{|l}{0.921} & \multicolumn{1}{l}{0.904} & \multicolumn{1}{l|}{0.862} & \multicolumn{1}{|l}{0.925} & \multicolumn{1}{l}{0.909} & \multicolumn{1}{l}{0.867} \\ 
			\multicolumn{1}{l|}{}  & {$n =2000$} & \multicolumn{1}{|l}{0.917} & \multicolumn{1}{l}{0.904} & \multicolumn{1}{l|}{0.862} & \multicolumn{1}{|l}{0.921} & \multicolumn{1}{l}{0.909} & \multicolumn{1}{l}{0.869} \\
			\hline
		\end{tabular}%
		
	\end{center}
	\par
	\medskip 
	\parbox{6.2in}{\footnotesize
		
		Notes: The Erd\"{o}s-R\'{e}nyi Random Graph 
		with probability equal to $p=\lambda /(n-1)$ was used. 
		The correlation between linked observations is set to be $\exp(-\rho D)$ where $D$ represents the length of the shortest path between the two indices of the observations on the graph.
		\bigskip}
\end{table}

\begin{table}[t]
	\caption{\small Empirical False Coverage Probability at 95\%}
	
	\begin{center}
		\small
		
		Independent Observations
		
		\begin{tabular}{cccccccc}
			\hline\hline
			\multicolumn{1}{l}{} & & \multicolumn{1}{l}{} & \multicolumn{1}{l}{U-Type} & \multicolumn{1}{l}{} & \multicolumn{1}{l}{} & \multicolumn{1}{l}{M-Type} & \multicolumn{1}{l}{} \\
			\multicolumn{1}{l}{} & & \multicolumn{1}{l}{$\bar \mu =0.06$} & \multicolumn{1}{l}{$\bar \mu =0.12$} & \multicolumn{1}{l}{$\bar \mu =0.18$} & \multicolumn{1}{|l}{$\bar \mu=0.06$} & \multicolumn{1}{l}{$\bar \mu =0.12$} & \multicolumn{1}{l}{$\bar \mu =0.18$}\\
			\hline 
			\multicolumn{1}{l|}{}  & \multicolumn{1}{l|}{$n = 500$} & \multicolumn{1}{|l}{0.829} & \multicolumn{1}{l}{0.526} & \multicolumn{1}{l|}{0.174} & \multicolumn{1}{|l}{0.864} & \multicolumn{1}{l}{0.680} & \multicolumn{1}{l}{0.425} \\ 
			\multicolumn{1}{l|}{Asymp.} & {$n =1000$} & \multicolumn{1}{|l}{0.745} & \multicolumn{1}{l}{0.244} & \multicolumn{1}{l|}{0.015} & \multicolumn{1}{|l}{0.815} & \multicolumn{1}{l}{0.506} & \multicolumn{1}{l}{0.190} \\ 
			\multicolumn{1}{l|}{}  & {$n =2000$} & \multicolumn{1}{|l}{0.554} & \multicolumn{1}{l}{0.032} & \multicolumn{1}{l|}{0.000} & \multicolumn{1}{|l}{0.716} & \multicolumn{1}{l}{0.257} & \multicolumn{1}{l}{0.030} \\
			\hline
			\multicolumn{1}{l|}{}  & \multicolumn{1}{l|}{$n = 500$} & \multicolumn{1}{|l}{0.870} & \multicolumn{1}{l}{0.585} & \multicolumn{1}{l|}{0.212} & \multicolumn{1}{|l}{0.868} & \multicolumn{1}{l}{0.688} & \multicolumn{1}{l}{0.434} \\ 
			\multicolumn{1}{l|}{Permut.} & {$n =1000$} & \multicolumn{1}{|l}{0.797} & \multicolumn{1}{l}{0.301} & \multicolumn{1}{l|}{0.022} & \multicolumn{1}{|l}{0.822} & \multicolumn{1}{l}{0.518} & \multicolumn{1}{l}{0.199} \\ 
			\multicolumn{1}{l|}{}  & {$n =2000$} & \multicolumn{1}{|l}{0.634} & \multicolumn{1}{l}{0.046} & \multicolumn{1}{l|}{0.000} & \multicolumn{1}{|l}{0.726} & \multicolumn{1}{l}{0.267} & \multicolumn{1}{l}{0.032} \\
			\hline
		\end{tabular}
	  \medskip
	  
		Dependency Graphs
		
		\begin{tabular}{cccccccc}
			\hline\hline
			\multicolumn{1}{l}{} & & \multicolumn{1}{l}{} & \multicolumn{1}{l}{U-Type} & \multicolumn{1}{l}{} & \multicolumn{1}{l}{} & \multicolumn{1}{l}{M-Type} & \multicolumn{1}{l}{} \\
			\multicolumn{1}{l}{} & & \multicolumn{1}{l}{$\bar \mu =0.06$} & \multicolumn{1}{l}{$\bar \mu =0.12$} & \multicolumn{1}{l}{$\bar \mu =0.18$} & \multicolumn{1}{|l}{$\bar \mu=0.06$} & \multicolumn{1}{l}{$\bar \mu =0.12$} & \multicolumn{1}{l}{$\bar \mu =0.18$}\\
			\hline 
			\multicolumn{1}{l|}{}  & \multicolumn{1}{l|}{$n = 500$} & \multicolumn{1}{|l}{0.833} & \multicolumn{1}{l}{0.542} & \multicolumn{1}{l|}{0.190} & \multicolumn{1}{|l}{0.867} & \multicolumn{1}{l}{0.689} & \multicolumn{1}{l}{0.437} \\ 
			\multicolumn{1}{l|}{Asymp.} & {$n =1000$} & \multicolumn{1}{|l}{0.744} & \multicolumn{1}{l}{0.248} & \multicolumn{1}{l|}{0.014} & \multicolumn{1}{|l}{0.819} & \multicolumn{1}{l}{0.513} & \multicolumn{1}{l}{0.194} \\ 
			\multicolumn{1}{l|}{}  & {$n =2000$} & \multicolumn{1}{|l}{0.558} & \multicolumn{1}{l}{0.032} & \multicolumn{1}{l|}{0.000} & \multicolumn{1}{|l}{0.719} & \multicolumn{1}{l}{0.260} & \multicolumn{1}{l}{0.030} \\
			\hline
			\multicolumn{1}{l|}{}  & \multicolumn{1}{l|}{$n = 500$} & \multicolumn{1}{|l}{0.874} & \multicolumn{1}{l}{0.602} & \multicolumn{1}{l|}{0.229} & \multicolumn{1}{|l}{0.872} & \multicolumn{1}{l}{0.697} & \multicolumn{1}{l}{0.446} \\ 
			\multicolumn{1}{l|}{Permut.} & {$n =1000$} & \multicolumn{1}{|l}{0.805} & \multicolumn{1}{l}{0.305} & \multicolumn{1}{l|}{0.021} & \multicolumn{1}{|l}{0.827} & \multicolumn{1}{l}{0.524} & \multicolumn{1}{l}{0.202} \\ 
			\multicolumn{1}{l|}{}  & {$n =2000$} & \multicolumn{1}{|l}{0.637} & \multicolumn{1}{l}{0.046} & \multicolumn{1}{l|}{0.000} & \multicolumn{1}{|l}{0.728} & \multicolumn{1}{l}{0.269} & \multicolumn{1}{l}{0.033} \\
			\hline
		\end{tabular}
		
	\end{center}
	\par
	\medskip 
	\parbox{6.2in}{\footnotesize
		
		Notes: We chose $\lambda = 3$ for the Erd\"{o}s-R\'{e}nyi (ER) Random Graph with $c=0.3$.
		\bigskip}
\end{table}

\subsection{Results}
\subsubsection{Finite Sample Size Properties}
First, let us report results on finite sample size properties. The results are shown in Tables 1 - 3. Table 1 presents the results from i.i.d. observations as a benchmark case. Table 2 uses simulated observations with dependency graphs where graphs are chosen to be from a single realization from two random graphs: Erd\"{o}s-Re\'{e}nyi random graph and Barab\'{a}si-Albert random graph. Finally, Table 3 shows the results from using network dependent observations.

First, for the U-type statistic approach, the permutation critical values perform conspicuously better than asymptotic critical values. Asymptotic critical values do not perform very well even in the case of independent observations. (See Table 1.) However, the contrast is much less stark for the M-type statistic approach.

Second, the U-type statistic approach exhibits more stable size properties than the M-type statistic approach. Interestingly, the performance does not seem to worsen much as the graph gets denser and the correlation stronger. This is perhaps because as one observation has more neighbors, the correlation between the observation and each neighbor tends to be weaker by the design of the data generating process.

Third, the randomized subsampling approach tends to over-reject the null hypothesis in the case of network dependence case (Table 3), where the overjection becomes severe as the correlation between linked observations gets stronger. 

Throughout the simulation study, the increase in the sample size does not necessarily show better size properties. This may be because the graph tends to have more nodes with more neighbors as the sample size becomes larger, and this may offset the improvement in size properties partially.

\subsubsection{Power Properties}

Let us turn to the power properties of the randomized tests. The results are shown in Table 4. The false coverage probabilities using asymptotic critical values are lower than those using permutation critical values. This is not surprising given that the asymptotic critical values exhibit lower coverage probabilities at the true value of $\mu_0=0$ than permutation critical values.

Interestingly, the case of independent observations show only very slightly lower false coverage probabilities than the case of dependency graphs show similar results. This demonstrates the robustness properties of the randomized subsampling approach. However, in the case of network dependent observations, the false coverage probabilities are higher. The performance of randomized subsampling approach in the simulation designs and the choice of $R$, $b_n$, and $n$ does not seem stable for network dependent observations.

Finally, the M-type statistic approach exhibits larger false coverage probability than the U-type statistic approach, despite its worse size distortion. This may reflect the inferior performance of the M-type statistic approach as theoretically shown in terms of rate-dominance with size control.

\section{Conclusion}
This paper proposes a randomized subsampling approach to perform inference with locally dependent data when the dependence ordering is not known. This paper first introduces the notion of local dependence that does not invoke any reference to the underlying dependence ordering, and pursues ordering-free inference based on M-type statistic and U-type statistic approaches. The main results include establishing conditions for the local dependence which ensure the asymptotic validity of the approaches, and introducing the notion of rate dominance with size control to formally compare the two approaches, and show that U-type statistic approach rate-dominates the M-type statistic approach when permutation-based critical values are used. In general, there is a tradeoff between size and power in the randomized subsampling approach. Of course, when we have a huge number of observations, choosing $R_n$ and $b_n$ much smaller than $n$ can improve the small sample size property without hurting much its power. Some theoretical results suggesting a good combination of $R_n$ and $b_n$ in general would be desirable.

\section{Appendix: Mathematical Proofs}

Recall the definition of $Z_{i,n} = \Sigma_n^{-1/2}(X_{i,n} - \mu_0)/\sqrt{d}$. We begin by providing a moment bound. The proof uses the recursive approach of a Doukhan-Portal type inequality (due to \cite{Doukhan/Portal:83:CRASPS}) which is derived for our set-up. (See e.g. Lemma 3 of \citet*{Andrews/Pollard:94:ISR} and Lemma 14 of \cite{Doukhan/Louhichi:99:SPA}.)
\begin{lemma}
	\label{moment bound2}
	Suppose that for some positive integer $q \ge 1$, and any positive sequence $a_n$, there exist $\bar C>0$ such that $b_n^k a_n^{-1} \lambda_n(k) \le \bar C$, for all $n \ge 1$ and for each $2 \le k \le 2q$.
	 
	Then there exists a constant $C_{d,q}>0$ that depends only on $d$, $q$ and $\bar C$ such that
	\begin{eqnarray*}
		\frac{1}{a_n |\Pi|} \sum_{\pi \in \Pi} \mathbf{E} \left(\left\|\frac{1}{\sqrt{b_n}} \sum_{i=1}^{b_n} Z_{\pi(i),n} \right\|^{2q} |\mathcal{C}_n \right)
		\le C_{d,q} \sum_{k=1}^{2q} \max_{1 \le i \le n} \mathbf{E}\left[\|Z_{i,n}\|^{2q}|\mathcal{C}_n \right],
	\end{eqnarray*}
	for all $n \ge 1$.
\end{lemma}

The poof of Lemma \ref{moment bound2} is found in Supplemental Note. Let us define
\begin{eqnarray}
\label{xi M xi U}
	\xi_{M,n}(\pi_r) &\equiv& \frac{1}{\sqrt{b_n}} \sum_{i=1}^{b_n} Z_{\pi_r(i),n}, \text{ and }\\
	\xi_{U,n}(\pi_r) &\equiv& \frac{1}{b_n} \sum_{i=1}^{b_n} \sum_{j=1, i \ne j}^{b_n} Z_{\pi_r(i),n}'Z_{\pi_r(j),n},
\end{eqnarray}
and let for each $\tau \in \{M,U\}$,
\begin{eqnarray*}
	S_{\tau,n}^*(\boldsymbol{\pi}) &\equiv& \frac{1}{\sqrt{R_n}} \sum_{r=1}^{R_n} \xi_{\tau,n}(\pi_r).
\end{eqnarray*}
We focus on the asymptotic properties of $S_{\tau,n}^*(\boldsymbol{\pi})$. For this, write (with $\tau \in \{M,U\}$)
\begin{eqnarray}
    \label{decomp}
	S_{\tau,n}^*(\boldsymbol{\pi}) = S_{\tau,A,n}^*(\boldsymbol{\pi}) + S_{\tau,B,n}^*(\boldsymbol{\pi}),
\end{eqnarray}
where
\begin{eqnarray*}
	S_{\tau,A,n}^*(\boldsymbol{\pi}) &=& \frac{1}{\sqrt{R_n}} \sum_{r=1}^{R_n} \left( \xi_{\tau,n}(\pi_r) - \mathbf{E}[\xi_{\tau,n}(\pi_r)|\mathscr{Z}_n] \right) \textnormal{ and }\\
	S_{\tau,B,n}^*(\boldsymbol{\pi}) &=& \frac{1}{\sqrt{R_n}} \sum_{r=1}^{R_n}  \mathbf{E}[\xi_{\tau,n}(\pi_r)|\mathscr{Z}_n].
\end{eqnarray*}
The following lemma gives the convergence rates of $S_{\tau,B,n}^*(\boldsymbol{\pi})$ for each $\tau \in \{M,U\}$.

\begin{lemma}
	\label{Tn_B}
	Suppose that Assumption \ref{assump: moment cond} holds. Then
	\begin{eqnarray*}	
		\mathbf{E}[||S_{M,B,n}^*(\boldsymbol{\pi})||^2] &=& O(R_n b_n (\bar \lambda_n(2)+n^{-1})), \text{ and } \\
		\mathbf{E}[(S_{U,B,n}^*(\boldsymbol{\pi}))^2] &=& O(R_n b_n^2\{\bar \lambda_n(4)+n^{-1}(\bar \lambda_n(2)+\bar \lambda_n(3))+n^{-2}\}).
	\end{eqnarray*}
\end{lemma}

\noindent \textbf{Proof: } For notational brevity, denote
\begin{eqnarray}
\label{zeta}
\zeta_{i,j,n} \equiv Z_{i,n}'Z_{j,n}, \textnormal{ and } \tilde \zeta_{i,j,n} \equiv Z_{i,n}Z_{j,n}'.	
\end{eqnarray}
Since we draw $\pi_r$'s i.i.d. from the uniform distribution on $\Pi$, we can rewrite
\begin{eqnarray*}
	S_{M,B,n}^*(\boldsymbol{\pi}) = \frac{\sqrt{R_n b_n}}{n} \sum_{i=1}^{n} Z_{i,n}.
\end{eqnarray*}
Hence for some constant $C>0$,
\begin{eqnarray*}
	\mathbf{E}[||S_{M,B,n}^*(\boldsymbol{\pi})||^2]
	&=& \frac{R_n b_n}{n^2} \sum_{i=1}^{n} \mathbf{E}[\zeta_{i,i,n}] + \frac{R_n b_n}{n^2} \sum_{i=1}^n \sum_{j=1: j \ne i}^n \mathbf{E}[\zeta_{i,j,n}]\\
	&\le& \frac{R_n b_n}{n^2} \sum_{i=1}^{n} \mathbf{E}[\zeta_{i,i,n}] + C R_n b_n \lambda_n(2)
	=O(R_n b_n (n^{-1} + \lambda_n(2))).
\end{eqnarray*}

Let us now turn to the second statement. We write
\begin{eqnarray*}
	S_{U,B,n}^*(\boldsymbol{\pi}) = \frac{\sqrt{R_n} (b_n-1)}{n(n-1)} \sum_{i=1}^n \sum_{j=1: j \ne i}^n \zeta_{i,j,n}. 
\end{eqnarray*}
Note that
\begin{eqnarray*}
	\mathbf{E}[(S_{U,B,n}^*(\boldsymbol{\pi}))^2] &=& R_n(b_n - 1)^2 \mathbf{E}\left[\frac{1}{n^2(n-1)^2} \sum_{i=1}^n \sum_{j=1: j \ne i}^n \zeta_{i,j,n}^2\right]+R_n(b_n - 1 )^2 A_n,
\end{eqnarray*}
where
\begin{eqnarray}
    \label{A_n}
	A_n =  \frac{1}{n^2(n-1)^2} \sum_{(i_1,j_1,i_2,j_2)} \mathbf{E}\left[ \zeta_{i_1,j_1,n}\zeta_{i_2,j_2,n} \right],
\end{eqnarray}
and the sum over $(i_1,j_1,i_2,j_2)$ includes 4-tuples of positive integers from $1$ to $n$ such that $i_1 \ne j_1$, $i_2 \ne j_2$, and $(i_1,j_1) \ne (i_2,j_2)$. The leading term in the decomposition above is $O(n^{-2} R_n b_n^2)$ by the moment conditions in Assumption \ref{assump: moment cond}.

Let us analyze $A_n$ which we write
\begin{eqnarray*}
	A_n = 4 B_{1,n}+B_{2,n},
\end{eqnarray*}
where
\begin{eqnarray*}
	B_{1,n} &=& \frac{1}{n^2(n-1)^2} \sum_{(i_1,j_1,j_2)^*} \mathbf{E}\left[\zeta_{i_1,j_1,n}
	\zeta_{i_1,j_2,n} \right], \textnormal{ and } \\
	B_{2,n} &=& \frac{1}{n^2(n-1)^2} \sum_{(i_1,j_1,i_2,j_2)^*} \mathbf{E}\left[\zeta_{i_1,j_1,n}\zeta_{i_2,j_2,n} \right],
\end{eqnarray*}
where the sum over $(i_1,j_1,j_2)^*$ is over all 3-tuples of distinct integers in $\{1,...,n\}$ and the sum over $(i_1,j_1,i_2,j_2)^*$ is over all 4-tuples of distinct integers in $\{1,...,n\}$. The factor 4 in front of $B_{1,n}$ appears because for each 4-tuple, say $(i_1',j_1',i_2',j_2')$, there are four ways to form a pair with one from  $(i_1',j_1')$ and the other from $(i_2',j_2')$.

We write
\begin{eqnarray}
\label{B_{1,n}}
	\quad \quad B_{1,n} &=& \frac{1}{n^2(n-1)^2} \sum_{(i_1,j_1,j_2)^*} \textnormal{tr}\left(\mathbf{E}[\tilde \zeta_{i_1,i_1,n}\tilde \zeta_{j_2,j_1,n}] -\mathbf{E}[\mathbf{E}[\tilde \zeta_{i_1,i_1,n}|\mathcal{C}_n] \mathbf{E} [\tilde \zeta_{j_2,j_1,n}|\mathcal{C}_n]] \right) \\
	&& +\frac{1}{n^2(n-1)^2} \sum_{(i_1,j_1,j_2)^*} \mathbf{E}\left[ \textnormal{tr} \left(\mathbf{E}[\tilde \zeta_{i_1,i_1,n}|\mathcal{C}_n] \mathbf{E} [\tilde \zeta_{j_2,j_1,n}|\mathcal{C}_n] \right)\right]. \notag
\end{eqnarray}
By Assumption \ref{assump: moment cond}(i), the last term is bounded by  $Cn^{-1}\mathbf{E}[\lambda_n(2)] = Cn^{-1} \bar \lambda_n(2)$, for some constant $C>0$ that does not depend on $n$.

We turn to the leading term on the right hand side of (\ref{B_{1,n}}). Note that
\begin{eqnarray*}
	\mathcal{P}(\{i_1,j_1,j_2\}) = \{\{\{i_1\},\{j_1,j_2\}\},\{\{i_1,j_1\},\{j_2\}\},\{\{i_1,j_2\},\{j_1\}\}\}.
\end{eqnarray*}
The trace in the leading sum in (\ref{B_{1,n}}) is bounded by $d \mathbf{E}[c_n(\{i_1\},\{j_1,j_2\})]$. As we can also write (using the fact that $\mathbf{E} \left[Z_{j_1,n}|\mathcal{C}_n \right]=0$)
\begin{eqnarray*}
	\mathbf{E}[\tilde \zeta_{i_1,i_1,n}\tilde \zeta_{j_2,j_1,n} |\mathcal{C}_n] = \mathbf{E}[\tilde \zeta_{i_1,i_1,n}\tilde \zeta_{j_2,j_1,n} |\mathcal{C}_n] -\mathbf{E}[Z_{i_1,n}Z_{i_1,n}'Z_{j_2,n}|\mathcal{C}_n]\mathbf{E}[Z_{j_1,n}'|\mathcal{C}_n],
\end{eqnarray*} 
the trace in the leading sum of (\ref{B_{1,n}}) is bounded by
\begin{eqnarray*}
	d \mathbf{E}[c_n(\{i_1,j_2\},\{j_1\})]+d C\mathbf{E}[c_n(\{j_2\},\{j_1\})],
\end{eqnarray*}
for some constant $C>0$, where $c_n(\{j_2\},\{j_1\})$ is due to $\mathbf{E} [\tilde \zeta_{j_2,j_1,n}|\mathcal{C}_n]$. Similarly, the same trace is also bounded by $d \mathbf{E}[c_n(\{i_1,j_1\},\{j_2\})]+d C \mathbf{E}[c_n(\{j_2\},\{j_1\})]$. Therefore, the leading sum on the right hand side of (\ref{B_{1,n}}) is bounded by $C n^{-1}(\mathbf{E}[\lambda_n(3)]+\mathbf{E}[\lambda_n(2)])=Cn^{-1}(\bar \lambda_n(2)+\bar \lambda_n(3))$, so that
\begin{eqnarray*}
	B_{1,n} \le C n^{-1}(\bar \lambda_n(2)+\bar \lambda_n(3)).
\end{eqnarray*}

Let us turn to $B_{2,n}$. By the definition of $\lambda_n(\cdot)$, we have
\begin{eqnarray*}
	B_{2,n} \le C \mathbf{E}[\lambda_n(4)]=C\bar \lambda_n(4),
\end{eqnarray*}
for some $C>0$. We conclude that
\begin{eqnarray*}
	R_n(b_n-1)^2 A_n = O(R_n b_n^2\{\bar \lambda_n(4) + n^{-1}(\bar \lambda_n(3)+\bar \lambda_n(2))\}),
\end{eqnarray*}
completing the proof. $\blacksquare$
\medskip

\begin{lemma}
	\label{Var}
	\noindent (i) Under Assumptions \ref{assump: local dep M} M-(i) and \ref{assump: moment cond}, $Var(S_{M,A,n}^*(\boldsymbol{\pi})|\mathscr{Z}_n) = I_d/d +o_P(1)$.
	
	\noindent (ii) Under Assumptions \ref{assump: local dep U} U-(i) and \ref{assump: moment cond},
	$Var(S_{U,A,n}^*(\boldsymbol{\pi})|\mathscr{Z}_n) = 1 + o_P(1).$
\end{lemma}
\noindent \textbf{Proof: } (i) Write
\begin{eqnarray*}
	Var(S_{M,A,n}^*(\boldsymbol{\pi})|\mathscr{Z}_n)
	= \frac{1}{n}\sum_{i=1}^n Z_{i,n}Z_{i,n}' - \overline{Z}_n \overline{Z}_n'.
\end{eqnarray*}
As for the leading term,
\begin{eqnarray}
\label{exp}
	&& \mathbf{E} \left(\left\|\frac{1}{n}\sum_{i=1}^n \left(Z_{i,n}Z_{i,n}' - \mathbf{E}[Z_{i,n}Z_{i,n}'|\mathcal{C}_n]\right) \right\|^2 \right) \\ \notag
	&=& \frac{1}{n^2}\sum_{i=1}^n  \textnormal{tr} \left(\mathbf{E}\left[(Z_{i,n}Z_{i,n}' - \mathbf{E}[Z_{i,n}Z_{i,n}'|\mathcal{C}_n])^2\right] \right) \\ \notag
	&+& \frac{1}{n^2}\sum_{i=1}^n \sum_{j=1,j \ne i}^n  \textnormal{tr} \left(\mathbf{E}\left[(Z_{i,n}Z_{i,n}' - \mathbf{E}[Z_{i,n}Z_{i,n}'|\mathcal{C}_n])(Z_{j,n}Z_{j,n}' - \mathbf{E}[Z_{j,n}Z_{j,n}'|\mathcal{C}_n])'\right] \right).  \\ \notag
\end{eqnarray}
The right hand side of the above equality is equal to $O(n^{-1}+\lambda_n(2))$. In the same way,
\begin{eqnarray}
\label{bar Z}
\overline{Z}_n = \frac{1}{n}\sum_{i=1}^n Z_{i,n} = O_P(n^{-1/2}+\sqrt{\lambda_n(2)}).
\end{eqnarray}
Therefore, $\overline{Z}_n \overline{Z}_n' = o_P(1)$, and
\begin{eqnarray*}
	Var(S_{M,A,n}^*(\boldsymbol{\pi})|\mathscr{Z}_n) 
	= \frac{1}{n}\sum_{i=1}^n \mathbf{E}[Z_{i,n}Z_{i,n}'|\mathcal{C}_n] + o_P(1) = I_d + o_P(1).
\end{eqnarray*}

\noindent (ii) Let us consider
\begin{eqnarray*}
	Var(S_{U,A,n}^*(\boldsymbol{\pi})|\mathscr{Z}_n) 
	&=& \frac{1}{n(n-1)} \sum_{i=1}^n \sum_{j=1:j \ne i}^n \zeta_{i,j,n}^2
	- \left(\frac{1}{n(n-1)} \sum_{i=1}^n \sum_{j=1:j \ne i}^n \zeta_{i,j,n}\right)^2\\
	&=& \frac{1}{n(n-1)} \sum_{i=1}^n \sum_{j=1:j \ne i}^n \zeta_{i,j,n}^2 +o_P(1), 
\end{eqnarray*}
because, as we saw in the proof of Lemma \ref{Tn_B},
\begin{eqnarray*}
	\frac{1}{n(n-1)} \sum_{i=1}^n \sum_{j=1:j \ne i}^n \zeta_{i,j,n}
	 = o_P(1).
\end{eqnarray*}
Now we write
\begin{eqnarray}
\label{expr}
	\frac{1}{n(n-1)} \sum_{i=1}^n \sum_{j=1:j \ne i}^n \zeta_{i,j,n}^2 =
	\frac{1}{n(n-1)} \sum_{i=1}^n \sum_{j=1:j \ne i}^n \mathbf{E}[\zeta_{i,j,n}^2|\mathcal{C}_n] +\Delta_n,
\end{eqnarray}
where
\begin{eqnarray*}
	\Delta_n = \frac{1}{n(n-1)} \sum_{i=1}^n \sum_{j=1:j \ne i}^n \left(\zeta_{i,j,n}^2 - \mathbf{E}[\zeta_{i,j,n}^2|\mathcal{C}_n]\right).
\end{eqnarray*}
Rewrite the leading term on the right hand side of (\ref{expr}) as
\begin{eqnarray*}
	1+\frac{1}{n(n-1)} \sum_{i=1}^n \sum_{j=1:j \ne i}^n \textnormal{tr} \left(\mathbf{E}[(Z_{i,n}Z_{i,n}'-I_d) (Z_{j,n}Z_{j,n}'-I_d)|\mathcal{C}_n] \right).
\end{eqnarray*}
The last term is bouned by $C \lambda_n(2)=o_P(1)$.

We turn to $\Delta_n$. Write
\begin{eqnarray*}
	\mathbf{E}[\Delta_n^2] &=&
	\frac{1}{n^2(n-1)^2} \sum_{(i_1,j_1,i_2,j_2)} \mathbf{E} \left[\zeta_{i_1,j_1,n}^2\zeta_{i_2,j_2,n}^2 \right]\\
	&-&\frac{1}{n^2(n-1)^2} \sum_{(i_1,j_1,i_2,j_2)} \mathbf{E}\left[\mathbf{E}[\zeta_{i_1,j_1,n}^2|\mathcal{C}_n] \mathbf{E}[\zeta_{i_2,j_2,n}^2|\mathcal{C}_n]\right],
\end{eqnarray*}
where the sum over $(i_1,j_1,i_2,j_2)$ is over all the 4-tuples of positive integers from $\{1,...,n\}$ such that 4-tuples of positive integers from $1$ to $n$ such that $i_1 \ne j_1$, $i_2 \ne j_2$, and $(i_1,j_1) \ne (i_2,j_2)$. Again, we write
\begin{eqnarray*}
	\mathbf{E}[\Delta_n^2] = \Delta_{1,n}+\Delta_{2,n},
\end{eqnarray*}
where
\begin{eqnarray*}
	\Delta_{1,n} &=& \frac{1}{n^2(n-1)^2} \sum_{(i_1,j_1,j_2)^*} \mathbf{E} \left[\zeta_{i_1,j_1,n}^2\zeta_{i_1,j_2,n}^2 \right]\\
	&-&\frac{1}{n^2(n-1)^2} \sum_{(i_1,j_1,j_2)^*} \mathbf{E}\left[\mathbf{E}[\zeta_{i_1,j_1,n}^2|\mathcal{C}_n] \mathbf{E}[\zeta_{i_1,j_2,n}^2|\mathcal{C}_n]\right]
\end{eqnarray*}
and
\begin{eqnarray*}
	\Delta_{2,n} &=& \frac{1}{n^2(n-1)^2} \sum_{(i_1,j_1,i_1,j_2)^*} \mathbf{E} \left[\zeta_{i_1,j_1,n}^2\zeta_{i_2,j_2,n}^2 \right]\\
	&-&\frac{1}{n^2(n-1)^2} \sum_{(i_1,j_1,i_2,j_2)^*} \mathbf{E}\left[\mathbf{E}[\zeta_{i_1,j_1,n}^2|\mathcal{C}_n] \mathbf{E}[\zeta_{i_2,j_2,n}^2|\mathcal{C}_n]\right].
\end{eqnarray*}
As in the proof of Lemma \ref{Tn_B}, the sum over $(i_1,j_1,j_2)^*$ is over all 3-tuples of distinct integers in $\{1,...,n\}$ and the sum over $(i_1,j_1,i_2,j_2)^*$ is over all 4-tuples of distinct integers in $\{1,...,n\}$. By Assumption \ref{assump: moment cond}, and by counting the number of all 3-tuples of distinct integers in $\{1,...,n\}$, we have
\begin{eqnarray*}
	\Delta_{1,n} = O(n^{-1}).
\end{eqnarray*}
As for $\Delta_{2,n}$, we write
\begin{eqnarray*}
	\Delta_{2,n} = \frac{1}{n^2(n-1)^2} \sum_{(i_1,j_1,i_1,j_2)^*} \mathbf{E}\left[Cov(\zeta_{i_1,j_1,n}^2, \zeta_{i_2,j_2,n}^2|\mathcal{C}_n)\right] \le C \mathbf{E}[\lambda_n(4)] = o(1).
\end{eqnarray*}
We conclude that $\Delta_n = o_P(1).$ $\blacksquare$
\medskip

Recall the definitions of $\xi_{M,n}(\pi_r)$ and $\xi_{U,n}(\pi_r)$ in (\ref{xi M xi U}).

\begin{lemma}
	\label{negligible_1}
	\noindent Suppose that the conditions of Lemma \ref{moment bound2} hold. Then for any $a \in \mathbf{R}^d$,
	\begin{eqnarray*}	
		 \mathbf{E}\left[|a'\xi_{M,n}(\pi_r) - \mathbf{E}[a'\xi_{M,n}(\pi_r)|\mathscr{Z}_n]|^3  \right]	&=& O(1), \text{ and }\\
		\mathbf{E}\left[|\xi_{U,n}(\pi_r) - \mathbf{E}[\xi_{U,n}(\pi_r)|\mathscr{Z}_n]|^3  \right]	&=& O(1).
	\end{eqnarray*}
\end{lemma}
\noindent \textbf{Proof: } We bound for some $C>0$,
\begin{eqnarray*}
	\mathbf{E}[|a'\xi_{M,n}(\pi_r) - \mathbf{E}[a'\xi_{M,n}(\pi_r)|\mathscr{Z}_n]|^3 ] 
	\le C \mathbf{E}\left[\left(\mathbf{E}[|a'\xi_{M,n}(\pi_r)|^4|\mathscr{Z}_n]\right)^{3/4} \right] \le C \left(\mathbf{E}[|a'\xi_{M,n}(\pi_r)|^4] \right)^{3/4}.
\end{eqnarray*}
Note that
\begin{eqnarray*}
	\mathbf{E}[|a'\xi_{M,n}(\pi_r)|^4] &\le& \frac{1}{|\Pi|} \sum_{\pi \in \Pi} \mathbf{E} \left[\frac{1}{\sqrt{b_n}} \sum_{i=1}^{b_n} a'Z_{\pi(i),n} \right]^4 \le C,
\end{eqnarray*}
for some $C>0$ by Lemma \ref{moment bound2}.

As for the second statement, similarly as before, we bound for some $C>0$,
\begin{eqnarray*}
	\mathbf{E}[|\xi_{U,n}(\pi_r) - \mathbf{E}[\xi_{U,n}(\pi_r)|\mathscr{Z}_n]|^3 ] 
	\le C \left(\mathbf{E}[\xi_{U,n}(\pi_r)^4] \right)^{3/4}.
\end{eqnarray*}
Note that
\begin{eqnarray*}
	\mathbf{E}[\xi_{U,n}(\pi_r)^4] &\le& \frac{1}{|\Pi|} \sum_{\pi \in \Pi} \mathbf{E} \left[\frac{1}{b_n} \sum_{i=1}^{b_n} \sum_{j=1, i \ne j}^{b_n} \zeta_{\pi(i),\pi(j),n} \right]^4.\\
\end{eqnarray*}
We bound the last term by
\begin{eqnarray*}
	 && \frac{2^3}{|\Pi|} \sum_{\pi \in \Pi} \mathbf{E} \left(\left(\frac{1}{\sqrt{b_n}} \sum_{i=1}^{b_n} Z_{\pi(i),n} \right)'\left(\frac{1}{\sqrt{b_n}} \sum_{i=1}^{b_n} Z_{\pi(i),n} \right) \right)^4\\
	&+& \frac{2^3}{|\Pi|} \sum_{\pi \in \Pi} \mathbf{E} \left(\frac{1}{b_n} \sum_{i=1}^{b_n} Z_{\pi(i),n}'Z_{\pi(i),n} \right)^4 
	 \equiv D_{1,n} +D_{2,n}.
\end{eqnarray*}
By Lemma \ref{moment bound2}, we have for some $C>0$, $D_{1,n} \le C$. As for $D_{2,n}$, we use Jensen's inequality and bound
\begin{eqnarray*}
	D_{2,n} \le \frac{2^3}{|\Pi|} \sum_{\pi \in \Pi} \frac{1}{b_n} \sum_{i=1}^{b_n} \mathbf{E}\left[Z_{\pi(i),n}'Z_{\pi(i),n} \right]^4 \le C,
\end{eqnarray*}
for some $C>0$, completing the proof. $\blacksquare$
\medskip

\begin{lemma}
	\label{Cov Mat}
	Suppose that Assumption \ref{assump: moment cond} holds, and that $\lambda_n(2) = o_P(1)$ as $n \rightarrow \infty$. Then
	\begin{eqnarray}
    \label{approx}
	\frac{1}{n}\sum_{i=1}^n (X_{i,n} - \overline{X}_n)(X_{i,n} - \overline{X}_n)' = \Sigma_n + O_P(n^{-1/2}+\sqrt{\lambda_n(2)}).	
	\end{eqnarray}
\end{lemma}
 
\noindent \textbf{Proof: } Let $D_n = \frac{1}{n}\sum_{i=1}^n \left(Z_{i,n}Z_{i,n}' - \mathbf{E}[Z_{i,n}Z_{i,n}'|\mathcal{C}_n]\right) - \overline{Z}_n\overline{Z}_n'$ and write
\begin{eqnarray*}
	\frac{1}{n}\sum_{i=1}^n (X_{i,n} - \overline{X}_n)(X_{i,n} - \overline{X}_n)' = \Sigma_n + \Sigma_n^{1/2} D_n \Sigma_n^{1/2}.
\end{eqnarray*}
The desired result follows because $D_n = O_P(n^{-1/2}+\sqrt{\lambda_n(2)})$. $\blacksquare$
\medskip

\begin{lemma}
	\label{Var_2}
	\noindent Suppose that Assumption \ref{assump: moment cond} holds. Then
	\begin{eqnarray*}
		Var\left(\frac{1}{\sqrt{R_n}} \sum_{r=1}^{R_n} \frac{1}{\sqrt{b_n}} \sum_{i=1}^{b_n} Z_{\pi_r(i),n} | \mathscr{Z}_n \right) = I_d/d + o_P(1).
	\end{eqnarray*}
\end{lemma}

\noindent \textbf{Proof: } We write
\begin{eqnarray*}
	&& Var\left(\frac{1}{\sqrt{R_n}} \sum_{r=1}^{R_n} \frac{1}{\sqrt{b_n}} \sum_{i=1}^{b_n} Z_{\pi_r(i),n} | \mathscr{Z}_n \right) \\
	&=& \frac{1}{R_n} \sum_{r=1}^{R_n} \mathbf{E} \left[\left(\frac{1}{\sqrt{b_n}} \sum_{i=1}^{b_n} (Z_{\pi_r(i),n} - \overline{Z}_n) \right)\left(\frac{1}{\sqrt{b_n}} \sum_{i=1}^{b_n} (Z_{\pi_r(i),n}  - \overline{Z}_n) \right)' | \mathscr{Z}_n \right].
\end{eqnarray*}\\
We write the last term as $E_{1,n} + E_{2,n}$, where
\begin{eqnarray*}
	E_{1,n} &=& \frac{1}{R_n} \sum_{r=1}^{R_n} \mathbf{E} \left[\frac{1}{b_n} \sum_{i=1}^{b_n} (Z_{\pi_r(i),n}-\overline{Z}_n)(Z_{\pi_r(i),n}-\overline{Z}_n)'| \mathscr{Z}_n \right] \textnormal{ and }\\
	E_{2,n} &=& \frac{1}{R_n} \sum_{r=1}^{R_n} \mathbf{E} \left[\frac{1}{b_n} \sum_{i,j=1, i \ne j}^{b_n} (Z_{\pi_r(i),n}-\overline{Z}_n)(Z_{\pi_r(j),n}-\overline{Z}_n)' | \mathscr{Z}_n \right].
\end{eqnarray*}
As for the first term, we write
\begin{eqnarray*}
	E_{1,n} &=& \frac{1}{n} \sum_{i=1}^{n} Z_{i,n}Z_{i,n}'-\overline{Z}_n\overline{Z}_n'\\
	&=& \frac{1}{n} \sum_{i=1}^{n} \left(Z_{i,n}Z_{i,n}' - \mathbf{E}[Z_{i,n}Z_{i,n}']\right)-\overline{Z}_n\overline{Z}_n'
	+ \frac{1}{n} \sum_{i=1}^{n} \mathbf{E}[Z_{i,n}Z_{i,n}'].
\end{eqnarray*}
The last term is $I_d$ and by (\ref{bar Z}),
\begin{eqnarray}
\label{bar Z2}
\overline{Z}_n\overline{Z}_n' = O_P(n^{-1} + \lambda_n(2)).
\end{eqnarray}
As for the leading term,
\begin{eqnarray*}
	&& \mathbf{E} \left\|\frac{1}{n} \sum_{i=1}^{n} \left(Z_{i,n}Z_{i,n}' - \frac{1}{d}I_d \right) \right\|^2\\
	&=& \frac{1}{n^2} \sum_{i,j=1, i \ne j}^{n} \textnormal{tr} \left( \mathbf{E} \left[\left(Z_{i,n}Z_{i,n}' - \frac{1}{d}I_d \right)\left(Z_{j,n}Z_{j,n}' - \frac{1}{d} I_d \right)' \right]\right)+O(n^{-1}).
\end{eqnarray*}
The leading term is bounded by $C\mathbf{E}[\lambda_n(2)] = C \bar \lambda_n(2)$. Hence
\begin{eqnarray*}
	E_{1,n} = \frac{1}{d} I_d + O_P(n^{-1/2} + \sqrt{\lambda_n(2)}).
\end{eqnarray*}
Let us turn to $E_{2,n}$ which we write as
\begin{eqnarray}
\label{second}
&& \frac{b_n-1}{n(n-1)} \sum_{i,j=1,i \ne j}^n \left(Z_{i,n} Z_{j,n}' - \mathbf{E}[Z_{i,n} Z_{j,n}'] \right) \\
&+& \frac{b_n-1}{n(n-1)} \sum_{i,j=1,i \ne j}^n \mathbf{E}[Z_{i,n} Z_{j,n}']+(b_n-1) B_n, \notag
\end{eqnarray}
where
\begin{eqnarray*}
	B_n &=& \overline{Z}_n\overline{Z}_n' - \frac{1}{n(n-1)} \sum_{i,j=1,i \ne j}^n \overline{Z}_nZ_{j,n}'- \frac{1}{n(n-1)} \sum_{i,j=1,i \ne j}^n Z_{i,n}\overline{Z}_n' \\
	&=& O_P(n^{-1} + \lambda_n(2)),
\end{eqnarray*}
using the same argument as in (\ref{bar Z2}).

The second term in (\ref{second}) is bounded by $b_n \lambda_n(2)$. As for the leading term in (\ref{second}), we can follow the same arguments in the proof of Lemma \ref{Tn_B} and show that it is equal to $o_P(1)$. Therefore, $E_{2,n} = o_P(1)$. $\blacksquare$
\medskip
	
\noindent \textbf{Proof of Theorem \ref{thm: Type U}: } 
\noindent (i) Let us consider the first statement. Recall the definition of $S_{M,A,n}^*(\boldsymbol{\pi})$ prior to Lemma \ref{Tn_B}. By Lemma \ref{Var}, we find that
\begin{eqnarray*}
	\mathbf{E}\left[|S_{M,A,n}^*(\boldsymbol{\pi})| |\mathscr{Z}_n\right] = O_P(1).
\end{eqnarray*}
Combining this with (\ref{approx}) and using (\ref{decomp}) and Lemma \ref{Tn_B}, we obtain that
\begin{eqnarray*}
	{S}_{M,n}(\mu_0;\boldsymbol{\pi}) = S_{M,A,n}^*(\boldsymbol{\pi}) +o_P(1).
\end{eqnarray*}
We focus on $S_{M,A,n}^*(\boldsymbol{\pi})$. Since $\{\xi_{M,n}(\pi_r)\}_{r=1}^{R_n}$ is a triangular array that is rowwise i.i.d. conditional on $\mathscr{Z}_n$, we deduce that by Berry-Esseen lemma (e.g. Theorem 3 of \citet*{Chow/Teicher:88:ProbTheory}, p.304), for some $C>0$,
\begin{eqnarray*}
	\left| P\left\{\frac{S_{M,A,n}^*(\boldsymbol{\pi})}{\sigma_{M,n}(\mathscr{Z}_n)} \le t |\mathscr{Z}_n\right\} - \Phi(t) \right|
	\le \frac{C}{R_n^{3/2}}\sum_{r=1}^{R_n} \mathbf{E}[|\xi_{M,n}(\pi_r) - \mathbf{E}[\xi_{M,n}(\pi_r)|\mathscr{Z}_n]|^3 |\mathscr{Z}_n],
\end{eqnarray*}
where $\sigma_{M,n}^2(\mathscr{Z}_n) = Var(S_{M,A,n}^*(\boldsymbol{\pi})|\mathscr{Z}_n)$. By Lemma \ref{negligible_1}, the expected value of the last bound has a rate $O(R_n^{-1/2})$. In view of Lemma \ref{Cov Mat}, this completes the proof of the first statement.

Let us turn to the second statement. From (\ref{S_M}), we write
\begin{eqnarray*}
	S_{M,n}(\overline{X}_n;\boldsymbol{\pi}) = \frac{1}{R_n b_n} \sum_{r_1=1}^{R_n} \sum_{r_2=1}^{R_n} \sum_{i_1=1}^{b_n} \sum_{i_2=1}^{b_n} Z_{\pi_{r_1}(i_1),n}' Z_{\pi_{r_2}(i_2),n} + B_{M,n} + o_P(1),
\end{eqnarray*}
where $o_P(1)$ is due to the estimation error in $\hat \Sigma$, and $B_{M,n}$ is as defined in (\ref{B_Mn}). By Lemma \ref{Var_2} and in view of the heuristics in Section \ref{sec: heuristics}, we find that
\begin{eqnarray*}
	B_{M,n} = O_P\left(R_n b_n (n^{-1/2} + n \lambda_n(2))^2 + \sqrt{R_n b_n/n}\right) = o_P(1).
\end{eqnarray*}
Therefore,
\begin{eqnarray*}
	S_{M,n}(\overline{X}_n;\boldsymbol{\pi}) = \tilde S_{M,n}(\mu_0;\boldsymbol{\pi}) + o_P(1),
\end{eqnarray*}
where $\tilde S_{M,n}(\mu_0;\boldsymbol{\pi})$ is the same as $S_{M,n}(\mu_0;\boldsymbol{\pi})$ except that $\hat \Sigma$ is replaced by $\Sigma_n$.
 
Hence the conditional distribution of $S_{M,n}(\overline{X}_n;\boldsymbol{\pi})$ given $\mathscr{Z}_n$ is the same as that of $S_{M,n}(\mu_0;\boldsymbol{\pi})$ up to a $o_P(1)$ term. From the proof of the first statement, we find that the conditional CDF of $S_{M,n}(\mu_0;\boldsymbol{\pi})$ given $\mathscr{Z}_n$ converges (uniformly over the evaluation points) in probability to $G_0$, which completes the proof.

\noindent (ii) The proof is almost the same as that of (i). $\blacksquare$
\medskip

\noindent \textbf{Proof of Theorem \ref{thm: local power analysis}: }
\noindent (i) Write $S_{M,n}(\bar \mu;\boldsymbol{\pi}) = (A_{M,n} + \delta)' \hat \Sigma^{-1} (A_{M,n} + \delta)$, where
\begin{eqnarray*}
	A_{M,n} = \frac{1}{\sqrt{d R_n b_n}} \sum_{r=1}^{R_n} \sum_{i=1}^{b_n} (X_{\pi_r(i)} - \mu_{\delta,M}). 
\end{eqnarray*}
Note that
\begin{eqnarray*}
	A_{M,n} = \frac{1}{\sqrt{d R_n b_n}} \sum_{r=1}^{R_n} \sum_{i=1}^{b_n} (X_{\pi_r(i)} - \overline X_n) + O_P\left(\frac{\sqrt{R_n b_n}}{\sqrt{n}} \right).
\end{eqnarray*}
The last $O_P$ term is $o_P(1)$ by Assumption \ref{assump: local dep M} M-(i). Thus we obtain the desired result from the proof of Theorem \ref{thm: Type U} and the continuous mapping theorem.

\noindent (ii) We write $S_{U,n}(\bar \mu;\boldsymbol{\pi}) = S_{U,n}(\mu_{\delta,U};\boldsymbol{\pi})+D_{U,n}$, where
\begin{eqnarray*}
	D_{U,n} &=& \frac{1}{\sqrt{R_n}} \sum_{r=1}^{R_n} \frac{1}{d b_n} \sum_{i=1}^{b_n} \sum_{j=1:j \ne i}^{b_n} (X_{\pi_r(i)} - \bar \mu)'\hat \Sigma^{-1}(X_{\pi_r(j)} - \bar \mu)\\
	&-& \frac{1}{\sqrt{R_n}} \sum_{r=1}^{R_n} \frac{1}{d b_n} \sum_{i=1}^{b_n} \sum_{j=1:j \ne i}^{b_n} (X_{\pi_r(i)} - \mu_{\delta,U})' \hat \Sigma^{-1}(X_{\pi_r(j)} - \mu_{\delta,U}). 
\end{eqnarray*}
We write $D_{U,n} = E_{1,U,n} +E_{2,U,n}$, where
\begin{eqnarray*}
	E_{1,U,n} &=& \left(\mu_{\delta,U} - \bar \mu\right)' \hat \Sigma^{-1} \frac{1}{\sqrt{R_n}} \sum_{r=1}^{R_n} \frac{1}{d b_n} \sum_{i=1}^{b_n} \sum_{j=1:j \ne i}^{b_n} \left(X_{\pi_r(i)}+X_{\pi_r(j)} - 2\mu_{\delta,U}\right), \textnormal{ and } \\
	E_{2,U,n} &=& \frac{\sqrt{R_n}b_n \left(\mu_{\delta,U} - \bar \mu\right)'\hat \Sigma^{-1}\left(\mu_{\delta,U} - \bar \mu\right)}{d}.
\end{eqnarray*}
We rewrite
\begin{eqnarray*}
	E_{1,U,n} &=& \left(\mu_{\delta,U} - \bar \mu\right)' \hat \Sigma^{-1}  \frac{1}{\sqrt{R_n}} \sum_{r=1}^{R_n} \frac{1}{d b_n} \sum_{i=1}^{b_n} \sum_{j=1:j \ne i}^{b_n} \left( X_{\pi_r(i)}+X_{\pi_r(j)} - 2\overline{X}_n \right)\\
	&+& \frac{2\sqrt{R_n}(b_n-1)\left(\mu_{\delta,U} - \bar \mu\right)' \hat \Sigma^{-1} \left(\overline{X}_n - \mu_{\delta,U}\right)}{d}.
\end{eqnarray*}
By Lemma \ref{Cov Mat}, and as we are under the local alternatives, the leading term is equal to $O_P(R_n^{-1/4})=o_P(1)$ and the last term is
\begin{eqnarray*}
	O_P\left(\left(\frac{\sqrt{R_n}b_n}{R_n^{1/4}b_n^{1/2}} \right) \left(\frac{1}{n^{1/2}} \right) \right) = O_P \left(\frac{R_n^{1/4}b_n^{1/2}}{n^{1/2}} \right)=o_P(1)
\end{eqnarray*}
by Assumption \ref{assump: local dep U} U-(i). As for $E_{2,U,n}$, we write it as
\begin{eqnarray*}
	\frac{\delta'\hat \Sigma^{-1} \delta}{d} = \frac{\delta' \Sigma_n^{-1} \delta}{d} +o_P(1),
\end{eqnarray*}
by (\ref{approx}). Hence we conclude that
\begin{eqnarray*}
	S_{U,n}(\bar \mu;\boldsymbol{\pi}) = S_{U,n}(\mu_{\delta,U};\boldsymbol{\pi}) + \frac{\delta' \Sigma_n^{-1} \delta}{d} +o_P(1).
\end{eqnarray*}
Using the same arguments in the proof of Theorem \ref{thm: Type U}, we obtain the desired result. $\blacksquare$
\medskip

\bibliographystyle{econometrica}
\bibliography{Local_Dependence_A7}

\newpage
\appendix
\begin{center}
	\Large Supplemental Note for ``Ordering-Free Inference from Locally Dependent Data" \medskip
	
	\normalsize
	
	Kyungchul Song\medskip
	
	\textit{Vancouver School of Economics, University of British Columbia}\medskip
	\medskip
\end{center}

\setcounter{section}{0}

Supplemental Note consists of four parts. The first part (Appendix A) explains extension of randomized subsampling inference to models with moment restrictions. The second part (Appendix B) is devoted to the proof of Lemmas \ref{lemma: dep graph} and \ref{lemma: random fields}. The third part (Appendix C) proves the moment inequality result in Lemma \ref{moment bound2}. The fourth part (Appendix D) provides the proof of Theorem \ref{thm: comparison}.
\medskip

\section{Inference from Moment Restrictions}

Suppose that we have a locally dependent triangular array of random vectors $\{X_{i,n}\}_{i=1}^n$, with $X_{i,n} \in \mathbf{R}^{d_x}$, and that there is a true parameter $\theta_0 \in \Theta \subset \mathbf{R}^{d_\theta}$ such that
\begin{eqnarray*}
	\mathbf{E}[g(X_{i,n};\theta_0)|\mathcal{C}_n] = 0,
\end{eqnarray*}
for all $i=1,...,n$, where $g(\cdot;\theta):\mathbf{R}^{d_x} \rightarrow \mathbf{R}^d$ is a given moment function. Here we do not assume that the moment restriction point-identifies $\theta_0$.

The development of randomized subsampling inference is built on the previous results on the testing on the population mean. For brevity, we focus only on the approach based on a U-type statistic.

The randomized subsampling approach developed for inference on the population mean applies in this set-up straightforwardly, by inverting the U-type statistic. First, define
\begin{eqnarray*}
	\hat \Sigma(\theta) \equiv \frac{1}{n} \sum_{i=1}^n \left(g_i(\theta) - \bar{g}(\theta) \right)\left(g_i(\theta) - \bar{g}(\theta) \right)',
\end{eqnarray*}
where $g_i(\theta) = g(X_{i,n};\theta)$ and $\bar{g}(\theta) = \frac{1}{n} \sum_{i=1}^n g_i(\theta)$. Let $\boldsymbol{\pi} = (\pi_1,...,\pi_R)$ be given as before. Define
\begin{eqnarray*}
	S_{U,n}(\theta;\boldsymbol{\pi}) \equiv \frac{1}{\sqrt{R_n}} \sum_{r=1}^{R_n} \frac{1}{db_n} \sum_{i,j =1: i \ne j}^{b_n} g_{\pi_r(i)}(\theta)'\hat \Sigma ^{-1}(\theta) g_{\pi_r(j)}(\theta),
\end{eqnarray*}
where $d$ refers to the number of the moment restrictions. Then, we define
\begin{eqnarray*}
	T_{U,n}(\theta;\boldsymbol{\pi}) &=& S_{U,n}(\theta;\boldsymbol{\pi}) - \frac{\sqrt{R_n} (b_n-1)}{n}.
\end{eqnarray*}
The normalization by $\hat \Sigma^{-1}(\theta)$ eliminates asymptotically the dependence among the moment restrictions in the limiting distribution. Again the construction of the test statistic and the critical values do not require knowledge of the dependence ordering of the triangular array $\{X_{i,n}\}_{i=1}^n.$

As for the permutation critical values, we draw $\boldsymbol \pi_l = (\pi_{1,l},...,\pi_{R_n,l})$, $l=1,...,L$, i.i.d., similarly as before, and define
\begin{eqnarray*}
	\tilde S_{U,n}(\theta;\boldsymbol \pi_l) = \frac{1}{\sqrt{R_n}} \sum_{r=1}^{R_n} \frac{1}{d b_n} \sum_{i,j =1: i \ne j}^{b_n}
	(g_{\pi_{r,l}(i)}(\theta) - \bar{g}(\theta))' \hat \Sigma ^{-1}(\theta) (g_{\pi_{r,l}(j)}(\theta) - \bar{g}(\theta)).
\end{eqnarray*}
Then we construct
\begin{eqnarray*}
	\tilde{c}_{U,\alpha}(\theta) = \inf \left\{c>0: \frac{1}{L} \sum_{l=1}^{L} 1\{\tilde S_{U,n}(\theta;\boldsymbol \pi_l)>c\} \le \alpha \right\}.
\end{eqnarray*}
Let us turn to the randomized confidence function. For a given integer $S \ge 1$, for each $s=1,...,S$, we let $\boldsymbol \pi_s = (\pi_{1,s},...,\pi_{R_n,s})$ as before. Define
\begin{eqnarray*}
	q_U(\theta;\alpha) &=& \frac{1}{S} \sum_{s=1}^S 1\{T_{U,n}(\theta;\boldsymbol \pi_s) \le \tilde{c}_{U,\alpha}(\theta) \}.
\end{eqnarray*}
We make the following assumption which is used to ensure that we can consistently estimate $\mathbf{E}[g_i(\theta)|\mathcal{C}_n]$ for each $\theta \in \Theta$ and each $i \in \{1,...,n\}$.
\begin{assumption}
	\label{assump: identical means}
	For each $\theta \in \Theta$, $\mathbf{E}[g_i(\theta)|\mathcal{C}_n]$ is identical a.s. across $i \in \{1,...,n\}$.
\end{assumption}
Let us study the asymptotic validity of the test. Define for each $\gamma\ge 0$,
\begin{eqnarray*}
	\Theta_0(\mathcal{C}_n) &=& \left\{\theta \in \Theta: \mathbf{E}[g_i(\theta)|\mathcal{C}_n] = 0 \right\} \textnormal{ and }\\
	\Theta_\gamma(\mathcal{C}_n) &=& \left\{\theta \in \Theta: ||\mathbf{E}[g_i(\theta)|\mathcal{C}_n]|| \ge \gamma \right\}.
\end{eqnarray*}
When $X_{i,n}$'s are i.i.d. (conditional on $\mathcal{C}_n$), the set $\Theta_0(\mathcal{C}_n)$ is an identified set for $\theta_0$ with respect to the conditional distribution of $X_{1,n}$ given $\mathcal{C}_n$. The set $\Theta_\gamma(\mathcal{C}_n)$ is the collection of $\theta$'s that are away from the identified set.

We make the following assumptions regarding the cross-sectional local dependence of $g_i(\theta)$'s across $i \in \{1,...,n\}$. 

\begin{assumption}
	\label{assump: local dep 2}
	(i) Assumption \ref{assump: local dep U} holds when we replace $(X_{i,n})_{i =1}^n$ by $(g_i(\theta))_{i \in \{1,...,n\}}$ for each $\theta \in \Theta$ such that $P\{\theta \in \Theta_0(\mathcal{C}_n)\} \rightarrow 1$.
	
	(ii) Assumption \ref{assump: moment cond} holds when we replace $(X_{i,n})_{i =1}^n$ by $(g_i(\theta))_{i=1}^n$ for each $\theta \in \Theta$.
\end{assumption}

We are prepared to present the asymptotic result for the randomized confidence function for $\theta_0$. The proof of this result can be obtained again by slightly modifying the proof of Theorem \ref{thm: Type U}.

\begin{corollary}
	\label{rand conf interval moment cond}
	\noindent Suppose that Assumptions \ref{assump: identical means}-\ref{assump: local dep 2} hold. Then for each $\theta \in \Theta$ and each $\gamma >0$,
	\begin{eqnarray*}
		q_U(\theta;\alpha) \underset{p}{\rightarrow} \left\{
		\begin{array}{l}				
			1-\alpha, \textit{ if } P\{\theta \in \Theta_0(\mathcal{C}_n)\} \rightarrow 1,\\
			0, \textit{ if } P\{\theta \in \Theta_\gamma(\mathcal{C}_n)\} \rightarrow 1,
		\end{array}
		\right.
	\end{eqnarray*}
	as $n,S \rightarrow \infty$.
\end{corollary}

\noindent Using this, we can construct confidence sets as in (\ref{CS}). More specfically, take $\beta \in (0,\alpha)$ and define
\begin{eqnarray}
\label{CS}
C_{U,\alpha} = \{\theta \in \Theta: q_U(\theta;\alpha-\beta) \ge 1 - \alpha\}.
\end{eqnarray}
This is the confidence set for $\theta_0$.

When the dimension of $\theta$ is high, computing the function can be cumbersome for permutation critical values $\tilde c(\theta)$, because one needs to compute the critical value for each $\theta$. Then we may consider the following profiling method. Suppose that we are interested in $\theta_1$ which is a subvector of $\theta$. Let $\theta_2$ be such that $\theta = (\theta_1,\theta_2)$. Define
\begin{eqnarray*}
	q_U(\theta_1;\alpha) = \inf_{\theta_2} \frac{1}{S} \sum_{s=1}^S 1\left\{T_{n}(\theta;\tilde \pi_s) \le \tilde c(\theta_1) \right\},
\end{eqnarray*}
where $\tilde c(\theta_1)$ is defined as 
\begin{eqnarray*}
	\tilde{c}(\theta_1) =  \inf \left\{c>0: \inf_{\theta_2} \frac{1}{L} \sum_{l=1}^{L} 1\{\tilde S_{n}(\theta;\tilde \pi_l)>c\} \le \alpha \right\}.
\end{eqnarray*}
The use of this profiling method in general gives conservative inference.

\section{Proof of Lemmas \ref{lemma: dep graph} and \ref{lemma: random fields}}

\noindent \textbf{Proof of Lemma \ref{lemma: dep graph}}: Whenever $\{\pi(1),...,\pi(k)\}$ does not constitute a connected component in the graph $G_n$, we have $c_n(\{\pi(1),...,\pi(k)\})=0$. Hence for any permutation $\pi$ such that $c_n(\{\pi(1),...,\pi(k)\})>0$, we have $\{\pi(i_1),...,\pi(i_k)\} = \{\pi(1),...,\pi(k)\}$ where $\pi(i_1)$ is chosen from $\{1,2,...,n\}$, and $\pi(i_2)$ is from the neighborhood of $\pi(i_1)$, and $\pi(i_3)$ from the neighborhood of $\pi(i_2)$, and so on. After placing $\pi(1),...,\pi(k)$ this way, we place $\pi(k+1),...,\pi(n)$ in the remaining $n-k$ places. In fact, any permutation $\pi \in \Pi$ such that $c_n(\{\pi(1),...,\pi(k)\})>0$ can be obtained in this way, except with a different ordering of $\pi(1),...,\pi(k)$ in this process. 

Thus, we have
\begin{eqnarray*}
	|\{\pi \in \Pi: c_n(\{\pi(1),...,\pi(k)\})>0 \}| \le C_k n d_n^{k-1}(n-k)!,
\end{eqnarray*}
where $C_k>0$ is a constant depending only on $k$. Thus,
\begin{eqnarray*}
	\lambda_n(k) \le \frac{C_k n d_n^{k-1}(n-k)}{n!},
\end{eqnarray*}
which gives the desired result. $\blacksquare$
\medskip

Proof of Lemma \ref{lemma: random fields} is a bit more involved. Let us introduce notation and an auxiliary lemma. For each nonnegative integer $m$ and $j_1,j_2 \in N$, let us define
\begin{eqnarray*}
	H_m'(j_1,j_2) = \{i \in N: \min\{d_\mu(i,j_1),d_\mu(i,j_2)\} < m+1 \},
\end{eqnarray*}
and let for $m \ge 1$,
\begin{eqnarray}
\label{Hm}
H_m(j_1,j_2) = H_m'(j_1,j_2) \setminus H_{m-1}'(j_1,j_2). 
\end{eqnarray}
Also we set $H_0(j_1,j_2) = H_0'(j_1,j_2) = \{j_1,j_2\}$. The sets $H_m(j_1,j_2)$, $m=0,1,2,...$ constitute a partition of $\{1,...,n\}$. Fix any integer $k \ge 1$ and define
\begin{eqnarray*}
	\Pi'_m[k] = \left\{\pi \in \Pi: \exists \{A_1,A_2\} \in \mathcal{P}(\{1,...,k\}), \textnormal{ s.t. } m \le  d_\mu(\pi(A_1),\pi(A_2)) < m+1 \right\},
\end{eqnarray*} 
where $\pi(A) = (\pi(j_1),...,\pi(j_r))$ when $A = (j_1,...,j_r)$, and let
\begin{eqnarray*}
	\Pi_m[k] = \Pi'_m[k] \setminus \bigcup_{j=m+1}^{\infty} \Pi'_j[k].
\end{eqnarray*}

\begin{lemma}
	\label{lemma: aux1}
	For each $m \ge 0$, $k \ge 1$ and $\pi \in \Pi_m[k]$, there exist $j_1,j_2 \in \{1,...,k\}$ and an integer $s_k>0$ such that $m \le d_\mu(\pi(j_1),\pi(j_2))< m+1$ and for each integer $0 \le s \le s_k$, $H_{sm}(\pi(j_1),\pi(j_2))\cap \{\pi(1),...,\pi(k)\} \ne \varnothing$ and for each integer $s > s_k$, $H_{sm}(\pi(j_1),\pi(j_2))\cap \{\pi(1),...,\pi(k)\} = \varnothing$.
\end{lemma}

\noindent \textbf{Proof:} For each $\pi \in \Pi_m[k]$, there exists a partition $(A_1,A_2)$ of $\{1,...,k\}$ such that
\begin{eqnarray*}
	m \le d_\mu(\pi(A_1),\pi(A_2))<m+1,
\end{eqnarray*}
but there exists no partition $(A_1',A_2')$ of $\{1,...,k\}$ such that $d_\mu(\pi(A_1'),\pi(A_2')) \ge m+1$. First note that by the definition of $\Pi_m[k]$, for each $\pi \in \Pi_m[k]$, there exist $j_1,j_2 \in \{1,...,k\}$ such that $m \le d_\mu(\pi(j_1),\pi(j_2)) <m+1$. Certainly, we cannot have $H_{sm}(\pi(j_1),\pi(j_2))\cap \{\pi(1),...,\pi(k)\} \ne \varnothing$ for infinite $s$'s, because $|\{\pi(1),...,\pi(k)\}| = k$. Let $\bar s$ be the smallest integer such that for all $s> \bar s$, $H_{sm}(\pi(j_1),\pi(j_2))\cap \{\pi(1),...,\pi(k)\} = \varnothing$. Let $s'$ be the largest integer such that for all $0 \le s \le s'$, $H_{sm}(\pi(j_1),\pi(j_2))\cap \{\pi(1),...,\pi(k)\} \ne \varnothing$. Then it must be that $0 \le s' \le \bar s$. For the lemma, it suffices to show that $s' = \bar s$. 

To the contrary, assume that $0 \le s'<\bar s$. We show that it contradicts that $\pi \in \Pi_m[k]$. First, by the definition of $s'$ and $\bar s$, we have $H_{s' m}(\pi(j_1),\pi(j_2))\cap \{\pi(1),...,\pi(k)\} \ne \varnothing$ and $H_{\bar s m}(\pi(j_1),\pi(j_2))\cap \{\pi(1),...,\pi(k)\} \ne \varnothing$, and there must exist $s_1$ such that $s'<s_1<\bar s$ and $H_{s_1m}(\pi(j_1),\pi(j_2))\cap \{\pi(1),...,\pi(k)\} = \varnothing$. We take a partition $(A_1',A_2')$ of $\{1,...,k\}$ such that
\begin{eqnarray*}
	\pi(A_1') &=& \bigcup_{s=0}^{s_1} H_{sm}(\pi(j_1),\pi(j_2)) \cap \{\pi(1),...,\pi(k)\}\\
	&=& \bigcup_{s=0}^{s'} H_{sm}(\pi(j_1),\pi(j_2)) \cap \{\pi(1),...,\pi(k)\},
\end{eqnarray*}
(by the choice of $s_1$) and
\begin{eqnarray*}
	\pi(A_2') &=& \bigcup_{j=s_1+1}^{\bar s} H_{sm}(\pi(j_1),\pi(j_2)) \cap \{\pi(1),...,\pi(k)\}.
\end{eqnarray*}
Since $s'<s_1<\bar s$, both $A_1'$ and $A_2'$ are not empty. Furthermore, by the definition of $H_{s_1m}(\pi(j_1),\pi(j_2))$, for all $i',j' \in \{1,...,k\}$ such that $\pi(i') \in H_{sm}(\pi(j_1),\pi(j_2))$ for some $s=0,1,...,s_1-1$ and $\pi(j') \in H_{sm}(\pi(j_1),\pi(j_2))$ for some $s=s_1+1,...,\bar s$, we have $d_\mu(\pi(i'),\pi(j')) \ge m+1$. Hence $d_\mu(\pi(A_1'),\pi(A_2')) \ge m+1$. This contradicts that $\pi \in \Pi_m[k]$. $\blacksquare$
\medskip	

\noindent \textbf{Proof of Lemma \ref{lemma: random fields}:} 
Fix $k \ge 2$. For each $m \ge 1$, let us first compute a bound for $|\Pi_m[k]|$. Let $\pi \in \Pi_m[k]$ and $j_1,j_2 \in \{1,...,k\}$ be such that $m \le d_\mu(\pi(j_1),\pi(j_2))<m+1$. Let $\bar s$ be the smallest integer such that for all $s> \bar s$, $H_{sm}(\pi(j_1),\pi(j_2))\cap \{\pi(1),...,\pi(k)\} = \varnothing$. Then by Lemma \ref{lemma: aux1}, for each integer $0 \le s \le \bar s$, $|H_{sm}(\pi(j_1),\pi(j_2))\cap \{\pi(1),...,\pi(k)\}| \ge 1$. We find a bound for $|\Pi_m[k]|$ by counting the number of $\pi$'s which satisfy this latter condition. By our setting of $H_0(\pi(j_1),\pi(j_2)) = \{\pi(j_1),\pi(j_2)\}$, we have
\begin{eqnarray*}
	|H_0(\pi(j_1),\pi(j_2))\cap \{\pi(1),...,\pi(k)\}| = 2.
\end{eqnarray*}
We consider two cases.

First, suppose that
\begin{eqnarray}
\label{bound22}
|H_{sm}(\pi(j_1),\pi(j_2))\cap \{\pi(1),...,\pi(k)\}| = 1, \text{ for each integer } 1 \le s \le \bar s.
\end{eqnarray}
By the way $H_m(j_1,j_2)$ is defined in (\ref{Hm}), each set $H_{sm}(\pi(j_1),\pi(j_2))$, $s=1,2,...$, excludes $\pi(j_1)$ and $\pi(j_2)$. Hence we must have $\bar s = k-2$, and there exist $i_1,...,i_{k-2}$ such that $\{i_1,...,i_{k-2}\} = \{1,...,k\}\setminus \{j_1,j_2\}$, and
\begin{eqnarray*}
	\pi(i_1) \in H_{m}(\pi(j_1),\pi(j_2)), \pi(i_2) \in H_{2m}(\pi(j_1), \pi(j_2)),...,\pi(i_{k-2}) \in H_{(k-2)m}(\pi(j_1),\pi(j_2)).
\end{eqnarray*}
A bound for the number of such permutations is computed as follows. First we place $\pi(j_1)$ in one of $n$ places, and then place $\pi(j_2)$ so that $m \le d_\mu(\pi(j_1),\pi(j_2))<m+1$. The total number of fixing $j_1$ and $j_2$ this way is bounded by $Cnm^{d-1}$ for some constant $C>0$. (See Lemma A.1 of \citet*{Jenish/Prucha:09:Supp}.) Now, we choose $\pi(i_1)$ from $H_{m}(\pi(j_1),\pi(j_2))$ and the number of choosing $\pi(i_1)$ this way is bounded by $Cm^{d-1}$, and choose $\pi(i_2)$ from $H_{2m}(\pi(j_1),\pi(j_2))$ and the number of choosing $\pi(i_1)$ this way is bounded by $C(2m)^{d-1}$. We keep choosing $\pi(i_2)$,...,$\pi(i_{k-2})$ this way. Then we choose the remaining $\pi(k+1),...,\pi(n)$. Hence the total number of the permutations $\pi$ which satisfy (\ref{bound22}) is bounded by
\begin{eqnarray}
\label{bound23}
&& Cnm^{d-1} m^{d-1} (2m)^{d-1} ... ((k-2)m)^{d-1} (n-k)! \\ \notag
&& \le C n m^{(k-1)(d-1)} ((k-2)!)^{d-1}(n-k)!, 
\end{eqnarray}
for some $C>0$ that does not depend on $n$.

Second, suppose that
\begin{eqnarray}
\label{bound24}
|H_{sm}(\pi(j_1),\pi(j_2))\cap \{\pi(1),...,\pi(k)\}| > 1, \text{ for some } 1 \le s \le \bar s.
\end{eqnarray}
Then it is not hard to show that the number of $\pi$'s with this property is bounded by the bound in (\ref{bound23}), because we can choose a bound for the number of choosing both $\pi(i)$ and $\pi(i')$, say, from the same $H_{sm}(\pi(j_1),\pi(j_2))$ in a way that the bound is smaller than the previous bound that comes from choosing $\pi(i)$ from $H_{sm}(\pi(j_1),\pi(j_2))$ and $\pi(i')$ from $H_{(s+1)m}(\pi(j_1),\pi(j_2))$. Thus the total number of the permutations $\pi$ which satisfy (\ref{bound24}) has the same bound in (\ref{bound23}).

Therefore,
\begin{eqnarray*}
	\frac{1}{|\Pi|} \sum_{\pi \in \Pi} c_n(\{\pi(1),...,\pi(k)\})
	&\le& \frac{1}{|\Pi|} \sum_{m=1}^{\infty} \sum_{\pi \in \Pi_m[k]} c_n(\{\pi(1),...,\pi(k)\}) \\
	&\le& \frac{1}{|\Pi|} \sum_{m=1}^{\infty} \sum_{\pi \in \Pi_m[k]} \bar{c}_{m,n}(\{\pi(1),...,\pi(k)\})\\
	&\le& \frac{C_{k,d}n(n-k)!}{n!} \sum_{m=1}^{\infty} m^{(k-1)(d-1)} \max_{\pi \in \Pi} \bar{c}_{m,n}(\{\pi(1),...,\pi(k)\}).
\end{eqnarray*}
Then it follows from (\ref{cond}) that for any fixed $k \ge 1$,
\begin{eqnarray*}
	\lambda_n(k) \le C_{k,d} n^{-k+1},	
\end{eqnarray*}
for some constant $C_{k,d}>0$. $\blacksquare$

\section{Proof of Lemma \ref{moment bound2}}

First, we introduce a basic inequality that involves permutations. For each $A \subset \{1,...,n\}$ and $\pi \in \Pi$, we write $\pi(A) = (\pi(i))_{i \in A}$.
\begin{lemma}
	\label{basic inequality}
	For some positive integers $k_1,k_2 \in \{1,...,n\}$, $n \ge 2$, let $f:\{1,...,n\}^{k_1} \rightarrow \mathbf{R}$ and $g:\{1,...,n\}^{k_2} \rightarrow \mathbf{R}$ be given nonnegative maps. Then for any disjoint nonempty subsets $A_1, A_2 \subset \{1,...,n\}$ such that $|A| = k_1$ and $|A_2|= k_2$,
	\begin{eqnarray}
	\label{ineq65}
    && \frac{1}{|\Pi|}\sum_{\pi \in \Pi} f(\pi(A_1))g(\pi(A_2))  \\ \notag
	&& \le \min\left\{(k_1 + 1)^{k_2},(k_2 + 1)^{k_1}\right\}\frac{1}{|\Pi|}\sum_{\pi \in \Pi} f(\pi(A_1))\frac{1}{|\Pi|} \sum_{\pi \in \Pi} g(\pi(A_2)).
	\end{eqnarray}
\end{lemma}

\noindent \textbf{Proof: } Let $N_{(k)}$ be the set of $k$-tuples $(i_1,...,i_k)$ such that $i_1,...,i_k$ are from $\{1,...,n\}$ and all $i_1,...,i_k$ are distinct. We write the left hand side of (\ref{ineq65}) by (letting $k = k_1 +k_2$)
\begin{eqnarray*}
	&& \frac{(n-k)!}{n!} \sum_{(i_1,...,i_{k_1+k_2}) \in N_{(k_1+k_2)}} f(i_1,...,i_{k_1}) g(i_{k_1+1},...,i_{k_1+k_2})\\
	&=& \frac{(n-k)!}{n!} \sum_{(i_1,...,i_{k_1}) \in N_{(k_1)}} f(i_1,...,i_{k_1}) \sum_{(j_1,...,j_{k_2}) \in N_{(k_2)}: \{j_1,...,j_{k_2}\} \cap \{i_1,...,i_{k_1}\} = \varnothing} g(j_1,...,j_{k_2}).	
\end{eqnarray*}
Since $f$ and $g$ are nonnegative, we bound the last term by
\begin{eqnarray*}
	&& \frac{(n-k)!}{n!} \frac{n!}{(n-k_1)!}\frac{n!}{(n-k_2)!} \\
	&& \times \frac{(n-k_1)!}{n!} \sum_{(i_1,...,i_{k_1}) \in N_{(k_1)}} f(i_1,...,i_{k_1})
	\times \frac{(n-k_2)!}{n!} \sum_{(j_1,...,j_{k_2}) \in N_{(k_2)}} g(j_1,...,j_{k_2}) \\
	&=& \frac{(n-k)!}{n!} \frac{n!}{(n-k_1)!}\frac{n!}{(n-k_2)!}
	\frac{1}{|\Pi|} \sum_{\pi \in \Pi} f(\pi(A_1)) \frac{1}{|\Pi|} \sum_{\pi \in \Pi} g(\pi(A_2)).
\end{eqnarray*}
Note that
\begin{eqnarray*}
	\frac{(n-k)!}{n!} \frac{n!}{(n-k_1)!}\frac{n!}{(n-k_2)!} 
	&=&  \frac{(n-k)!}{(n-k_1)!}\frac{n!}{(n-k_2)!} \\
	&=& \prod_{j=0}^{k_2-1}\frac{n-j}{n-j-k_1} \le \left(\frac{n-k_2+1}{n-k_1-k_2 + 1}\right)^{k_2},
\end{eqnarray*}
where the last bound is obtained by replacing $j$ by $k_2-1$. Exchanging the roles of $k_1$ and $k_2$, and noting that $n \ge k_1 + k_2$, we obtain the upper bound. $\blacksquare$
\medskip

Let us introduce some notation. Let $W=\{W_{i,n}\}_{i=1}^n$ be a given triangular array of random variables.  For any $k \ge 1$ and $r = (r_1,...,r_k) \subset \{0,1,...,n\}^k$, we let
\begin{eqnarray*}
	B(k,r) \equiv \frac{1}{|\Pi|} \sum_{\pi \in \Pi} \left|\mathbf{E}[W_{\pi(1),n}^{r_1}W_{\pi(2),n}^{r_2}...W_{\pi(k),n}^{r_k}|\mathcal{C}_n] \right|.
\end{eqnarray*}
The following lemma is obtained by using a Doukhan-Portal type inequality (\cite{Doukhan/Portal:83:CRASPS}).
\begin{lemma}
	\label{moment bound1}
	Let $\{W_{i,n}\}_{i=1}^n$ be a triangular array of random variables having $\lambda_n(\cdot)$ as its $\lambda$-coefficient for some $\sigma$-field $\mathcal{C}_n$, and define for integers $n,k\ge 1$,
	\begin{eqnarray*}
		L_n(k)	\equiv \max_{1 \le i \le n} \mathbf{E}[|W_{i,n}|^k|\mathcal{C}_n].
	\end{eqnarray*} 
	
	Then for any integers $k,q \ge 1$, there exists constant $C_{k,q}>0$ such that $C_{k,q}$ depends only on $k$ and $q$, and for any $r=(r_1,...,r_k) \subset \{0,...,n\}^k$ with $r_1+...+r_k = q$,
	\begin{eqnarray}
	\label{bound45}
	\quad
	B(k,r) \le C_{k,q} \sum_{A \subset \{1,...,k\}} L_n\left(\sum_{j=1: j \notin A}^k r_j\right) \prod_{j \in A} B(1,r_j)  \max_{(v_1,...,v_k) \in J(k - |A|)} \prod_{j=1}^k \lambda_n(v_j),
	\end{eqnarray}	
	where $J_k(s) = \{(v_1,...,v_k) \in \{0,...,k\}^k: v_1+...+v_k = s\}$, and the product $\prod_{j \in A} B(1,r_j)$ is taken to be 1 if $A$ is empty.
\end{lemma}

\noindent \textbf{Proof: } Suppose that $k=1$. Then the inequality of the lemma trivially holds. Suppose that $k \ge 2$. For any vector $s = (s_1,...,s_k) \in \{1,...,n\}^k$ of distinct integers, we let 
\begin{eqnarray*}
	\Psi(s,r) \equiv \mathbf{E}\left[W_{s_1,n}^{r_1}W_{s_2,n}^{r_2}...W_{s_k,n}^{r_k}|\mathcal{C}_n\right].
\end{eqnarray*}
For each $\{A_1,A_2\} \in \mathcal{P}(\{1,...,k\})$, we let $\Pi(A_1,A_2) \subset \Pi$ be the collection of $\pi$'s such that
\begin{eqnarray*}
	c_n(\{\pi(1),...,\pi(k)\}) = c_n(\pi(A_1),\pi(A_2)),
\end{eqnarray*}
i.e., the collection of $\pi$'s such that $(\pi(A_1),\pi(A_2)) \in \mathcal{P}(\{\pi(1),...,\pi(k)\})$ is a minimizer of $c_n(A_1',A_2')$ over all $(A_1',A_2') \in \mathcal{P}(\{\pi(1),...,\pi(k)\})$. Using the fact $Cov(X,Y) = \mathbf{E}[XY] - \mathbf{E}[X]\mathbf{E}[Y]$, we bound
\begin{eqnarray*}
	B(k,r) \le \frac{1}{|\Pi|} \sum_{\{A_1,A_2\} \in \mathcal{P}(\{1,...,k\})} \sum_{\pi \in \Pi(A_1,A_2)}  |\Psi(\pi(A_1),r_{A_1})||\Psi(\pi(A_2),r_{A_2})| + R(k,r),
\end{eqnarray*}
where $r_A = (r_j)_{j \in A}$, $A \subset \{1,...,n\}$, and
\begin{eqnarray*}
	R(k,r) = \frac{1}{|\Pi|} \sum_{\{A_1,A_2\} \in \mathcal{P}(\{1,...,k\})} \sum_{\pi \in \Pi(A_1,A_2)} \left| Cov\left(\prod_{j \in A_1} W_{\pi(j),n}^{r_j}, \prod_{j \in A_2} W_{\pi(j),n}^{r_j}|\mathcal{C}_n\right)\right|.
\end{eqnarray*}

Note that by the choice of $\Pi(A_1,A_2)$,
\begin{eqnarray*}
	R(k,r) &\le& \frac{L_n(k)}{|\Pi|} \sum_{\{A_1,A_2\} \in \mathcal{P}(\{1,...,k\})} \sum_{\pi \in \Pi(A_1,A_2)} c_n(\{\pi(1),...,\pi(k)\})\\
	&=& \frac{L_n(k)}{|\Pi|} \sum_{\pi \in \Pi}c_n(\{\pi(1),...,\pi(k)\}) = L_n(k) \lambda_n(k).
\end{eqnarray*}
Using the bound $|\mathcal{P}(\{1,...,k\})| \le 2^k$, and Lemma \ref{basic inequality},
\begin{eqnarray*}
	&& \frac{1}{|\Pi|} \sum_{\{A_1,A_2\} \in \mathcal{P}(\{1,...,k\})} \sum_{\pi \in \Pi(A_1,A_2)}  |\Psi(\pi(A_1),r_{A_1})||\Psi(\pi(A_2),r_{A_2})|\\
	&\le&  (k+1)^k \sum_{\{A_1,A_2\} \in \mathcal{P}(\{1,...,k\})} \frac{1}{|\Pi|}\sum_{\pi \in \Pi} |\Psi(\pi(A_1),r_{A_1})|\frac{1}{|\Pi|}\sum_{\pi \in \Pi} |\Psi(\pi(A_2),r_{A_2})\\
	&=& (k+1)^k \sum_{k_1=1}^{k-1} \sum_{\{A_1,A_2\} \in \mathcal{P}(\{1,...,k\}):|A_1| = k_1} B(k_1,r_{A_1})B(k-k_1,r_{A_2}).
\end{eqnarray*}
We conclude that
\begin{eqnarray}
\label{recursive}
B(k,r) &\le& (k+1)^k\sum_{k_1=1}^{k-1} \sum_{\{A_1,A_2\} \in \mathcal{P}(\{1,...,k\}):|A_1| = k_1} B(k_1,r_{A_1})B(k-k_1,r_{A_2})\\ \notag
&& + L_n(k) \lambda_n(k).
\end{eqnarray}

We prove the inequality (\ref{bound45}) by induction. Define
\begin{eqnarray}
\label{bound12} \quad 
\overline B(k,r) \equiv \sum_{A \subset \{1,...,k\}} L_n\left(\sum_{j=1: j \notin A}^k r_j\right) \prod_{j \in A} B(1,r_j) \max_{(v_1,...,v_k) \in J(k-|A|)} \prod_{j=1}^k \lambda_n(v_j).
\end{eqnarray}
As we saw before, the inequality (\ref{bound45}) holds for $k=1$. Suppose that it holds for $k \ge 1$. Take arbitrary $r' = (r_1,...,r_{k+1}) \in \{0,...,n\}^{k+1}$ such that $r_1+...+r_{k+1} = q$. By (\ref{recursive}),
\begin{eqnarray*}
	B(k+1,r') &\le& (k+2)^{k+1}\sum_{k_1=1}^{k} \sum_{\{A_1,A_2\} \in \mathcal{P}(\{1,...,k+1\}):|A_1| = k_1} B(k_1,r_{A_1}')B(k+1-k_1,r_{A_2}')\\
	&& + L_n(k+1) \lambda_n(k+1).
\end{eqnarray*}
By the hypothesis of induction, we have
\begin{eqnarray}
\label{bound33}
B(k_1,r_{A_1}')B(k+1-k_1,r_{A_2}') \le C_{k,q} \overline B(k_1,r_{A_1}') \overline B(k+1 - k_1,r_{A_2}'),
\end{eqnarray}
for some constant $C_{k,q}$ that depends only on $k,q$. For each $A \subset \{1,...,n\}$ and $j =1,...,|A|$, let $r_{j,A}$ denote the $j$-th entry of $r_A$. As for the last bound in (\ref{bound33}), note that
\begin{eqnarray*}
	&& \overline B(k_1,r_{A_1}') \overline B(k+1 - k_1,r_{A_2}') \\
	&=& \sum_{A \subset \{1,...,k_1\}} \sum_{A' \subset \{1,...,k+1 - k_1\}} L_n\left(\sum_{j=1: j \notin A}^{k_1} r_j\right)L_n\left(\sum_{j=1: j \notin A'}^{k+1-k_1} r_j\right)  \left\{\prod_{j \in A}\prod_{j' \in A'} B(1,r_{j,A_1}')B(1,r_{j',A_2}') \right.\\
	&& \left. \times  \max_{(v_1,...,v_{k_1}) \in J(k_1 - |A|)}\max_{(v_1',...,v_{k+1 - k_1}') \in J(k + 1 - k_1 - |A'|)} \prod_{j=1}^{k_1} \prod_{j'=1}^{k+1-k_1} \lambda_n(v_j)\lambda_n(v_{j'}) \right\}\\
	&\le& C_{k,q}'\sum_{A \subset \{1,...,k+1\}} L_n\left(\sum_{j=1: j \notin A}^{k+1} r_j\right)\left\{\prod_{j \in A} B(1,r_j')  \max_{(v_1,...,v_{k+1}) \in J(k + 1 - |A|)} \prod_{j'=1}^{k+1} \lambda_n(v_{j'}) \right\}\\
	&=& C_{k,q}'\overline B(k+1,r'),
\end{eqnarray*}
for some constant $C_{k,q}'>0$ that depends only on $k,q$. Thus the proof is complete. $\blacksquare$
\medskip

\noindent \textbf{Proof of Lemma \ref{moment bound2}: } For brevity, we assume that $d=1$. The proof for the general case of $d \ge 1$ is similar. Define for positive integers $q$ and $k \in \{1,...,2q\}$,
\begin{eqnarray}
\label{An(r)}
A_n(k) = \{\mathbf{i} \in \{1,...,b_n\}^{2q}: \mathbf{i} \textnormal{ has exactly } k \textnormal{ different numbers.} \}. 
\end{eqnarray}
Let us write 
\begin{eqnarray}
\label{ineq5}
&& \frac{1}{a_n|\Pi|} \sum_{\pi \in \Pi} \mathbf{E} \left(\left\|\frac{1}{\sqrt{b_n}} \sum_{i=1}^{b_n} Z_{\pi(i),n} \right\|^{2q} |\mathcal{C}_n\right)\\
&\le& \frac{1}{a_n b_n^q} \sum_{k=1}^{2q} \sum_{(i_1,...,i_q,j_1,...,j_q) \in A_{n}(k)} \left| \frac{1}{|\Pi|} \sum_{\pi \in \Pi} \mathbf{E} \left[Z_{\pi(i_1),n}Z_{\pi(j_1),n}...Z_{\pi(i_q),n}Z_{\pi(j_q),n} |\mathcal{C}_n \right] \right|. \notag
\end{eqnarray}
For each $\mathbf{i} \equiv (i_1,...,i_q,j_1,...,j_q) \in A_n(k)$ having $(s_1,...,s_k)$ consisting of uniquely distinct integers in $\mathbf{i}$, with each $s_j$ appearing in $\mathbf{i}$ $r_j$ times, we let $r(\mathbf{i})$ be $(r_1,...,r_k)$ with such $r_j$'s. Using Lemma \ref{moment bound1} and recalling the definition (\ref{bound12}) (with $\{Z_{i,n}\}$ replacing $\{W_{i,n}\}$ there), we bound the term on the right hand side of (\ref{ineq5}) by
\begin{eqnarray}
\label{bound4}
\quad \quad
\frac{C_q}{a_n b_n^q} \sum_{k=1}^{2q} \sum_{\mathbf{i} \in A_{n}(k)} \overline B(k,r(\mathbf{i})),
\end{eqnarray}
for some constant $C_q>0$ that depends only on $q$ and $\bar C$. Note that
\begin{eqnarray*}
	\overline B(k,r(\mathbf{i})) &=& \sum_{A \subset \{1,...,k\}: |A| \le q} L_n\left(\sum_{j=1: j \notin A}^k r_j\right) \left\{\prod_{j \in A} B(1,r_j(\mathbf{i})) \max_{(v_1,...,v_k) \in J(k-|A|)} \prod_{j=1}^k  \lambda_n(v_j) \right\}\\
	&\le& L_n(2q) \sum_{A \subset \{1,...,k\}: |A| \le q} \max_{(v_1,...,v_k) \in J(k-|A|)} \prod_{j=1}^k  \lambda_n(v_j),
\end{eqnarray*}
where $r_j(\mathbf{i})$ denotes the $j$-th entry of $r(\mathbf{i})$. The restriction $|A| \le q$ in the sum over $A$ follows because for each $A \subset \{1,2,...,k\}$ with $|A|>q$, we have $\prod_{j \in A} B(1,r_j)=0$. This because for such $A$, there exists $j \in A$ such that $r_j  = 1$, i.e., there exists an integer, say, $j$, in $A$ which appears only once in the vector $\mathbf{i}$, and for such $j$, $B(1,r_j)= 0$ because $\mathbf{E}[Z_{i,n}|\mathcal{C}_n]=0$ for all $i \in \{1,...,n\}$. Hence
\begin{eqnarray*}
	&& \frac{1}{a_n b_n^q} \sum_{k = 2}^{2q} \sum_{\mathbf{i} \in A_{n}(k)} \overline B(k,r(\mathbf{i}))\\
	&\le&  L_n(2q) \sum_{k = 2}^{2q} \sum_{\mathbf{i} \in A_{n}(k)} \sum_{A \subset \{1,...,k\}: |A| \le q} b_n^{|A|-q} \max_{(v_1,...,v_k) \in J(k-|A|)} \prod_{j=1}^k  a_n^{-1} b_n^{v_j} \lambda_n(v_j)\\
	&\le& C_{k,q}'' L_n(2q),
\end{eqnarray*}
where $C_{k,q}''$ is a constant that depends only on $k,q$ and $\bar C$. The first inequality uses the fact that $|A_n(k)| \le b_n^k$. $\blacksquare$

\section{Proof of Theorem \ref{thm: comparison}}
\begin{lemma}
	\label{lemm: A}
	Suppose that $A$ and $B$ are $d \times d$ positive definite matrices such that $\|A - B\| \le c$, for some $0<c<\lambda_{\min}(B)/\sqrt{d}$, where $\|\cdot\|$ denotes the Frobenius norm, i.e, $\|A\| = \sqrt{\text{tr}(A A')}$. Then,
	\begin{eqnarray*}
		\| A^{-1} - B^{-1}\| \le \frac{c d \| A - B \|}{\lambda_{\min}(B)(\lambda_{\min}(B) - c \sqrt{d})}.
	\end{eqnarray*}
\end{lemma}

\noindent \textbf{Proof:} Since $A^{-1} - B^{-1} = A^{-1}(B - A) B^{-1}$, we obtain that
\begin{eqnarray*}
	\| A^{-1} - B^{-1}\| &\le& \left( \| A^{-1} - B^{-1} \| + \| B^{-1}\| \right) \| A - B\| \| B^{-1}\|\\
	&\le& c \sqrt{d} \lambda_{\min}(B) \| A^{-1} - B^{-1} \| +c d \lambda_{\min}^2(B),
\end{eqnarray*}
because $\|B^{-1}\|^2 \le d \lambda_{\min}^{-2}(B)$. $\blacksquare$
\medskip

\begin{lemma}
	\label{lemm: B}
	Suppose that there exist constants $M>0$, $v \in (0,1)$, and $\overline C>0$ and integers $q \ge 1$ and $s \ge 0$ such that $\max_{1 \le i \le n} \sup_{P \in \mathscr{P}_n} \mathbf{E}[\|X_{i,n}\|^{2q(s+1)}] \le M$ and $n^{k-vq} \sup_{P \in \mathscr{P}_n} \lambda_n(k) \le \overline C$ for all $n \ge 1$, and all $2 \le k \le 2q$.
	
	Then for any $0\le s' \le s$ and $M_1>0$ such that 
	\begin{eqnarray*}
		\max_{1 \le i \le n} \sup_{P \in \mathscr{P}_n} \mathbf{E}[\|X_{i,n}\|^{s'+2}] \le M_1
	\end{eqnarray*}
for all $n \ge 1$, there exists a constant $C>0$ such that
\begin{eqnarray*}
	\sup_{P \in \mathscr{P}_n} P\left\{ \frac{1}{n}\sum_{i=1}^n \|X_{i,n}\|^{s'+2} \ge 2 M_1  \right\} \le C n^{-q(1-v)}, \text{ for all } n \ge 1.
\end{eqnarray*}
\end{lemma}
	
\noindent \textbf{Proof:} Note that
\begin{eqnarray*}
	P\left\{\frac{1}{n}\sum_{i=1}^n \|X_{i,n}\|^{s'+2} \ge 2 M_1  \right\} 
	\le P\left\{\frac{1}{n}\sum_{i=1}^n \left( \|X_{i,n}\|^{s'+2} - \mathbf{E} \|X_{i,n}\|^{s'+2}  \right) \ge  M_1  \right\}.
\end{eqnarray*}
By Markov's inequality, the last probability is bounded by
\begin{eqnarray*}
	\frac{n^{-q(1-v)}}{M_1} \mathbf{E}\left[ \left| \frac{1}{n^{(1+v)/2}}\sum_{i=1}^n \left( \|X_{i,n}\|^{s'+2} - \mathbf{E} \|X_{i,n}\|^{s'+2} \right)\right|^{2q}\right] \le \frac{C n^{-q(1-v)}}{M_1},
\end{eqnarray*}
by Lemma \ref{moment bound2} (taking $b_n = n$ and $a_n = n^{vq}$ there), where $C$ depends only on $d,q,\bar C$ and $M$. $\blacksquare$

\begin{lemma}
	\label{lemm: C}
	Suppose that there exist constants $M>0$ and $\overline C>0$ such that
	\begin{eqnarray*}
		\max_{1 \le i \le n} \sup_{P \in \mathscr{P}_n} \mathbf{E}[\|X_{i,n}\|^4] \le M
	\end{eqnarray*}
	and $n \sup_{P \in \mathscr{P}_n} \lambda_n(2) \le \overline C$ for all $n \ge 1$.
	
	Then there exists $C>0$ such that for all $ \varepsilon >0$ and all $n \ge 1$,
	\begin{eqnarray*}
		 \sup_{P \in \mathscr{P}_n} P\left\{\|\hat \Sigma - \Sigma_n\| > \varepsilon \right\} \le \frac{C}{n \varepsilon}.
	\end{eqnarray*}
\end{lemma}

\noindent \textbf{Proof:} As in the proof of Lemma \ref{Cov Mat}, we write
\begin{eqnarray*}
	\hat \Sigma - \Sigma_n = \Sigma_n^{1/2} D_n \Sigma_n^{1/2},
\end{eqnarray*}
where $D_n = \frac{1}{n}\sum_{i=1}^n \left(Z_{i,n}Z_{i,n}' - \mathbf{E}[Z_{i,n}Z_{i,n}'|\mathcal{C}_n]\right) - \overline{Z}_n\overline{Z}_n'$. Note that
\begin{eqnarray*}
	 && \sup_{P \in \mathscr{P}_n} \mathbf{E}\left[ \left\| \frac{1}{n}\sum_{i=1}^n \left(Z_{i,n}Z_{i,n}' - \mathbf{E}[Z_{i,n}Z_{i,n}'|\mathcal{C}_n]\right) \right\|^2 \right]\\
	&\le& C   \max_{1 \le i \le n} \sup_{P \in \mathscr{P}_n}  \mathbf{E}\|Z_{i,n}\|^4 \left(n^{-1} +\sup_{P \in \mathscr{P}_n} \lambda_n(2)\right).
\end{eqnarray*}
We obtain a similar bound for $\overline Z_n \overline Z_n'$. $\blacksquare$.

\begin{lemma}
	\label{lemm: D}
	Under the conditions of Theorem \ref{thm: comparison}, the statements below hold. 
	
	(i) There exist $M'>0$ and $C>0$ such that for all $0 \le s' \le s$,
	\begin{eqnarray}
	\label{bound233}
		\sup_{P \in \mathscr{P}_n} P\left\{ \frac{1}{|\Pi|}\sum_{\pi \in \Pi} \|M_n(\overline X_n;\pi)\|^{s'+2} \ge M'  \right\} \le C n^{-1}, \text{ for all } n \ge n'.
	\end{eqnarray}

    (ii) There exist $M'>0$ and $C>0$ such that for all $0 \le s' \le s$,
\begin{eqnarray*}
	\sup_{P \in \mathscr{P}_n} P\left\{ \frac{1}{|\Pi|}\sum_{\pi \in \Pi} \|U_n(\overline X_n;\pi)\|^{s'+2} \ge M'  \right\} \le C n^{-1}, \text{ for all } n \ge n'.
\end{eqnarray*}
\end{lemma}

\noindent \textbf{Proof:} (i) We bound the probability on the right hand side of (\ref{bound233}) by
\begin{eqnarray*}
    && P\left\{ \frac{1}{|\Pi|}\sum_{\pi \in \Pi} \left\|\frac{1}{\sqrt{d b_n}}\sum_{i=1}^{b_n} \hat \Sigma^{-1/2} (X_{\pi(i),n} - \overline X_n)\right\|^{s'+2} \ge M'  \right\} \\
	&\le&  P\left\{ d^{-(s'+2)/2} b_n^{(s'+2)/2}\frac{1}{n} \sum_{i=1}^n  \left\| \hat \Sigma^{-1/2} (X_{i,n} - \overline X_n)\right\|^{s'+2} \ge M'  \right\} \\
	&\le& 2 P\left\{ 2^{s'+1} d^{-(s'+2)/2} b_n^{(s'+2)/2}\frac{1}{n} \sum_{i=1}^n  \left\| \hat \Sigma^{-1/2} X_{i,n}\right\|^{s'+2} \ge \frac{M'}{2}  \right\} \le 2A_{1n} + 2A_{2n},
\end{eqnarray*}
where
\begin{eqnarray*}
	A_{1n} &=& P\left\{ \frac{1}{n} \sum_{i=1}^n  \left\| \Sigma_n^{-1/2} X_{i,n}\right\|^{s'+2} \ge M'/4  \right\}, \text{ and }\\
	A_{2n} &=& P\left\{ \frac{1}{n} \sum_{i=1}^n  \left\| (\hat \Sigma^{-1/2} -\Sigma_n^{-1/2}) X_{i,n}\right\|^{s'+2} \ge M'/4  \right\}.
\end{eqnarray*}
We bound $A_{1n}$ by
\begin{eqnarray*}
	 P\left\{ \frac{1}{n} \sum_{i=1}^n  \|\Sigma_n^{-1/2}\|^{s'+2} \left\|  X_{i,n}\right\|^{s'+2} \ge M'/4  \right\}
	 \le P\left\{ \frac{1}{n} \sum_{i=1}^n  \left\|  X_{i,n}\right\|^{s'+2} \ge \frac{M'}{4d^{s'+2}(\lambda_{\min}(\Sigma_n^{-1/2}))^{s'+2}} \right\},
\end{eqnarray*}
where $M'''$ is a constant that depends only on $M''$ and $\inf_{n \ge 1}\inf_{P \in \mathscr{P}_n} \lambda_{\min}(\Sigma_n)$. This probability is bounded by $Cn^{-1}$ by Lemma \ref{lemm: B} (with $v = 1-(1/q)$), if we choose $M'$ large enough so that
\begin{eqnarray*}
	\frac{M'}{4d^{s'+2}(\inf_{P \in \mathscr{P}_n}\lambda_{\min}(\Sigma_n^{-1/2}))^{s'+2}} > 2M_1.
\end{eqnarray*}

Now, let us turn to $A_{2n}$. We bound for any constant $c>0$ such that $c<\lambda_{\min}(\Sigma_n^{1/2})/\sqrt{d}$,
\begin{eqnarray*}
	A_{2n} &\le& P\left\{\left(\frac{1}{n}\sum_{i=1}^n\|X_{i,n}\|^{s+2} \right)\|\hat \Sigma^{-1/2} - \Sigma_n^{-1/2}\|^{s+2} \ge M'' \right\}\\
	&\le& P\left\{\left(\frac{1}{n}\sum_{i=1}^n\|X_{i,n}\|^{s+2} \right) c^{s+2} \ge M''  \right\} + P\mathscr{E}_n^c(c) \le C n^{-1} +  P\mathscr{E}_n^c(c),
\end{eqnarray*}
for some constant $C>0$ by Lemma \ref{lemm: B} again, where $\mathscr{E}_n^c(c) \equiv \{\|\hat \Sigma^{-1/2} - \Sigma_n^{-1/2}\| \le c\}$. By choosing $c$ small enough and using Lemmas \ref{lemm: A} amd \ref{lemm: C}, we can bound the last probability by $Cn^{-1}$ for some constant $C>0$.\medskip

\noindent (ii) First, we observe that
\begin{eqnarray*}
	&& \frac{1}{n^2}\sum_{i,j=1}^n \left|tr\left((X_{\pi(i)} - \overline X_n)(X_{\pi(j)} - \overline X_n)'\hat \Sigma ^{-1}\right)\right|^{s+2}\\
	&=& \frac{1}{n^2}\sum_{i,j=1}^n \left(\|X_{i,n}\|\|X_{j,n}\| +(\|X_{i,n}\| + \|X_{j,n}\|)\|\overline X_n\| + \|\overline X_n\|^2 \right)^{s+2}\|\hat \Sigma^{-1}\|^{s+2}\\
	&\le& 4^{s+1} \left(\frac{1}{n}\sum_{i=1}^n \|X_{i,n}\|^{s+2}\right)^2 \| \hat \Sigma^{-1}\|^{s+2}.
\end{eqnarray*}

We bound 
\begin{eqnarray*}
	&& P\left\{ \frac{1}{|\Pi|}\sum_{\pi \in \Pi} \|U_n(\overline X_n;\pi)\|^{s'+2} \ge M'  \right\} \\
	&\le&  P\left\{ \left(\frac{b_n(b_n-1)}{d}\right)^{s+2} \frac{1}{n(n-1)}\sum_{i,j=1,i \ne j}^n \left|tr\left((X_{\pi(i)} - \overline X_n)(X_{\pi(j)} - \overline X_n)'\hat \Sigma ^{-1}\right) \right|^{s+2} \ge M'  \right\}\\
	&\le& P\left\{ 2\left(\frac{b_n(b_n-1)}{d}\right)^{s+2}4^{s+1} \left(\frac{1}{n}\sum_{i=1}^n \|X_{i,n}\|^{s+2}\right)^2 \| \hat \Sigma^{-1}\|^{s+2} \ge \frac{n(n-1)M'}{2n^2}  \right\} \le B_{1n} + B_{2n},
\end{eqnarray*}
where
\begin{eqnarray*}
	B_{1n} &=& P\left\{ 2\left(\frac{b_n(b_n-1)}{d}\right)^{s+2}8^{s+1} \left(\frac{1}{n}\sum_{i=1}^n \|X_{i,n}\|^{s+2}\right)^2 \|\Sigma^{-1}\|^{s+2} \ge \frac{n(n-1)M'}{4n^2}  \right\}, \text{ and }\\
	B_{2n} &=& P\left\{ 2\left(\frac{b_n(b_n-1)}{d}\right)^{s+2}8^{s+1} \left(\frac{1}{n}\sum_{i=1}^n \|X_{i,n}\|^{s+2}\right)^2 \|\hat \Sigma^{-1} - \Sigma^{-1}\|^{s+2} \ge \frac{n(n-1)M'}{4n^2}  \right\}.
\end{eqnarray*}
We can show that both $B_{1n}$ and $B_{2n}$ are bounded by $Cn^{-1}$ using the same arguments as in the proof of (i).
$\blacksquare$
\medskip

Let us define
\begin{eqnarray*}
	\hat \Omega_{M,n} = \frac{1}{|\Pi|}\sum_{\pi \in \Pi} M_n(\overline X_n;\pi)M_n(\overline X_n;\pi)', \text{ and }
	\hat \Omega_{U,n} = \frac{1}{|\Pi|}\sum_{\pi \in \Pi} U_n^2(\overline X_n;\pi).
\end{eqnarray*}
\begin{lemma}
	\label{lemm: E}
	Suppose that the conditions of Theorem \ref{thm: comparison} hold. Then, for any $\varepsilon_1>0$,
	\begin{eqnarray*}
		\sup_{P \in \mathscr{P}_n} P\left\{\lambda_{\min}(\hat \Omega_{M,n}) \le \varepsilon_1 \right\} &\le& M' n^{-1}, \text{ and }\\
		\sup_{P \in \mathscr{P}_n} P\left\{\lambda_{\min}(\hat \Omega_{U,n}) \le \varepsilon_1 \right\} &\le& M' n^{-1},
	\end{eqnarray*}
for all $n \ge n'$.
\end{lemma}

\noindent \textbf{Proof: } Note that for any $\varepsilon_1>0$,
\begin{eqnarray*}
	P\left\{\lambda_{\min}(\hat \Omega_{M,n}) \le \varepsilon_1 \right\}
	&\le& P\left\{\frac{1}{\varepsilon_1} \le \|\hat \Omega_{M,n}\| \right\} \\
	&\le& P\left\{\frac{1}{\varepsilon_1} \le \frac{1}{|\Pi|}\sum_{\pi \in \Pi} \|M_n(\overline X_n; \pi)\|^2 \right\} \le Cn^{-1}, 
\end{eqnarray*}
for some constant $C>0$ by taking sufficiently small $\varepsilon_1$, by Lemma \ref{lemm: D}(i). We can deal with the second statement similarly. $\blacksquare$
\medskip

\begin{lemma}
	\label{lemm: Smootheness}
	Suppose that the conditions of Theorem \ref{thm: comparison} hold. Then,
	\begin{eqnarray*}
		&& \sup_{t \in \mathbf{R}}\sup_{P \in \mathscr{P}_n}\left|P\{S_{M,n}(\overline X_n;\pi) \le t |\mathscr{Z}_n\}
		- G_{M,n}(t)\right| = O_P(R_n^{-(2s-3)/4}), \text{ and }\\
		&& \sup_{t \in \mathbf{R}}\sup_{P \in \mathscr{P}_n}\left|P\{S_{U,n}(\overline X_n;\pi) \le t|\mathscr{Z}_n\}
		- G_{M,n}(t)\right| = O_P(R_n^{-(2s-3)/4}),
	\end{eqnarray*}
	as $n \rightarrow \infty$, where $G_{M,n}: \mathbf{R} \rightarrow [0,1]$ and $G_{U,n}: \mathbf{R} \rightarrow [0,1]$ are random function such that for each $t \in \mathbf{R}$, $G_{M,n}(t)$ and $G_{U,n}(t)$ are $\mathscr{Z}_n$-measurable, and there exist $C>0$ and $n_1\ge 1$ such that for each $\eta>0$ and $n \ge n_1$,
	\begin{eqnarray*}
		\sup_{t \in \mathbf{R}}\sup_{P \in \mathscr{P}_n}|\mathbf{E}[G_{M,n}(t) - G_{M,n}(t + \eta)]| + |\mathbf{E}[G_{U,n}(t) - G_{U,n}(t + \eta)]| \le C (\eta + R_n^{-(2s-3)/4} + n^{-1}).
	\end{eqnarray*}
\end{lemma}

\noindent \textbf{Proof: } There exists $M>0$ such that the complement of the event that
\begin{eqnarray*}
	\lambda_{\min}(\hat \Omega_{M,n}) \le M, \text{ and } \lambda_{\min}(\hat \Omega_{U,n}) &\le& M,
\end{eqnarray*}
and
\begin{eqnarray*}
	\frac{1}{|\Pi|}\sum_{\pi \in \Pi} \|M_n(\overline X_n;\pi)\|^{s'+2} \le M, \text{ and } 
		\frac{1}{|\Pi|}\sum_{\pi \in \Pi} \|U_n(\overline X_n;\pi)\|^{s'+2} \le M
\end{eqnarray*}
has probability vanishing to zero at the rate of $O(n^{-1})$ uniformly over $P \in \mathscr{P}_n$, by Lemmas \ref{lemm: D} and \ref{lemm: E}. The result then follows from Theorem 3.2 and Corollary 3.1 of \cite{Song:18:WP}. (We take $\delta = (2s-3)/4$, $a = 2s-2$ there.) $\blacksquare$

\begin{lemma}
	\label{lemm: Validity}
	Suppose that the conditions of Theorem \ref{thm: comparison} hold. Then
	\begin{eqnarray*}
		\sup_{t \in \mathbf{R}}\sup_{P \in \mathscr{P}_n}\left|P\{T_{M,n}(\mu;\pi) > c_{M,\alpha}\} - \alpha \right| &=& O\left(\sqrt{\frac{R_n b_n}{n}}\right), \text{ and }\\
		\sup_{t \in \mathbf{R}}\sup_{P \in \mathscr{P}_n}\left|P\{T_{U,n}(\mu;\pi) > c_{U,\alpha}\} - \alpha \right| &=& O\left(\frac{\sqrt{R_n} b_n}{n} + \sqrt{\frac{b_n}{n}}\right),
	\end{eqnarray*}
as $n,L \rightarrow \infty$. Furthermore, the terms $O(\sqrt{R_n b_n/n})$ and $O(\sqrt{R_n} b_n/n + \sqrt{b_n/n})$ on the right hand side of the equations are not $o(\sqrt{R_n b_n/n})$ and $o(\sqrt{R_n} b_n/n + \sqrt{b_n/n})$.
\end{lemma}

\noindent \textbf{Proof: } As for the first result, we write $P\{T_{M,n}(\mu;\pi) > c_{M,\alpha}|\mathscr{Z}_n\} - \alpha$ as 
\begin{eqnarray*}
     && P\left\{S_{M,n}(\overline X_n;\pi) > c_{M,\alpha} + B_{M,n} + \frac{R_n b_n}{n}|\mathscr{Z}_n\right\} - P\left\{S_{M,n}(\overline X_n;\pi) > c_{M,\alpha}|\mathscr{Z}_n\right\} \\
     &=& O_P\left( B_{M,n} + \frac{R_n b_n}{n} + R_n^{-(2s-3)/4}\right),
\end{eqnarray*}
by Lemma \ref{lemm: Smootheness}. Note that
\begin{eqnarray*}
	B_{M,n} +\frac{R_n b_n}{n} = \frac{R_n b_n}{n} \left(1- n \overline Z_n' \overline Z_n\right) - 2A_{M,n}.
\end{eqnarray*}
As for $A_{M,n}$, we note that
\begin{eqnarray*}
	\mathbf{E}[A_{M,n}^2] = \mathbf{E}\left[ \|\overline Z_n\|^2 \mathbf{E}\left[ \left\| \sum_{r=1}^{R_n}\sum_{i=1}^{b_n}(Z_{\pi(i)} - \overline Z_n)\right\|^2|\mathscr{Z}_n\right]\right].
\end{eqnarray*}
Using the same arguments in the proof of Lemma \ref{Tn_B}, we find that there exists a constant $C>0$ such that for all $n \ge 1$,
\begin{eqnarray*}
	 \mathbf{E}\left[ \left\| \sum_{r=1}^{R_n}\sum_{i=1}^{b_n}(Z_{\pi(i)} - \overline Z_n)\right\|^2|\mathscr{Z}_n\right] \le C R_n b_n,
\end{eqnarray*}
and $\mathbf{E}\left[\|\overline Z_n\|^2\right] \le C (n^{-1} + \lambda_n(2))$. By the assumptions of the lemma,
\begin{eqnarray*}
	\sup_{P \in \mathscr{P}_n} \mathbf{E}[A_{M,n}^2] = O\left( \frac{R_n b_n}{n} \right).
\end{eqnarray*}
Since $\sup_{P \in \mathscr{P}_n}|1 - n \mathbf{E}[\overline Z_n' \overline Z_n]| = O(1)$, 
\begin{eqnarray*}
	\sup_{P \in \mathscr{P}_n} \mathbf{E}\left[\left|B_{M,n} + \frac{R_n b_n}{n}\right| \right] = O\left(\sqrt{\frac{R_n b_n}{n}}\right).
\end{eqnarray*}

As for the second result, similarly as before, we write $P\{T_{U,n}(\mu;\pi) > c_{M,\alpha}\} - \alpha$ as 
\begin{eqnarray*}
	&& P\left\{S_{U,n}(\overline X_n;\pi) > c_{U,\alpha} + B_{M,n} + \frac{R_n b_n}{n}|\mathscr{Z}_n\right\} - P\left\{S_{U,n}(\overline X_n;\pi) > c_{U,\alpha}|\mathscr{Z}_n\right\} \\
	&=& O_P\left( B_{U,n} + \frac{\sqrt{R_n}(b_n-1)}{n} + R_n^{-(2s-3)/4}\right),
\end{eqnarray*}
by Lemma \ref{lemm: Smootheness}. We write
\begin{eqnarray*}
	B_{U,n} +\frac{\sqrt{R_n} (b_n-1)}{n} = \frac{\sqrt{R_n}(b_n-1)}{n} \left(1- n \overline Z_n' \overline Z_n\right) - 2A_{U,n},
\end{eqnarray*}
where
\begin{eqnarray*}
	A_{U,n} = \frac{\overline Z_n'\sqrt{b_n}(b_n-1)}{b_n} \frac{1}{\sqrt{R_n b_n}} \sum_{r=1}^{R_n}\sum_{i=1}^{b_n} (Z_{\pi_r(i)} - \overline Z_n).
\end{eqnarray*}
Using similar arguments as before, we find that
\begin{eqnarray*}
	\sup_{P \in \mathscr{P}_n} \mathbf{E}[A_{U,n}^2] = O\left(\frac{b_n}{n}\right),
\end{eqnarray*}
and obtain the first result of the lemma. Given the arguments in the proof of Theorem \ref{thm: Type U}, it is not hard to show that the terms $O(\sqrt{R_n b_n/n})$ and $O(\sqrt{R_n} b_n/n + \sqrt{b_n/n})$ in the lemma are not $o(\sqrt{R_n b_n/n})$ and $o(\sqrt{R_n} b_n/n + \sqrt{b_n/n})$. Details are omitted. $\blacksquare$
\medskip

\noindent \textbf{Proof of Theorem \ref{thm: comparison}: } Define
\begin{eqnarray*}
	\omega_n = \sqrt{\frac{R_{M,n} b_{M,n}}{n}} = \frac{\sqrt{R_{U,n}} b_{U,n}}{n}+\sqrt{\frac{b_{U,n}}{n}}.
\end{eqnarray*}
Then by Lemma \ref{lemm: Validity}, both tests $(T_{M,n},c_{M,\alpha})$ and $(T_{U,n},c_{U,\alpha})$ are asymptotically exact at the rate $\omega_n$. From the local power results in Theorem \ref{thm: local power analysis}, the rates of the two tests are given by $\delta_n^*(T_{M,n},c_{M,\alpha}) = 1/(\sqrt{n}\omega_n)$ and $\delta_n^*(T_{U,n},c_{U,\alpha}) = 1/(R_{U,n}^{1/4}\sqrt{n}\omega_n)$. Since $R_{U,n} \rightarrow \infty$, we obtain that the test $(T_{U,n},c_{U,\alpha})$ rate-dominates $(T_{M,n},c_{M,\alpha})$ with size control at $\omega_n$. $\blacksquare$
\end{document}